\documentclass[12pt]{article}
\newcommand{\mez}{\hspace{+0.5cm}}
\newcommand{\mz}{\hspace{+0.25cm}}
\newcommand{\m}{\hspace*{-0.30mm}}
\newcommand{\n}{\hspace*{-0.20mm}}
\newcommand{\be}{\begin{equation}}
\newcommand{\ee}{\end{equation}}
\newcommand{\la}{\langle}
\newcommand{\ra}{\rangle}

\usepackage{bm,amssymb,color,graphicx}
\usepackage{amssymb,amsmath,amsfonts,amsthm,mathtools}
\definecolor{orange}{rgb}{1,0.5,0}

\begin{document}

\begin{centering}
\vspace*{+4.00cm}
{\Large \bf {\color{blue} An Introduction to Renormalization in Atomic Physics}}\\
\vspace*{+0.50cm}
{\bf Lecture notes for the $\bm 2^{\bf nd}$ International Summer School on Advanced Quantum Mechanics}\\ {\bf (Prague, September 02-11, 2021)}\\
\vspace*{+1.00cm}
{\color{blue} {\sl Milan \v{S}indelka}}\\
\vspace*{+0.30cm}
{\it Laser Plasma Department, Institute of Plasma Physics of the Czech Academy of Sciences,\\
U Slovanky 2525/1a, 18200 Prague 8, Czech Republic\\ {\tt sindelka@ipp.cas.cz} }\\
\vspace*{+1.00cm}
{\it (\today)}\\
\vspace*{+1.50cm}
\end{centering}

\mez 
The purpose of these notes is to provide a pedagogical introduction to the concept of renormalization in atomic physics.
We study quantum dynamics of a model of a nonrelativistic single electron atom coupled to the quantum radiation field in the dipole approximation. An interaction between
the electron and the radiation field is regularized, using a suitable cutoff prescription which eliminates the coupling of the 
electron to the highly ultraviolet (UV) field modes. Subsequently, we analyze the corresponding Heisenberg picture equations of motion, and focus on behavior of physical observables in the limit of a pointlike electron (i.e., in the limit when the above mentioned UV
cutoff is gradually lifted to infinity). We identify the radiation reaction force acting on the electron, and show that this force
actually diverges towards infinity when gradually removing the UV cutoff.
Such an analysis leads ultimately to an unique renormalization prescription for the electron mass, as well as to determination of the Abraham-Lorentz force acting on the electron. Dependence of the just sketched renormalization procedure upon the spatial dimension is highlighted. Finally, we present an example calculation of the atomic level shifts by means of the renormalized perturbation theory,
and formulate also the renormalized mean field theory.

\newpage

{\color{blue} {\large \bf A. Basic theoretical setup}}
\vspace*{+0.20cm}
\begin{itemize}
\item \underline{\bf A.1 Introducing our model problem (preliminaries)}\\
      Our theoretical model lives in the $d$-dimensional Euclidean space ${\mathbb R}^d$\m.\\
      We shall consider the cases of $d=2,3$.\\
      {\sl Subsystem \#1:}\\
      Model of a nonrelativistic atom containing a single electron. Defined by the usual Hamiltonian
      \be \label{H-A-def}
         \hat{H}_{\m{\rm A}} \; = \; \frac{\hat{\bm p}^2}{2\,m} \; + \; V\m(\hat{\bm x}) \mez .
      \ee
      Explanations:\\
      $m$ = physical mass of the electron.\\
      $\hat{\bm x}$ and $\hat{\bm p}$ = standard canonical observables\\
      \phantom{$\hat{\bm x}$ and $\hat{\bm p}$ = }corresponding to the position and momentum of the electron.\\
      $V\m({\bm x})$ = an external scalar potential acting on the electron.\\
      \phantom{$V\m({\bm x})$ = }We assume that $V\m({\bm x})$ is strictly bound (diverging to $+\infty$ at spatial infinities).\\
      Note that the electron spin is ignored in our subsequent considerations.\\
      {\sl Subsystem \#2:}\\
      Quantized radiation field (continuous ensemble of harmonic oscillators).\\ Defined by the Hamiltonian
      \be \label{H-R-def}
         \hat{H}_{{\rm R}} \; = \; \int_{{\mathbb R}^d} \m {\rm d}^d\n k \, \sum_{\wp} \, \hbar\omega_{k} \; \hat{a}_{{\bm k}\wp}^\dagger \, \hat{a}_{{\bm k}\wp} \mez .
      \ee
      Explanations:\\
      $({\bm k}\wp)$ = modes of the radiation field, carrying modal frequencies $\omega_{k}=c|{\bm k}|$ and polarization $\wp$.\\
      \phantom{$({\bm k}\wp)$ =} $[\,$The electromagnetic field modes are transverse. No longitudinal photons.\\
      \phantom{$({\bm k}\wp)$ = $[\,$}Thus, for each ${\bm k} \in {\mathbb R}^d$\m,
      there exists just a single polarization in the case of $d=2$,\\
      \phantom{$({\bm k}\wp)$ = $[\,$}and two linearly independent mutually orthogonal polarizations in the case of $d=3$.$\,]$\\
      $\hat{a}_{{\bm k}\wp}$ = modal annihilators, characterized by their standard commutation properties
      \be \label{hat-a-commutators}
         \mez \Bigl[ \, \hat{a}_{{\bm k}\wp} \, , \, \hat{a}_{{\bm k}'\n\wp'\n}\, \Bigr] \; = \; \hat{0} \mz , \mz
         \Bigl[ \, \hat{a}_{{\bm k}\wp} \, , \, \hat{a}_{{\bm k}'\n\wp'\n}^\dagger\, \Bigr] \; = \; \delta^d\m({\bm k}-{\bm k}'\n) \; \delta_{\wp\wp'\n} \; \hat{1} \mz , \mz
         \Bigl[ \, \hat{a}_{{\bm k}\wp} \, , \, \hat{\bm x} \, \Bigr] \; = \;\Bigl[ \, \hat{a}_{{\bm k}\wp} \, , \, \hat{\bm p} \, \Bigr] \; = \; \hat{0} \mz .
      \ee
      \phantom{$\hat{a}_{{\bm k}\wp}$ = }The field mode $({\bm k}\wp)$ contains a zero point energy (ZPE) equal to $\hbar\omega_k/2$,\\
      \phantom{$\hat{a}_{{\bm k}\wp}$ = }this contribution has been discarded in (\ref{H-R-def}),\\
      \phantom{$\hat{a}_{{\bm k}\wp}$ = }in spite of the fact that the total ZPE $\int_{{\mathbb R}^d} \m {\rm d}^d\n k \, \sum_{\wp} \, \hbar\omega_{k}/2$ is infinite.\\
      $\hat{\bm A}({\bm y})$ = the Coulomb gauge vector potential, given by its standard modal expansion
      \be \label{hat-A-def}
         \hat{\bm A}({\bm y}) \; = \; \int_{{\mathbb R}^d} \m {\rm d}^d\n k \, \sum_\wp \, \frac{1}{2\,\pi} \, \sqrt{\frac{\hbar\,c^2}{\omega_k}} \;\,
         e^{+i{\bm k}\bm\cdot{\bm y}} \, \hat{a}_{{\bm k}\wp} \; {\bm \varepsilon}_{{\bm k}\wp} \; + \; {\rm c.c.} \mez .
      \ee
      ${\bm \varepsilon}_{{\bm k}\wp}$ = unit polarization vector of the field mode $({\bm k}\wp)$.\\
      \phantom{${\bm \varepsilon}_{{\bm k}\wp}$ = }Vectors ${\bm \kappa}={\bm k}/k$ and
      $\{{\bm \varepsilon}_{{\bm k}\wp}\}_\wp$ form an orthonormal basis set of ${\mathbb R}^d$\m. Closure property
      \be \label{kappa-closure}
         {\bm \kappa} \, {\bm \kappa}^T \; + \;
         \sum_{\wp} \, {\bm \varepsilon}_{{\bm k}\wp} \, {\bm \varepsilon}_{{\bm k}\wp}^{T} \; = \; 
         {\mathbb I} \mez .
      \ee
      \phantom{${\bm \varepsilon}_{{\bm k}\wp}$ = }Here ${\mathbb I}$ stands of course for the $d$-by-$d$ unit matrix.\\
      {\sl Subsystems \#1 and \#2 put together:}\\
      When the subsystems \#1 and \#2 are mutually uncoupled (non-interacting),\\
      the total Hamiltonian of our problem is of course just a sum
      \be \label{H-free-def}
         \hat{H}_{\rm free} \; = \; \hat{H}_{\m{\rm A}} \; + \; \hat{H}_{{\rm R}} \mez .
      \ee
\item \underline{\bf A.2 Introducing our model problem (coupling and related matters)}\\
      {\sl Interacting subsystems \#1 and \#2:}\\
      Nonrelativistic single electron atom coupled\\ to the quantum radiation field in the dipole approximation.\\
      Schr\"{o}dinger picture. The minimum coupling Hamiltonian
      \be \label{H}
         {\color{blue}
         \hat{H}_t \; = \; \frac{1}{2\,m_o^t(\alpha)} \left( \hat{\bm p} \, - \, \frac{q_t}{c}
         \int_{{\mathbb R}^d} \m \varrho({\bm y}) \; \hat{\bm A}({\bm y}) \; {\rm d}^d\n y \right)^{\m\m\m\m 2}
         + \; V\m(\hat{\bm x}) \; + \; \hat{H}_{{\rm R}} \mez .}
      \ee
      Explanations:\\
      $q_t$ = electric charge of the electron, varying infinitely slowly (adiabatically) in time.\\
      \phantom{$q_t$ = }Vanishing in the infinite past, and reaching its physical value $q$ at finite times.\\
      \phantom{$q_t$ = }Stated mathematically,
      \be \label{dot-q-t-zero}
         \bm\dot{q}_t \; \equiv \; \frac{{\rm d}q_t}{{\rm d}t} \; \to \; 0 \mez ;
      \ee
      \be \label{q-t-infinite-past}
         q_{t\to-\infty} \; = \; 0 \mez .
      \ee
      $\varrho({\bm y})$ = charge density profile of the electron, we shall assume an unit normalized Gaussian
      \be \label{varrho-def}
         \varrho({\bm y}) \; = \; \left(\frac{\alpha}{\pi}\right)^{\m\m\frac{d}{2}} e^{-\alpha{\bm y}^2} \mez . \mez [\,\alpha>0\,]
      \ee
      $\alpha$ = positive real parameter serving as an UV regulator\\
      \phantom{$\alpha$ = }(this terminology will be clarified shortly, see equation (\ref{int-varrho-A}) below).\\
      \phantom{$\alpha$ = }$[\,$Later on we shall be concerned with the limit of $\alpha\to+\infty$
      for which $\varrho({\bm y}) \to \delta^d\n({\bm y})$.\\
      \phantom{$\alpha$ = }\phantom{$[\,$}The limit of $\alpha\to+\infty$ will be taken only after the adiabatic limit
      concerning $q_t$.$\,]$\\
      $m_o^t(\alpha)$ = bare mass of the electron,\\
      \phantom{$m_o^t(\alpha)$ = }depending in an as yet unspecified manner upon the UV regulator $\alpha$.\\
      \phantom{$m_o^t(\alpha)$ = }Varying infinitely slowly (adiabatically) with time,\\
      \phantom{$m_o^t(\alpha)$ = }again in an as yet unspecified manner,\\
      \phantom{$m_o^t(\alpha)$ = }and reducing to the physical electron mass $m$ in the infinite past.\\
      \phantom{$m_o^t(\alpha)$ = }Stated mathematically,
      \be \label{dot-m-alpha-zero}
         \bm\dot{m}_o^t(\alpha) \; \equiv \; \frac{{\rm d}\,m_o^t(\alpha)}{{\rm d}t} \; \to \; 0 \mez ;
      \ee
      \be \label{m-alpha-infinite-past}
         m_o^{t \to -\infty}\n(\alpha) \; = \; m \mez .
      \ee
      \phantom{$m_o^t(\alpha)$ = }$[\,$Later on, the adiabatic temporal variations of $m_o^t(\alpha)$ will be linked to those of $q_t$,\\
      \phantom{\phantom{$m_o^t(\alpha)$ = }$[\,$}and the $\alpha$-dependence of $m_o^t(\alpha)$ will be fixed by a renormalization prescription.\\ \phantom{\phantom{$m_o^t(\alpha)$ = }$[\,$}You may have a look at equations (\ref{m-ren-d=2}) and (\ref{m-ren-d=3}) below.\\
      \phantom{\phantom{$m_o^t(\alpha)$ = }$[\,$}Let us conveniently point out in advance\\
      \phantom{\phantom{$m_o^t(\alpha)$ = }$[\,$}that the just mentioned renormalization prescription is constructed in such a way\\
      \phantom{\phantom{$m_o^t(\alpha)$ = }$[\,$}as to restore physical meaningfulness of our model\\
      \phantom{\phantom{$m_o^t(\alpha)$ = }$[\,$}in the highly nontrivial limit of $\alpha \to +\infty$.$\,]$\\
      Direct calculation utilizing (\ref{varrho-def}) and (\ref{hat-A-def}) yields explicitly
      \begin{eqnarray} \label{int-varrho-A}
         \int_{{\mathbb R}^d} \m \varrho({\bm y}) \; \hat{\bm A}({\bm y}) \; {\rm d}^d\n y
         & = & \int_{{\mathbb R}^d} \m {\rm d}^d\n k \, \sum_\wp \, \frac{1}{2\,\pi} \,
         \sqrt{\frac{\hbar\,c^2}{\omega_k}} \;\, e^{-\frac{{\bm k}^2}{4\,\alpha}} \, \hat{a}_{{\bm k}\wp}
         \; {\bm \varepsilon}_{{\bm k}\wp} \; + \; {\rm c.c.} \mez .
      \end{eqnarray}
      Formula (\ref{int-varrho-A}) shows that the highly UV field modes\\
      (i.e., those modes for which the factor $e^{-\frac{{\bm k}^2}{4\,\alpha}}$ is essentially zero)\\
      are actually decoupled (cut off) from the electronic degrees of freedom.\\
      This explains why is the parameter $\alpha$ named as an UV regulator.\\
      The limit of $\alpha \to +\infty$ implies removing gradually the UV cutoff.\\
      Note also that properties (\ref{q-t-infinite-past}) and (\ref{m-alpha-infinite-past}) imply having
      \be \label{H-t-infinite-past}
         \hat{H}_{t \to -\infty} \; = \; \hat{H}_{\rm free} \; = \; (\ref{H-free-def}) \mez .
      \ee
\item \underline{\bf A.3 Why such a model system ?}\\
      {\color{blue} The ultimate goal of these lecture notes is to provide an accessible introduction to renormalization.}\\
      Our chosen model defined by the Hamiltonian (\ref{H}) lends itself very well for this purpose:
      \vspace*{-0.20cm}
      \begin{itemize}
      \item[$\star$] The model is relatively simple, this enables us\\
                     to {\color{blue} pursue an analytic and non-perturbative treatment of mass renormalization},\\
                     without being distracted by tedious technical elaborations.\\
                     $[\,$Perturbative or numerical treatments become inevitable in fully relativistic QED,\\
                     \phantom{$[\,$}as well as in other QFT theories, or in various theories of condensed matter physics.\\
                     \phantom{$[\,$}Moreover, extremely tedious technicalities do emerge along the way\\
                     \phantom{$[\,$}both in QFT and in condensed matter physics.\\
                     \phantom{$[\,$}For a newcomer, it is often not easy to understand the essence of renormalization\\
                     \phantom{$[\,$}when presentation of the subject is obscured by computational technicalities.$\,]$
      \item[$\star$] The model is closely related to atomic physics, it can even be used\\
                     as the first approximation to explore a selected class of QED phenomena in atomic systems\\
                     (such as QED theory of harmonic generation, not shown in these lecture notes).
      \end{itemize}
\end{itemize}

\vspace*{+0.20cm}

{\color{blue} {\large \bf B. Heisenberg picture quantum dynamics}}
\vspace*{-0.20cm}
\begin{itemize}
\item \underline{\bf B.1 Basic elaborations}\\
      We wish to derive the Heisenberg equations of motion for:
      \vspace*{-0.20cm}
      \begin{itemize}
      \item[$\circ$] the electronic degrees of freedom $\hat{\bm x}(t)$, $\hat{\bm p}(t)$ ;
      \item[$\circ$] the field degrees of freedom $\hat{\bm A}(t,{\bm y})$,
                     these are equivalently represented by $\hat{a}_{{\bm k}\wp}\n(t)$.
      \end{itemize}
      \vspace*{-0.20cm}
      Proceeding in the standard way, one gets
      \begin{eqnarray}
         \label{Heisenberg-x}
         \frac{{\rm d}}{{\rm d}t} \, \hat{\bm x}(t) & = & \frac{1}{i\,\hbar} \, \left[ \, \hat{\bm x}(t) \, , \hat{H}_t \, \right]
         \; = \; \frac{1}{m_o^t(\alpha)} \left( \hat{\bm p}(t) \, - \, \frac{q_t}{c} \int_{{\mathbb R}^d} \m \varrho({\bm y}) \; \hat{\bm A}(t,{\bm y}) \; {\rm d}^d\n y \right) \mez ;\\
         \label{Heisenberg-p}
         \frac{{\rm d}}{{\rm d}t} \, \hat{\bm p}(t) & = & \frac{1}{i\,\hbar} \, \left[ \, \hat{\bm p}(t) \, , \hat{H}_t \, \right]
         \; = \; -\,\bm\nabla\,V\m(\hat{\bm x}(t)) \mez ;\\
         \label{Heisenberg-a}
         {\color{blue} \frac{{\rm d}}{{\rm d}t} \, \hat{a}_{{\bm k}\wp}\n(t) } & {\color{blue} = } & {\color{blue} \frac{1}{i\,\hbar} \, \left[ \, \hat{a}_{{\bm k}\wp}\n(t)
         \, , \hat{H}_t \, \right] \; = \; - \, i \, \omega_k \; \hat{a}_{{\bm k}\wp}\n(t) \; + \; i \, \frac{q_t}{2\,\pi} \,
         \frac{1}{\sqrt{\hbar\omega_k}} \; e^{-\frac{{\bm k}^2}{4\,\alpha}} \,
         \Bigl( {\bm \varepsilon}_{{\bm k}\wp} \m\bm\cdot\m \bm\dot{\hat{\bm x}}(t) \Bigr) \mez . }
      \end{eqnarray}
      Here $\bm\dot{\hat{\bm x}}(t)=\frac{{\rm d}}{{\rm d}t} \, \hat{\bm x}(t)=(\ref{Heisenberg-x})$.\\
      In (\ref{Heisenberg-x}) one encounters the Heisenberg picture vector potential
      \be \label{hat-A-t-y}
         \hat{\bm A}(t,{\bm y}) \; = \; \int_{{\mathbb R}^d} \m {\rm d}^d\n k \, \sum_\wp \, \frac{1}{2\,\pi} \, \sqrt{\frac{\hbar\,c^2}{\omega_k}} \;\,
         e^{+i{\bm k}\bm\cdot{\bm y}} \, \hat{a}_{{\bm k}\wp}\n(t) \; {\bm \varepsilon}_{{\bm k}\wp} \; + \; {\rm c.c.} \mez .
      \ee
      Different velocity components $\bm\dot{\hat{x}}_{j}\n(t)$ and $\bm\dot{\hat{x}}_{j'}\n(t)$
      do not commute with each other for $j \neq j'$\m.\\
      In (\ref{Heisenberg-a}) we have taken advantage of a general operator identity
      \be \label{anticommutator-identity}
         \Bigl[ \, \hat{f} , \hat{g}^2 \, \Bigr] \; = \;
         \Bigl\{ \, \Bigl[ \, \hat{f} , \hat{g} \, \Bigr] , \hat{g} \, \Bigr\} \mez ;
      \ee
      and of the commutation property
      \be
         \left[ \, \hat{a}_{{\bm k}\wp}\n(t)
         \, , \left( \hat{\bm p}(t) \, - \, \frac{q_t}{c} \int_{{\mathbb R}^d} \m \varrho({\bm y}) \; \hat{\bm A}(t,{\bm y}) \; {\rm d}^d\n y \right) \, \right] \; = \;
         -\,\frac{q_t}{2\,\pi} \; \sqrt{\frac{\hbar}{\omega_k}} \;
         e^{-\frac{{\bm k}^2}{4\,\alpha}} \; {\bm \varepsilon}_{{\bm k}\wp} \mez ;
      \ee
      which follows immediately from (\ref{hat-a-commutators}) and (\ref{int-varrho-A}), (\ref{hat-A-t-y}).
\item \underline{\bf B.2 The electric field}\\
      For our subsequent purposes,\\
      it is convenient to introduce at this point the Heisenberg picture electric field operator
      \be \label{hat-E-t-y}
         \hat{\bm E}(t,{\bm y}) \; = \; -\,\frac{1}{c} \; \partial_t \, \hat{\bm A}(t,{\bm y}) \mez .
      \ee
      Let us determine explicitly the modal expansion of $\hat{\bm E}(t,{\bm y})$\\
      by plugging into (\ref{hat-E-t-y}) the formulas (\ref{hat-A-t-y}) and (\ref{Heisenberg-a}).
      One obtains a preliminary outcome
      \begin{eqnarray}
         \label{hat-E-t-y-explicit-take-1} \hat{\bm E}(t,{\bm y}) & = &
         i \, \int_{{\mathbb R}^d} \m {\rm d}^d\n k \, \sum_\wp \, \frac{1}{2\,\pi} \, \sqrt{\hbar\,\omega_k} \;\,
         e^{+i{\bm k}\bm\cdot{\bm y}} \, \hat{a}_{{\bm k}\wp}\n(t) \; {\bm \varepsilon}_{{\bm k}\wp} \; + \; {\rm c.c.} \\
         & + & \frac{q_t}{2\,\pi^2} \, \int_{{\mathbb R}^d} \m {\rm d}^d\n k \;\, \omega_k^{-1} \;
         e^{-\frac{{\bm k}^2}{4\,\alpha}} \, \sin\m\Bigl({\bm k}\m\n\bm\cdot\m{\bm y}\Bigr) \,
         \sum_\wp \, {\bm \varepsilon}_{{\bm k}\wp} \m \left({\bm \varepsilon}_{{\bm k}\wp} \m\bm\cdot\m
         \bm\dot{\hat{\bm x}}(t)\right) \mez . \nonumber
      \end{eqnarray}
      The closure property (\ref{kappa-closure}) enables us to rewrite the second line of (\ref{hat-E-t-y-explicit-take-1}) as
      \be \label{hat-E-t-y-explicit-take-2}
         \frac{q_t}{2\,\pi^2} \, \int_{{\mathbb R}^d} \m {\rm d}^d\n k \;\, \omega_k^{-1} \; e^{-\frac{{\bm k}^2}{4\,\alpha}} \,
         \sin\m\Bigl({\bm k}\m\n\bm\cdot\m\n{\bm y}\Bigr) \, \Bigl( {\mathbb I} \, - \, {\bm \kappa} \, {\bm \kappa}^T \Bigr)
         \, \bm\dot{\hat{\bm x}}(t) \mez .
      \ee
      However, an integrand of (\ref{hat-E-t-y-explicit-take-2}) is an odd function of ${\bm k}$, hence the whole term (\ref{hat-E-t-y-explicit-take-2})
      vanishes for all ${\bm y}$.\\ Correspondingly, one has simply
      \begin{eqnarray}
         \label{hat-E-t-y-explicit-take-3} \hat{\bm E}(t,{\bm y}) & = &
         i \, \int_{{\mathbb R}^d} \m {\rm d}^d\n k \, \sum_\wp \, \frac{1}{2\,\pi} \, \sqrt{\hbar\,\omega_k} \;\,
         e^{+i{\bm k}\bm\cdot{\bm y}} \, \hat{a}_{{\bm k}\wp}\n(t) \; {\bm \varepsilon}_{{\bm k}\wp} \; + \; {\rm c.c.} \mez .
      \end{eqnarray}
      Coupling of the radiation field to our model atom enters inside (\ref{hat-E-t-y-explicit-take-3}) solely through
      $\hat{a}_{{\bm k}\wp}\n(t)$.
\item \underline{\bf B.3 Newton-Heisenberg equation of motion for $\hat{\bm x}(t)$}\\
      Now we are ready to write down the Newton-Heisenberg equation of motion for $\hat{\bm x}(t)$.\\
      Combination of (\ref{dot-m-alpha-zero}), (\ref{Heisenberg-x}), (\ref{Heisenberg-p}), (\ref{hat-E-t-y}) yields immediately
      \be \label{eom-hat-x-t-take-1}
         {\color{blue} m_o^t(\alpha) \; \frac{{\rm d}^2}{{\rm d}t^2} \, \hat{\bm x}(t) \; = \; -\,\bm\nabla V\m(\hat{\bm x}(t)) \; + \;
         q_t \m \int_{{\mathbb R}^d} \m \varrho({\bm y}) \; \hat{\bm E}(t,{\bm y}) \; {\rm d}^d\n y \mez .}
      \ee
      The just derived Newton-Heisenberg equation of motion (\ref{eom-hat-x-t-take-1})\\
      possesses an appearance which is much expected (based on physics intuition).\\
      Indeed, on the r.h.s.~of (\ref{eom-hat-x-t-take-1}) one recognizes the scalar potential term and the electric force term.\\
      In passing we note that an additional elaboration (not shown here) would provide\\
      the Heisenberg-Maxwell wave equation for $\hat{\bm A}(t,{\bm y})$ or $\hat{\bm E}(t,{\bm y})$. Such a wave equation\\
      turns out to be inhomogeneous, containing a source term associated with\\ the electric current generated by motion of the electron.
\item \underline{\bf B.4 An intermediate summary and additional reflections}\\
      Quantum dynamics of our studied system\\
      (= model of a nonrelativistic atom coupled to the quantized radiation field)\\
      is described by the following Heisenberg equations of motion:
      \vspace*{-0.20cm}
      \begin{itemize}
      \item[] Eq.~(\ref{eom-hat-x-t-take-1}) \dotfill governing quantum dynamics of the electron; $\phantom{MMMMMMMMMi}$
      \item[] Eq.~(\ref{Heisenberg-a}) \dotfill governing quantum dynamics of the radiation field. \hspace*{+3.00cm}
      \end{itemize}
      \vspace*{-0.20cm}
      Equations (\ref{eom-hat-x-t-take-1}) \& (\ref{Heisenberg-a}) are mutually coupled,\\
      and need to be solved starting from certain properly specified initial conditions:
      \be \label{ics-x}
         \hat{\bm x}(t_0) \; = \; \hat{\bm x} \mez , \mez \bm\dot{\hat{\bm x}}(t_0) \; = \; (\ref{Heisenberg-x})|_{t=t_0}
         \; = \; \frac{1}{m_o^{t_0}\n(\alpha)} \left( \hat{\bm p}(t_0) \, - \, \frac{q_{t_0}}{c} \int_{{\mathbb R}^d} \m \varrho({\bm y}) \;
         \hat{\bm A}(t_0,{\bm y}) \; {\rm d}^d\n y \right) \mez ;
      \ee
      and      
      \be \label{ics-a}
         \hat{a}_{{\bm k}\wp}\n(t_0) \mez .
      \ee
      Here $t_0$ stands for an arbitrary initial time instant.\\
      Let us hereafter push $t_0 \to -\infty$, taking advantage of the fact that\\
      our model atom and the radiation field become mutually uncoupled in the infinite past,\\
      see above equations (\ref{q-t-infinite-past}), (\ref{m-alpha-infinite-past}), (\ref{H-t-infinite-past}).\\
      Consistently with property (\ref{H-t-infinite-past}) we identify
      $\hat{\bm x}(t_0 \to -\infty)$, $\hat{\bm p}(t_0 \to -\infty)$, and $\hat{a}_{{\bm k}\wp}\n(t_0 \to -\infty)$\\
      with the corresponding interaction picture operators. This amounts to set
      \begin{eqnarray}
         \label{ics-final-x}
         {\color{blue} \hat{\bm x}(t \to -\infty) } & {\color{blue} = } & {\color{blue} 
         e^{+\frac{i}{\hbar}\hat{H}_{\rm free}t} \; \hat{\bm x} \; e^{-\frac{i}{\hbar}\hat{H}_{\rm free}t} \; = \;
         e^{+\frac{i}{\hbar}\hat{H}_{\m{\rm A}}t} \; \hat{\bm x} \; e^{-\frac{i}{\hbar}\hat{H}_{\m{\rm A}}t} \mez ;}\\
         \label{ics-final-p}
         {\color{blue} \hat{\bm p}(t \to -\infty) } & {\color{blue} = } & {\color{blue} 
         e^{+\frac{i}{\hbar}\hat{H}_{\rm free}t} \; \hat{\bm p} \; e^{-\frac{i}{\hbar}\hat{H}_{\rm free}t} \; = \;
         e^{+\frac{i}{\hbar}\hat{H}_{\m{\rm A}}t} \; \hat{\bm p} \; e^{-\frac{i}{\hbar}\hat{H}_{\m{\rm A}}t} \mez \,\n;}\\
         \label{ics-final-a}
         {\color{blue} \hat{a}_{{\bm k}\wp}\n(t \to -\infty) } & {\color{blue} = } & {\color{blue} 
         e^{+\frac{i}{\hbar}\hat{H}_{\rm free}t} \; \hat{a}_{{\bm k}\wp} \; e^{-\frac{i}{\hbar}\hat{H}_{\rm free}t} \; = \;
         e^{+\frac{i}{\hbar}\hat{H}_{{\rm R}}t} \; \hat{a}_{{\bm k}\wp} \; e^{-\frac{i}{\hbar}\hat{H}_{{\rm R}}t} \; = \;
         \hat{a}_{{\bm k}\wp} \; e^{-i\omega t} \mez .}
      \end{eqnarray}
      The r.h.s.~of (\ref{ics-final-x}), (\ref{ics-final-p}), (\ref{ics-final-a}) contains of course\\
      the Schr\"{o}dinger picture operators $\hat{\bm x}$, $\hat{\bm p}$, and $\hat{a}_{{\bm k}\wp}$ of subsection A.1.\\
      Note also that
      \be \label{dot-x-infinite-past}
         {\color{blue} m\,\bm\dot{\hat{\bm x}}(t \to -\infty) \; = \; \hat{\bm p}(t \to -\infty) \mez .}
      \ee
      {\color{magenta} {\sl So far, so good.}}\\
      All the elaborations presented above in Section B are absolutely standard,\\
      conventionally looking, and yield intuitively plausible resulting formulas.\\
      Recall, however, that our starting Hamiltonian (\ref{H}) is parameterized by the UV regulator $\alpha$\\
      which controls the width of the charge density profile $\varrho({\bm y})=(\ref{varrho-def})$ of the electron,\\
      and which thus decouples the highly UV field modes from the electronic motions.\\
      Our ultimate interest consists in exploring physical properties of the studied system\\
      (= model of a nonrelativistic atom coupled to the quantized radiation field)\\
      in the limit of $\alpha \to +\infty$, for which $\varrho({\bm y}) \to \delta^d\n({\bm y})$,\\
      and for which thus all the UV field modes are coupled to the electronic degrees of freedom in (\ref{H}).\\
      One would perhaps be tempted to anticipate at this point that the sought limit should be trivial.\\
      One would perhaps be tempted to set $m_o^t(\alpha)=m$\\
      $[\,$i.e., to identify the bare mass $m_o^t(\alpha)$ with the physical mass $m$ of the electron$\,]$,\\
      to reduce (\ref{eom-hat-x-t-take-1}) simply into
      \be \label{eom-hat-x-t-take-1-wrong}
         m \; \frac{{\rm d}^2}{{\rm d}t^2} \, \hat{\bm x}(t) \; = \;
         -\,\bm\nabla V\m(\hat{\bm x}(t)) \; + \; q_t \m \; \hat{\bm E}(t,{\bm 0}) \mez ;
      \ee
      and to reduce (\ref{Heisenberg-a}) simply into
      \begin{eqnarray}
         \label{Heisenberg-a-wrong}
         \frac{{\rm d}}{{\rm d}t} \, \hat{a}_{{\bm k}\wp}\n(t) & = &
         - \, i \, \omega_k \; \hat{a}_{{\bm k}\wp}\n(t) \; + \; i \, \frac{q_t}{2\,\pi} \,
         \frac{1}{\sqrt{\hbar\omega_k}} \; \Bigl( {\bm \varepsilon}_{{\bm k}\wp} \m\bm\cdot\m \bm\dot{\hat{\bm x}}(t) \Bigr) \mez .
      \end{eqnarray}
      Most importantly, the just mentioned speculative anticipations turn out to be totally wrong.\\
      {\color{blue} The limit of $\alpha \to +\infty$ is actually highly nontrivial.}\\
      {\color{blue} An appropriate theoretical analysis of the limit of $\alpha \to +\infty$ gives rise }\\
      {\color{blue} to the story of mass renormalization and to the concept of Abraham-Lorentz force.}\\
      {\color{magenta} {\sl The adventure is just beginning!}}
\end{itemize}

\vspace*{+0.20cm}

{\color{blue} {\large \bf C. Electric field of the radiation reaction}}
\vspace*{-0.20cm}
\begin{itemize}
\item \underline{\bf C.1 Formal solution of equation (\ref{Heisenberg-a})}\\
      Importantly, the field part of the studied problem lends itself to an analytic treatment.\\
      Indeed, the Heisenberg equation of motion (\ref{Heisenberg-a}) possesses the following formal solution:
      \begin{eqnarray} \label{hat-a-Heisenberg-explicit}
         \hspace*{-1.00cm} \hat{a}_{{\bm k}\wp}\n(t) & = & \hat{a}_{{\bm k}\wp} \, e^{-i \omega_k t} \\
         \hspace*{-1.00cm} & + & \frac{i}{2\,\pi} \, \frac{1}{\sqrt{\hbar\omega_k}} \; e^{-\frac{{\bm k}^2}{4\,\alpha}} \m
         \int_{-\infty}^{t} \m {\rm d}t' \;\, q_{t'} \; e^{-i \omega_k (t-t'\n)} \,
         \Bigl( {\bm \varepsilon}_{{\bm k}\wp} \m\bm\cdot\m \bm\dot{\hat{\bm x}}(t'\n) \Bigr) \mez . \nonumber
      \end{eqnarray}
      The initial condition
      $\hat{a}_{{\bm k}\wp}\n(t \to -\infty) = \hat{a}_{{\bm k}\wp} \; e^{-i\omega t}$ of equation (\ref{ics-final-a})
      has been incorporated, {\sl cf.}~also (\ref{q-t-infinite-past}).
\item \underline{\bf C.2 The electric field}\\
      Having in hand formula (\ref{hat-a-Heisenberg-explicit}), we may plug it into (\ref{hat-E-t-y-explicit-take-3})\\
      and evaluate explicitly the corresponding Heisenberg picture electric field $\hat{\bm E}(t,{\bm y})$.\\ One has      
      \be \label{hat-E-t-y-explicit-splitting}
         \hat{\bm E}(t,{\bm y}) \; = \; \hat{\bm E}_0\n(t,{\bm y}) \; + \; \hat{\bm E}_{\rm RR}\n(t,{\bm y}) \mez ;
      \ee
      where by definition
      \begin{eqnarray}
         \label{hat-E-0-t-y-def} \hat{\bm E}_0\n(t,{\bm y}) & = &
         i \, \int_{{\mathbb R}^d} \m {\rm d}^d\n k \, \sum_\wp \, \frac{1}{2\,\pi} \, \sqrt{\hbar\,\omega_k} \;\,
         e^{-i(\omega_k t-{\bm k}\bm\cdot{\bm y})} \, \hat{a}_{{\bm k}\wp} \; {\bm \varepsilon}_{{\bm k}\wp}
         \; + \; {\rm c.c.} \mz ;
      \end{eqnarray}
      and
      \begin{eqnarray} \label{hat-E-RR-t-y-def-prelim}
         \hspace*{-1.00cm} \hat{\bm E}_{\rm RR}\n(t,{\bm y}) & = &
         -\,\frac{1}{2\,\pi^2} \, \int_{-\infty}^{t} \m {\rm d}t' \int_{{\mathbb R}^d} \m {\rm d}^d\n k \; q_{t'} \,
         \cos\Bigl( \omega_k(t-t'\n)-{\bm k}\m\bm\cdot\m {\bm y} \Bigr) \;
         e^{-\frac{{\bm k}^2}{4\,\alpha}} \, \sum_{\wp} \, {\bm \varepsilon}_{{\bm k}\wp} \m
         \left({\bm \varepsilon}_{{\bm k}\wp} \m\bm\cdot\m \bm\dot{\hat{\bm x}}(t'\n)\right) \mz .
      \end{eqnarray}
      The closure property (\ref{kappa-closure}) implies
      \begin{eqnarray} \label{hat-E-RR-t-y-def-T}
         \hspace*{-1.00cm} \hat{\bm E}_{\rm RR}\n(t,{\bm y}) & = &
         -\,\frac{1}{2\,\pi^2} \, \int_{-\infty}^{t} \m {\rm d}t' \int_{{\mathbb R}^d} \m {\rm d}^d\n k \; q_{t'} \,
         \cos\Bigl( \omega_k(t-t'\n)-{\bm k}\m\bm\cdot\m {\bm y} \Bigr) \;
         e^{-\frac{{\bm k}^2}{4\,\alpha}} \, \Bigl( {\mathbb I} \, - \, {\bm \kappa} \, {\bm \kappa}^T \Bigr) \,
         \bm\dot{\hat{\bm x}}(t'\n) \mz .
      \end{eqnarray}
      Physical interpretation of the sum (\ref{hat-E-t-y-explicit-splitting}):\\
      $\hat{\bm E}_0\n(t,{\bm y})=(\ref{hat-E-0-t-y-def})$ is just the interaction picture
      counterpart of the Heisenberg electric field (\ref{hat-E-t-y-explicit-take-3}).\\
      Meaning that $\hat{\bm E}_0\n(t,{\bm y})$ evolves in time freely,\\
      as if there was no charged particle (electron) generating a current.\\
      $\hat{\bm E}_{\rm RR}\n(t,{\bm y})=(\ref{hat-E-RR-t-y-def-T})$ corresponds
      to the {\color{blue} electric field of the radiation reaction (RR)}.\\
      Meaning that $\hat{\bm E}_{\rm RR}\n(t,{\bm y})$ describes the radiation\\
      which has been emitted/absorbed by our charged particle (electron) during its motion\\
      at the time instants prior to $t$.
\item \underline{\bf C.3 Equation of motion for the electron revisited}\\
      Let us return to the Newton-Heisenberg equation of motion (\ref{eom-hat-x-t-take-1}) for the electron.\\
      After incorporating the splitting (\ref{hat-E-t-y-explicit-splitting}) one gets
      \be \label{eom-hat-x-t-take-2}
         m_o^t(\alpha) \; \frac{{\rm d}^2}{{\rm d}t^2} \, \hat{\bm x}(t) \; = \; -\,\bm\nabla V\m(\hat{\bm x}(t)) \; + \;
         \hat{\bm F}_0^{\rm el}(t) \; + \; \hat{\bm F}_{\rm RR}^{\rm el}\n(t) \mez ;
      \ee
      where by definition
      \be \label{electric-force-E0}
         \hat{\bm F}_0^{\rm el}\n(t) \; = \;
         q_t \m \int_{{\mathbb R}^d} \m \varrho({\bm y}) \; \hat{\bm E}_0\n(t,{\bm y}) \; {\rm d}^d\n y \mez ;
      \ee
      and
      \be \label{electric-force-ERR}
         \hat{\bm F}_{\rm RR}^{\rm el}\n(t) \; = \;
         q_t \m \int_{{\mathbb R}^d} \m \varrho({\bm y}) \; \hat{\bm E}_{\rm RR}\n(t,{\bm y}) \; {\rm d}^d\n y \mez .
      \ee
      Entity $\hat{\bm F}_0^{\rm el}(t)=(\ref{electric-force-E0})$ represents the interaction picture contribution\\
      to the electric force acting on the electron. Formulas (\ref{hat-E-0-t-y-def}) and (\ref{varrho-def}) imply having
      \be \label{bm-F-0-el-t}
         \hat{\bm F}_0^{\rm el}\n(t) \; = \; q_t \;\, i \, \int_{{\mathbb R}^d} \m {\rm d}^d\n k \, \sum_\wp \,
         \frac{1}{2\,\pi} \, \sqrt{\hbar\,\omega_k} \; e^{-\frac{{\bm k}^2}{4\,\alpha}} \, e^{-i\omega_k t} \,
         \hat{a}_{{\bm k}\wp} \; {\bm \varepsilon}_{{\bm k}\wp} \; + \; {\rm c.c.} \mez ;
      \ee
      {\sl cf.}~similar integration performed in equation (\ref{int-varrho-A}). If so, then
      \be \label{electric-force-E0-expand} 
         \hat{\bm F}_0^{\rm el}\n(t) \; = \; q_t \; \hat{\bm E}_0\n(t,{\bm 0}) \; + \; {\cal O}(\alpha^{-1}) \mez ;
      \ee
      with the ${\cal O}(\alpha^{-1})$ correction becoming insignificant for large enough $\alpha$.\\
      Note that $\hat{\bm E}_0\n(t,{\bm 0})=(\ref{hat-E-0-t-y-def})|_{{\bm y}={\bm 0}}$ is a well defined interaction picture operator\\
      whose time evolution is explicitly known and independent upon $\hat{\bm x}(t)$.\\
      Entity $\hat{\bm F}_{\rm RR}^{\rm el}(t)=(\ref{electric-force-ERR})$ represents the RR contribution\\
      to the Heisenberg picture electric force acting on the electron.\\
      After plugging (\ref{hat-E-RR-t-y-def-T}) into (\ref{electric-force-ERR}) one finds that
      \be \label{electric-force-ERR-T}
         \hat{\bm F}_{\rm RR}^{\rm el}\n(t) \; = \; -\,\frac{q_t}{2\,\pi^2} \, \int_{-\infty}^{t} \m {\rm d}t'
         \int_{{\mathbb R}^d} \m {\rm d}^d\n k \; q_{t'} \, \cos\Bigl(\omega_k(t-t'\n)\Bigr) \; e^{-\frac{{\bm k}^2}{2\,\alpha}}
         \, \Bigl( {\mathbb I} \, - \, {\bm \kappa} \, {\bm \kappa}^T \Bigr) \, \bm\dot{\hat{\bm x}}(t'\n) \mez .
      \ee
      The r.h.s.~of our Newton-Heisenberg equation of motion (\ref{eom-hat-x-t-take-2}) for the electron\\
      is now fully specified.\\
      Our next step will consist in evaluating explicitly the electric RR force
      $\hat{\bm F}_{\rm RR}^{\rm el}\n(t)=(\ref{electric-force-ERR-T})$\\
      while assuming that the UV regulator $\alpha$ approaches its limiting value $\alpha\to+\infty$.\\
      This amounts to integrate analytically both over ${\bm k} \in {\mathbb R}^d$ and over $t'\m \in (-\infty,t)$\\
      in the regime of $\alpha \to +\infty$.\\
      Since both technical details and the ultimate outcome\\
      depend crucially upon the spatial dimension $d$ of the problem,\\
      it is inevitable to discuss the cases of $d=2$ and $d=3$ separately,\\
      this is done in the two subsequent paragraphs.\\
      We shall begin with the $d=3$ case which is simpler to deal with than the $d=2$ case.
\item \underline{\bf C.4 An explicit evaluation of the electric RR force for $\bm d \bm = \bm 3$}\\
      Integration over ${\bm k} \in {\mathbb R}^3$ is conveniently performed in the spherical polar coordinates $(k,\varphi_{\bm \kappa},\vartheta_{\bm \kappa})$.\\
      Geometric identity
      \be \label{auxiliary-identity-d=3}
         \int_{0}^{2\pi} \m {\rm d}\varphi_{\bm \kappa} \, \int_{0}^{\pi} \m \sin\vartheta_{\bm \kappa} \; {\rm d}\vartheta_{\bm \kappa} \;\,
         {\bm \kappa} \, {\bm \kappa}^T \; = \; \frac{4\,\pi}{3} \; {\mathbb I}
      \ee
      converts equation (\ref{electric-force-ERR-T}) into
      \begin{eqnarray}
         \label{electric-force-ERR-take-2-d=3-T} \hat{\bm F}_{\rm RR}^{\rm el}\n(t) \; = \;
         - \, \frac{4}{3} \, \frac{q_t}{\pi} \, \int_{-\infty}^{t} \m {\rm d}t' \int_{0}^{\infty} \m k^2 \, {\rm d}k \;
         \cos\Bigl(kc(t-t'\n)\Bigr) \; e^{-\frac{k^2}{2\,\alpha}} \; q_{t'} \, \bm\dot{\hat{\bm x}}(t'\n) \mez .
      \end{eqnarray}
      Define now conveniently an auxiliary function
      \be \label{rho-alpha-def}
         \rho_\alpha\n\Bigl(c(t-t'\n)\Bigr) \; = \;
         \frac{1}{\pi} \, \int_{0}^{\infty}\m {\rm d}k \; \cos\Bigl(kc(t-t'\n)\Bigr) \; e^{-\frac{k^2}{2\,\alpha}}
         \; = \; \sqrt{\frac{\alpha}{2\,\pi}} \; e^{-\frac{\alpha}{2}\bigl(c(t-t'\n)\bigr)^2} \mez .
      \ee
      $[\,$Note that $\rho_\alpha(c(t-t'\n))$ approaches $\delta(c(t-t'\n))$ in the limit of $\alpha \to +\infty$.$\,]$\\
      Formula (\ref{electric-force-ERR-take-2-d=3-T}) simplifies accordingly, we have
      \begin{eqnarray}
         \label{electric-force-ERR-A-0-take-3}
         \hat{\bm F}_{{\rm RR}}^{\rm el}\n(t) & = & \frac{4}{3} \, \frac{q_t}{c^2} \, \int_{-\infty}^{t} \m {\rm d}t' \;
         \Bigl( \partial_{t't'} \, \rho_\alpha\n(c(t-t'\n)) \Bigr) \; q_{t'} \, \bm\dot{\hat{\bm x}}(t'\n) \mez .
      \end{eqnarray}
      In the regime of $\alpha \to +\infty$, the term $\Bigl( \partial_{t't'} \, \rho_\alpha\n(c(t-t'\n)) \Bigr)$ falls off rapidly to zero\\
      unless $t'$ is very close to $t$. This observation enables us to proceed further:
      \vspace*{-0.20cm}
      \begin{itemize}
      \item[$*$] The adiabatic switching property (\ref{dot-q-t-zero}) enables us to replace
                 $q_{t'}$ by $q_{t}$ in (\ref{electric-force-ERR-A-0-take-3}).
      \item[$*$] The electron velocity $\bm\dot{\hat{\bm x}}(t'\n)$ can be expanded into power series around $t'\m=t$.\\
                 Such that
                 \be
                    \hspace*{-1.00cm}
                    \bm\dot{\hat{\bm x}}(t'\n) \; = \; \bm\dot{\hat{\bm x}}(t) \; + \; \bm\ddot{\hat{\bm x}}(t) \, (t'\n-t) \; + \;
                    \frac{1}{2} \, \dddot{\hat{\bm x}}(t) \, (t'\n-t)^2 \; + \; {\cal O}\Bigl((t'\n-t)^3\Bigr) \mez .
                 \ee
      \end{itemize}
      \vspace*{-0.20cm}
      Equation (\ref{electric-force-ERR-A-0-take-3}) becomes thus
      \begin{eqnarray}
         \label{electric-force-ERR-A-0-take-4}
         \hat{\bm F}_{{\rm RR}}^{\rm el}\n(t) & = & \frac{4}{3} \, \frac{q_t^2}{c^2} \, \int_{-\infty}^{t} \m {\rm d}t' \;
         \Bigl( \partial_{t't'} \, \rho_\alpha\n(c(t-t'\n)) \Bigr) \nonumber\\
         & & \left\{ \bm\dot{\hat{\bm x}}(t) \; + \; \bm\ddot{\hat{\bm x}}(t) \, (t'\n-t) \; + \;
         \frac{1}{2} \, \dddot{\hat{\bm x}}(t) \, (t'\n-t)^2 \; + \; {\cal O}\Bigl((t'\n-t)^3\Bigr) \right\} \; = \nonumber\\
         & = & \frac{4}{3} \, \frac{q_t^2}{c^2} \, \int_{0}^{+\infty} \m {\rm d}\tau \;
         \Bigl( \partial_{\tau\tau} \, \rho_\alpha\n(c\tau) \Bigr) \;
         \left\{ \bm\dot{\hat{\bm x}}(t) \; - \; \bm\ddot{\hat{\bm x}}(t) \; \tau\; + \;
         \frac{1}{2} \, \dddot{\hat{\bm x}}(t) \; \tau^2 \; + \; {\cal O}(\tau^3) \right\} \mez .
      \end{eqnarray}
      Recall that $\rho_\alpha\n(c\tau)$ is given by above prescription (\ref{rho-alpha-def}).\\ The integrals
      $\int_{0}^{+\infty} \m \tau^n \, \partial_{\tau\tau} \, \rho_\alpha\n(c\tau) \; {\rm d}\tau$
      are doable analytically for all $n \in \{ 0,1,2,3,\bm\cdots\}$.\\ One gets a sequence of outcomes
      \begin{eqnarray}
         \int_{0}^{+\infty} \m {\rm d}\tau \; \Bigl( \partial_{\tau\tau} \, \rho_\alpha\n(c\tau) \Bigr) \; \tau^0 & = & 0 \\
         \int_{0}^{+\infty} \m {\rm d}\tau \; \Bigl( \partial_{\tau\tau} \, \rho_\alpha\n(c\tau) \Bigr) \; \tau^1 & = &
         \sqrt{\frac{\alpha}{2\,\pi}} \\
         \int_{0}^{+\infty} \m {\rm d}\tau \; \Bigl( \partial_{\tau\tau} \, \rho_\alpha\n(c\tau) \Bigr) \; \tau^2 & = &
         \frac{1}{c} \\
         \int_{0}^{+\infty} \m {\rm d}\tau \; \Bigl( \partial_{\tau\tau} \, \rho_\alpha\n(c\tau) \Bigr) \; \tau^3 & = &
         \frac{3}{c^2} \, \sqrt{\frac{2}{\pi\,\alpha}} \\
         \bm\vdots \hspace*{+2.00cm} & & \hspace*{+1.00cm} \bm\vdots \nonumber
      \end{eqnarray}
      Correspondingly, our electric RR force (\ref{electric-force-ERR-A-0-take-4}) takes the following final appearance,
      \begin{eqnarray}
         \label{electric-force-ERR-d=3-final} {\color{blue} \hat{\bm F}_{{\rm RR}}^{\rm el}\n(t) } & {\color{blue} = } &
         {\color{blue} - \, \frac{4}{3} \, \frac{q_t^2}{c^2} \, \sqrt{\frac{\alpha}{2\,\pi}} \; \bm\ddot{\hat{\bm x}}(t) \; + \;
         \frac{2}{3} \, \frac{q_t^2}{c^3} \, \dddot{\hat{\bm x}}(t) \; + \; {\cal O}(\alpha^{-1/2}) \mez . \mez [\,d=3\,] }
      \end{eqnarray}
      Most importantly, the first term on the r.h.s.~of (\ref{electric-force-ERR-d=3-final}) is proportional to $\alpha^{+1/2}$\m.\\
      {\color{blue} Hence our just evaluated electric RR force $\hat{\bm F}_{{\rm RR}}^{\rm el}\n(t)$\\
      possesses an infinite magnitude in the limit of $\alpha \to +\infty$.}\\
      This highly unsettling conclusion is bringing us already to the essence of renormalization.\\
      We will carefully analyze the implications of formula (\ref{electric-force-ERR-d=3-final}) later on in Section D.
\item \underline{\bf C.5 An explicit evaluation of the electric RR force for $\bm d \bm = \bm 2$}\\
      Integration over ${\bm k} \in {\mathbb R}^2$ is conveniently performed in the polar coordinates
      $(k,\varphi_{\bm \kappa})$.\\ Geometric identity
      \be \label{auxiliary-identity-d=2}
         \int_{0}^{2\pi} \m {\rm d}\varphi_{\bm \kappa} \;\, {\bm \kappa} \, {\bm \kappa}^T \; = \; \pi \; {\mathbb I}
      \ee
      converts equation (\ref{electric-force-ERR-T}) into
      \be \label{electric-force-ERR-T-take-2-d=2}
         \hat{\bm F}_{\rm RR}^{\rm el}\n(t) \; = \; -\,\frac{q_t}{2\,\pi} \, \int_{-\infty}^{t} \m {\rm d}t'
         \int_{0}^{\infty} \m k \; {\rm d}k \; \cos\Bigl(kc(t-t'\n)\Bigr) \; e^{-\frac{k^2}{2\,\alpha}} \;
         q_{t'} \, \bm\dot{\hat{\bm x}}(t'\n) \mez .
      \ee
      Define now conveniently an auxiliary function
      \be \label{Xi-alpha-def}
         \Xi_\alpha\n\Bigl(c(t-t'\n)\Bigr) \; = \;
         \int_{0}^{\infty}\m {\rm d}k \; \sin\Bigl(kc(t-t'\n)\Bigr) \; e^{-\frac{k^2}{2\,\alpha}} \mez .
      \ee
      $[\,$Note that $\Xi_\alpha(c(t-t'\n))=(\ref{Xi-alpha-def})$ can be defined even for any negative $c(t-t'\n)$.$\,]$\\
      Formula (\ref{electric-force-ERR-T-take-2-d=2}) simplifies accordingly, we have
      \be \label{electric-force-ERR-T-take-3-d=2}
         \hat{\bm F}_{\rm RR}^{\rm el}\n(t) \; = \; \frac{q_t}{2\,\pi\,c} \, \int_{-\infty}^{t} \m {\rm d}t' \,
         \Bigl( \partial_{t'} \, \Xi_\alpha\n(c(t-t'\n)) \Bigr) \, q_{t'} \, \bm\dot{\hat{\bm x}}(t'\n) \mez .
      \ee
      Proceeding further, we integrate in (\ref{electric-force-ERR-T-take-3-d=2}) by parts,\\
      while taking advantage of properties (\ref{dot-q-t-zero})-(\ref{q-t-infinite-past}),
      and while noting also that $\Xi_\alpha\n(0)=0$.\\
      One gets an intermediate outcome
      \be \label{electric-force-ERR-T-take-4-d=2}
         \hat{\bm F}_{\rm RR}^{\rm el}\n(t) \; = \; - \, \frac{q_t}{2\,\pi\,c} \, \int_{-\infty}^{t} \m {\rm d}t' \;\,
         \Xi_\alpha\n(c(t-t'\n)) \; q_{t'} \, \bm\ddot{\hat{\bm x}}(t'\n) \mez .
      \ee
      Next, let us make a small digression, and gather two auxiliary insights:
      \vspace*{-0.20cm}
      \begin{itemize}
      \item[{\sl 1)}] Direct calculation yields explicitly
      \be \label{f-alpha-take-2}
         \Xi_\alpha\n(c(t-t'\n)) \; = \; -i \; \sqrt{\frac{\pi\alpha}{2}} \;\, {\rm erf}\m\m\left( i \, \sqrt{\frac{\alpha}{2}} \; c(t-t'\n) \m \right) \,
         e^{-\frac{\alpha}{2}(c(t-t'\n))^2} \mez .
      \ee
      We shall actually not use the formula (\ref{f-alpha-take-2}) directly in our mainstream considerations.\\
      However, we shall notice that
      \be \label{f-alpha-take-3}
         \Xi_\alpha\n(c(t-t'\n)) \; = \; \frac{1}{c(t-t'\n)} \; + \; {\cal O}(\alpha^{-1}) \mez ;
      \ee
      valid as long as $c(t-t'\n) \neq 0$.\\
      Convergence of $\Xi_\alpha\n(c(t-t'\n))=(\ref{f-alpha-take-2})$ to its leading order approximation $\frac{1}{c(t-t'\n)}$ is not uniform.\\
      $[\,$This can be seen most easily by plotting $\Xi_\alpha\n(c(t-t'\n))=(\ref{f-alpha-take-2})$ for some fixed finite $\alpha$.$\,]$
      \item[{\sl 2)}] Our massive charged particle (electron) possesses the Compton wavelength
      \be \label{lambda-Compton}
         \lambda \; = \; \frac{\hbar}{mc} \mez .
      \ee
      Here $m$ stands for the physical mass,\\
      and not for the {\sl bare} mass $m_o^t(\alpha)$ appearing in the Hamiltonian $\hat{H}=(\ref{H})$.\\
      The associated Compton time equals then to
      \be
         \tilde{t} \; = \; \frac{\lambda}{c} \; = \; \frac{\hbar}{mc^2} \mez .
      \ee
      In addition, let $\eta>0$ be any dimensionless constant. We define for convenience a cutoff time
      \be
         t_c \; = \; \eta\,\tilde{t} \mez .
      \ee
      \end{itemize}
      Now we are ready to return to (\ref{electric-force-ERR-T-take-4-d=2}) and set
      \be \label{F-alpha-take-1}
         {\cal F}_\alpha(t-t'\n) \; = \; \int_{t_c}^{t-t'\n} \m\m \Xi_\alpha(c\tau) \; {\rm d}\tau \mez ;
      \ee
      such that
      \be
         - \, \partial_{t'} \, {\cal F}_\alpha(t-t'\n) \; = \; \Xi_\alpha(c(t-t'\n)) \mez .
      \ee
      The electric RR force $\hat{\bm F}_{\rm RR}^{\rm el}\n(t)=(\ref{electric-force-ERR-T-take-4-d=2})$ is subsequently transformed into
      \begin{eqnarray} \label{F-RR-analysis-take-3}
         \hat{\bm F}_{\rm RR}^{\rm el}\n(t) & = & \frac{q_t}{2\,\pi\,c} \, \int_{-\infty}^{t} \m {\rm d}t' \, q_{t'} \, \bm\ddot{\hat{\bm x}}(t'\n) \;
         \partial_{t'} \, {\cal F}_\alpha(t-t'\n) \; = \nonumber\\
         & = & \frac{q_t^2}{2\,\pi\,c} \; \bm\ddot{\hat{\bm x}}(t) \, {\cal F}_\alpha(0) \; - \;
         \frac{q_t}{2\,\pi\,c} \, \int_{-\infty}^{t} \m {\rm d}t' \, q_{t'} \, \dddot{\hat{\bm x}}(t'\n) \; {\cal F}_\alpha(t-t'\n) \mez .
      \end{eqnarray}
      In the second line we have integrated by parts while exploiting (\ref{dot-q-t-zero})-(\ref{q-t-infinite-past}).\\
      For large values of $\alpha$, one can take advantage of an expansion (\ref{f-alpha-take-3}), and write correspondingly
      \be \label{F-alpha-take-2}
         {\cal F}_\alpha(t-t'\n) \; = \; \frac{1}{c} \, \ln\m\n\left(\frac{t-t'}{t_c}\right) \; + \; {\cal O}(\alpha^{-1}) \mez . \mez
         [\,(t-t'\n) > 0\,]
      \ee     
      Formula (\ref{F-alpha-take-2}) enables us to write down another important intermediate outcome
      \begin{eqnarray} \label{F-RR-analysis-take-4}
         \hat{\bm F}_{\rm RR}^{\rm el}\n(t) & = & \frac{q_t^2}{2\,\pi\,c} \; \bm\ddot{\hat{\bm x}}(t) \, {\cal F}_\alpha(0) \; - \;
         \frac{q_t}{2\,\pi\,c^2} \, \int_{-\infty}^{t} \m q_{t'} \, \dddot{\hat{\bm x}}(t'\n) \, \ln\m\m\left(\frac{t-t'}{t_c}\right) {\rm d}t' \\
         & + & {\rm an}\;\,\alpha{\rm -dependent \;\, correction \;\, vanishing \;\, as} \;\, \alpha \to +\infty \mez \mez . \nonumber
      \end{eqnarray}
      Observe that the second (logarithmic) term of (\ref{F-RR-analysis-take-4}) is independent upon the UV regulator $\alpha$.\\
      The reader may be puzzled here by the singularities of $\ln(t-t'\n)$\\
      which occur at $t'\n \to t$ and $t'\n \to -\infty$.\\
      Fortunately, the just mentioned singularities are anticipated to be benign:\\
      \phantom{mz}Stated briefly, the $t'\n \to t$ singularity is weak enough to be integrable,\\
      \phantom{mz}as long as the fourth time derivative $\ddddot{\hat{\bm x}}(t'\n)$ remains bounded in magnitude.\\
      \phantom{mz}Furthermore, the slow divergence of $\ln(t-t'\n)$ at $t'\n \to -\infty$ is overcompensated\\
      \phantom{mz}by an oscillatory behavior of $\dddot{\hat{\bm x}}(t'\n)$. Time oscillations of $\bm\dot{\hat{\bm x}}(t'\n)$,
      $\bm\ddot{\hat{\bm x}}(t'\n)$, $\dddot{\hat{\bm x}}(t'\n)$ etc.~do take place,\\
      \phantom{mz}since in our model the electron moves in a restricted spatial region\\
      \phantom{mz}determined by a strictly bound potential $V\m({\bm x})$, see A.1.\\
      \phantom{mz}A more careful mathematical treatment of the just discussed singularities of $\ln(t-t'\n)$\\
      \phantom{mz}is relegated to {\sl Appendix A}.\\
      In the second line of (\ref{F-RR-analysis-take-4}) we wrote just "an $\alpha-$dependent correction vanishing as $\alpha \to +\infty$"\m\m,\\ this is sufficient for our purposes.\\
      $[\,$Writing hastily ${\cal O}(\alpha^{-1})$ seems to be unsafe, since in equation (\ref{F-alpha-take-2})\\
      \phantom{$[\,$}the convergence of ${\cal F}_\alpha(t-t'\n)$ to its leading order approximation
      $\frac{1}{c} \, \ln\m\n\left(\frac{t-t'}{t_c}\right)$ is not uniform.$\,]$\\
      Proceeding further in our elaborations, we shall focus on analyzing the first term of equation (\ref{F-RR-analysis-take-4}),\\
      or, equivalently, the quantity
      \be \label{F-alpha-take-3}
         {\cal F}_\alpha(0) \; = \; \int_{t_c}^{0} \m\m \Xi_\alpha(c\tau) \; {\rm d}\tau \; = \; \int_{t_c}^{0} \m {\rm d}\tau
         \int_{0}^{+\infty} \m {\rm d}k \; \sin\Bigl(kc\tau\Bigr) \; e^{-\frac{k^2}{2\,\alpha}} \mez .
      \ee
      An integration over $\tau$ can be performed first, this yields
      \be \label{F-alpha-take-4}
         {\cal F}_\alpha(0) \; = \; \int_{0}^{+\infty} \m {\rm d}k \;\, e^{-\frac{k^2}{2\,\alpha}} \; \frac{\Bigl( \cos(k\,c\,t_c) \, - \, 1 \Bigr)}{kc} \mez .
      \ee
      The integral (\ref{F-alpha-take-4}) obviously diverges to $-\infty$ for $\alpha \to +\infty$.\\
      A more explicit insight is gained by calculating the $\alpha$-derivative
      \begin{eqnarray} \label{F-alpha-take-5}
         \frac{{\rm d}}{{\rm d}\alpha} \, {\cal F}_\alpha(0) & = &
         \int_{0}^{+\infty} \m {\rm d}k \;\, e^{-\frac{k^2}{2\,\alpha}} \; \frac{k^2}{2\,\alpha^2} \; \frac{\Bigl( \cos(k\,c\,t_c) \, - \, 1 \Bigr)}{kc}
         \; = \nonumber\\
         & = & \frac{i\,t_c}{2} \, \sqrt{\frac{\pi}{2\,\alpha}} \;\, {\rm erf}\m\m\left(i\,\sqrt{\frac{\alpha}{2}}\,(c\,t_c)\right) e^{-\frac{\alpha}{2}(ct_c)^2} \; = \nonumber\\
         & = & -\,\frac{1}{2\,c\,\alpha} \; - \; \frac{1}{2\,c^3\,t_c^2\,\alpha^2} \; + \; {\cal O}(\alpha^{-3}) \mez .
      \end{eqnarray}
      Formula (\ref{F-alpha-take-5}) implies that, inevitably,
      \be \label{F-alpha-take-6}
         {\cal F}_\alpha(0) \; = \; -\,\frac{1}{2\,c} \, \ln \alpha \; + \; \zeta(\eta) \; + \; \frac{1}{2\,c^3\,t_c^2\,\alpha} \; + \; {\cal O}(\alpha^{-2}) \mez .
      \ee
      Here $\zeta(\eta)$ stands for an as yet undetermined finite constant, which is equal formally to
      \begin{eqnarray} \label{zeta-gamma-def}
         \zeta(\eta) & = & \lim_{\alpha \to +\infty} \left\{ {\cal F}_\alpha(0) \; + \; \frac{1}{2\,c} \, \ln \alpha \right\} \; = \nonumber\\
         & = & \lim_{\alpha \to +\infty} \, \left\{ \int_{0}^{+\infty} \m {\rm d}k \;\, e^{-\frac{k^2}{2\,\alpha}} \; \frac{\Bigl( \cos(k\,c\,\eta\,\tilde{t}\,)
         \, - \, 1 \Bigr)}{kc} \; + \; \frac{1}{2\,c} \, \ln \alpha \right\} \mez .
      \end{eqnarray}
      The dependence of $\zeta(\eta)$ upon $\eta$ can be found via differentiating (\ref{zeta-gamma-def}) with respect to $\eta$.\\ One gets
      \be \label{d-zeta-d-gamma}
         \frac{{\rm d}\zeta(\eta)}{{\rm d}\eta} \; = \; -\,\tilde{t}\,\lim_{\alpha \to +\infty} \, \Xi_\alpha(c\,t_c) \; = \; \frac{(-1)}{\eta\,c} \mez .
      \ee      
      In (\ref{d-zeta-d-gamma}) we have exploited the definition (\ref{Xi-alpha-def}) and the asymptotic behavior (\ref{f-alpha-take-3}).\\
      Relation (\ref{d-zeta-d-gamma}) reveals having simply
      \be \label{zeta-gamma-finres}
         \zeta(\eta) \; = \; \zeta \; - \; \frac{1}{c} \, \ln \eta \mez ;
      \ee      
      with
      \be \label{zeta-def}
         \zeta \; \equiv \; \zeta(1) \; = \; \lim_{\alpha \to +\infty} \, \left\{ \int_{0}^{+\infty} \m {\rm d}k \;\, e^{-\frac{k^2}{2\,\alpha}} \;
         \frac{\Bigl( \cos\m\m\left(\frac{\hbar k}{mc}\right) \, - \, 1 \Bigr)}{kc} \; + \; \frac{1}{2\,c} \, \ln \alpha \right\} \mez .
      \ee
      For our present purposes, it is not necessary to work out a more explicit prescription for $\zeta=(\ref{zeta-def})$.\\
      It is sufficient to regard $\zeta$ as some well defined finite number depending upon $(m,c,\hbar)$.\\
      After plugging (\ref{zeta-gamma-finres}) into (\ref{F-alpha-take-6}) one arrives towards a finalized $\alpha$-expansion of
      ${\cal F}_\alpha(0)$, namely,
      \be \label{F-alpha-take-7}
         {\cal F}_\alpha(0) \; = \; -\,\frac{1}{2\,c} \, \ln\Bigl(\eta^2\alpha\Bigr) \; + \; \zeta \; + \; \frac{1}{2\,c^3\,\eta^2\,\tilde{t}^2\,\alpha} \; + \; {\cal O}(\alpha^{-2}) \mez .
      \ee
      Our sought RR force $\hat{\bm F}_{\rm RR}^{\rm el}\n(t)=(\ref{F-RR-analysis-take-4})$ equals then to
      \begin{eqnarray} \label{F-RR-analysis-take-5}
         \hat{\bm F}_{\rm RR}^{\rm el}\n(t) & = & -\, \frac{q_t^2}{4\,\pi\,c^2} \; \bm\ddot{\hat{\bm x}}(t) \, \ln \alpha 
         \; - \;  \frac{q_t^2}{2\,\pi\,c^2} \; \bm\ddot{\hat{\bm x}}(t) \, \ln \eta
         \; + \;  \frac{q_t^2}{2\,\pi\,c} \; \bm\ddot{\hat{\bm x}}(t) \, \zeta \\
         & - &  \frac{q_t}{2\,\pi\,c^2} \, \int_{-\infty}^{t} \m q_{t'} \, \dddot{\hat{\bm x}}(t'\n) \; \ln\m\m\left(\frac{t-t'}{\eta\,\tilde{t}}\right) {\rm d}t' \nonumber\\
         & + & {\rm an}\;\,\alpha{\rm -dependent \;\, correction \;\, vanishing \;\, as} \;\, \alpha \to +\infty \mez . \nonumber
      \end{eqnarray}
      Additionally, an identity
      \begin{eqnarray}
         & & - \; \frac{q_t}{2\,\pi\,c^2} \, \int_{-\infty}^{t} \m q_{t'} \, \dddot{\hat{\bm x}}(t'\n) \; \ln\m\m\left(\frac{t-t'}{\eta\,\tilde{t}}\right) {\rm d}t'
         \; = \nonumber\\ & = & - \; \frac{q_t}{2\,\pi\,c^2} \, \int_{-\infty}^{t} \m q_{t'} \, \dddot{\hat{\bm x}}(t'\n) \; \ln\m\m\left(\frac{t-t'}{\tilde{t}}\right)
         {\rm d}t' \; + \; \frac{q_t^2}{2\,\pi\,c^2} \, \bm\ddot{\hat{\bm x}}(t') \, \ln \eta
      \end{eqnarray}
      shows that the two terms proportional to $\ln \eta$ cancel each other in (\ref{F-RR-analysis-take-5}).\\
      Thereby our electric RR force $\hat{\bm F}_{\rm RR}^{\rm el}\n(t)$ turns out to be independent upon $\eta$,\\
      and takes the final, simple, compelling appearance
      \begin{eqnarray} \label{F-RR-analysis-take-6}
         {\color{blue} \hat{\bm F}_{\rm RR}^{\rm el}\n(t) } & {\color{blue} = } &
         {\color{blue} -\,\frac{q_t^2}{4\,\pi\,c^2} \; \bm\ddot{\hat{\bm x}}(t) \, \ln\m\m\left(\frac{\alpha}{\alpha_0}\right)
         \; - \; \frac{q_t}{2\,\pi\,c^2} \, \int_{-\infty}^{t} \m q_{t'} \, \dddot{\hat{\bm x}}(t'\n) \; \ln\m\m\left(\frac{t-t'}{\tilde{t}}\right) {\rm d}t' }\\
         & {\color{blue} + } & {\color{blue} {\rm an}\;\,\alpha{\rm -dependent \;\, correction \;\, vanishing \;\, as} \;\, \alpha \to +\infty \hspace*{+2.00cm} .} \nonumber
      \end{eqnarray}
      We have conveniently set here
      \be
         \zeta \; = \; \frac{1}{2\,c} \, \ln \alpha_0 \mez , \mez {\rm i.e.} \mez \alpha_0 \; = \; e^{2\,c\,\zeta} \mez .
      \ee
      The passage from (\ref{electric-force-ERR-T-take-2-d=2}) to (\ref{F-RR-analysis-take-6}) was painfully technical and tedious,
      but certainly worthy of an effort.\\
      One may observe that the just derived electric RR force $\hat{\bm F}_{\rm RR}^{\rm el}\n(t)=(\ref{F-RR-analysis-take-6})$ is given as a sum over:\\
      {\it i)}\phantom{\it ii} a local term proportional to $\bm\ddot{\hat{\bm x}}(t)$, {\color{blue} whose prefactor diverges logarithmically with $\alpha \to +\infty$};\\
      {\it ii)}\phantom{\it i} an $\alpha$-independent nonlocal term, contanining $\dddot{\hat{\bm x}}(t'\n)$ $[\,t'\n<t\,]$ and a memory kernel $\ln\m\m\left(\frac{t-t'}{\tilde{t}}\right)$;\\
      {\it iii)} other contributions which however fall off to zero as $\alpha \to +\infty$.\\
      {\color{blue} Due to {\it i)}, our just evaluated electric RR force $\hat{\bm F}_{{\rm RR}}^{\rm el}\n(t)$\\
      possesses an infinite magnitude in the limit of $\alpha \to +\infty$.}\\
      This highly unsettling conclusion\\ is bringing us already to the essence of renormalization, much as in C.4.\\
      We will carefully analyze the implications of formula (\ref{F-RR-analysis-take-6}) later on in Section D.
\item \underline{\bf C.6 Equation of motion for the electron revisited again}\\
      Let us return to the Newton-Heisenberg equation of motion (\ref{eom-hat-x-t-take-2}) for the electron,\\
      and incorporate our above derived expressions (\ref{electric-force-E0-expand}) and (\ref{electric-force-ERR-d=3-final})
      or (\ref{F-RR-analysis-take-6})\\
      which determine the electric forces $\hat{\bm F}_0^{\rm el}\n(t)$ and $\hat{\bm F}_{\rm RR}^{\rm el}\n(t)$.\\
      In the case of $d=2$, one gets explicitly
      \begin{eqnarray} \label{eom-hat-x-t-take-3-d=2}
         {\color{blue} m_o^t(\alpha) \; \frac{{\rm d}^2}{{\rm d}t^2} \, \hat{\bm x}(t) } & {\color{blue} = } &
         {\color{blue} -\,\bm\nabla V\m(\hat{\bm x}(t)) \; + \; q_t \; \hat{\bm E}_0\n(t,{\bm 0}) }\\
         & {\color{blue} - } & {\color{blue} \frac{q_t^2}{4\,\pi\,c^2} \; \bm\ddot{\hat{\bm x}}(t) \, \ln\m\m\left(\frac{\alpha}{\alpha_0}\right)
         \; - \; \frac{q_t}{2\,\pi\,c^2} \, \int_{-\infty}^{t} \m q_{t'} \, \dddot{\hat{\bm x}}(t'\n) \; \ln\m\m\left(\frac{t-t'}{\tilde{t}}\right) {\rm d}t' } \nonumber\\
         & {\color{blue} + } & {\color{blue} {\rm an}\;\,\alpha{\rm -dependent \;\, correction \;\, vanishing \;\, as} \;\, \alpha \to +\infty \hspace*{+1.50cm} . \mez [\,d=2\,]} \nonumber
      \end{eqnarray}
      In the case of $d=3$, one gets explicitly
      \begin{eqnarray} \label{eom-hat-x-t-take-3-d=3}
         {\color{blue} m_o^t(\alpha) \; \frac{{\rm d}^2}{{\rm d}t^2} \, \hat{\bm x}(t) } & {\color{blue} = } &
         {\color{blue} -\,\bm\nabla V\m(\hat{\bm x}(t)) \; + \; q_t \; \hat{\bm E}_0\n(t,{\bm 0}) }\\
         & {\color{blue} - } & {\color{blue} \frac{4}{3} \, \frac{q_t^2}{c^2} \, \sqrt{\frac{\alpha}{2\,\pi}} \;
         \bm\ddot{\hat{\bm x}}(t) \; + \; \frac{2}{3} \, \frac{q_t^2}{c^3} \, \dddot{\hat{\bm x}}(t) \; + \;
         {\cal O}(\alpha^{-1/2}) \mez . \mez [\,d=3\,] } \nonumber
      \end{eqnarray}
      Formulas (\ref{eom-hat-x-t-take-3-d=2})-(\ref{eom-hat-x-t-take-3-d=3}) represent final outcome of the somewhat technical Section C.\\
      The electric RR force of equation (\ref{eom-hat-x-t-take-3-d=2}) takes a qualitatively different appearance\\
      than the electric RR force of equation (\ref{eom-hat-x-t-take-3-d=3}).\\
      Both in (\ref{eom-hat-x-t-take-3-d=2}) and (\ref{eom-hat-x-t-take-3-d=3}), the electric RR force term diverges for
      $\alpha \to +\infty$.\\
      {\color{magenta} {\sl We are done with all the technical preparations,\\ and ready to jump into the matters of mass renormalization now!}}
\end{itemize}
    
\vspace*{+0.20cm}

{\color{blue} {\large \bf D. Mass renormalization}}
\begin{itemize}
\item \underline{\bf D.1 Preliminaries}\\
      The concept or mass renormalization will be introduced here via examining the $\alpha \to +\infty$ limit\\
      of the Newton-Heisenberg equations of motion (\ref{eom-hat-x-t-take-3-d=2}) and (\ref{eom-hat-x-t-take-3-d=3})
      for $\hat{\bm x}(t)$.\\
      Recall that for $\alpha \to +\infty$ the charge density $\varrho({\bm y})=(\ref{varrho-def})$
      approaches $\delta^d\n({\bm y})$.\\
      This is equivalent to saying that our electron approaches the limit of a charged mass point.\\
      Pushing $\alpha \to +\infty$ elevates towards infinity\\
      the UV cutoff in our "electron -- radiation field" coupling (\ref{int-varrho-A}), see A.2.\\
      In other words, for $\alpha \to +\infty$, all the radiation field modes\\
      become affected by the electronic motion and vice versa.\\
      As we have just explicitly seen above in C.4 and C.5,\\
      coupling the electron to all the UV field modes\\
      causes an UV divergence of the corresponding electric RR force,\\
      see again formulas (\ref{electric-force-ERR-d=3-final}), (\ref{F-RR-analysis-take-6})
      or the equations of motion (\ref{eom-hat-x-t-take-3-d=2}), (\ref{eom-hat-x-t-take-3-d=3}) themselves.\\
      {\color{blue} Our Newton-Heisenberg equations of motion (\ref{eom-hat-x-t-take-3-d=2}) and (\ref{eom-hat-x-t-take-3-d=3})\\
      become inevitably unphysical in the $\alpha \to +\infty$ limit, unless\\
      the so far unspecified bare mass term $m_o^t(\alpha)$
      is adjusted in such an ingenuous manner\\
      as to overcompensate the UV divergent part of the electric RR force.}\\
      We shall implement exactly this kind of trick below, separately for $d=2$ and $d=3$.
\item \underline{\bf D.2 Mass renormalization in the case of $\bm d \bm = \bm 2$}\\
      The Newton-Heisenberg equation of motion (\ref{eom-hat-x-t-take-3-d=2}) can be equivalently redisplayed as follows:
      \begin{eqnarray} \label{eom-hat-x-t-take-3-d=2-take-2}
         \hspace*{-1.50cm} \left\{ \, m_o^t(\alpha) \; + \; \frac{q_t^2}{4\,\pi\,c^2} \; \ln\m\m\left(\frac{\alpha}{\alpha_0}\right) \, \right\} \frac{{\rm d}^2}{{\rm d}t^2} \, \hat{\bm x}(t) & = &
         -\,\bm\nabla V\m(\hat{\bm x}(t)) \; + \; q_t \; \hat{\bm E}_0\n(t,{\bm 0}) \\
         \hspace*{-1.50cm} & - & \frac{q_t}{2\,\pi\,c^2} \, \int_{-\infty}^{t} \m q_{t'} \, \dddot{\hat{\bm x}}(t'\n) \; \ln\m\m\left(\frac{t-t'}{\tilde{t}}\right) {\rm d}t' \nonumber\\
         \hspace*{-1.50cm} & + & {\rm an}\;\,\alpha{\rm -dependent \;\, correction \;\, vanishing \;\, as} \;\, \alpha \to +\infty
         \mz . \nonumber
      \end{eqnarray}
      Showing that, for large values of $\alpha$,\\
      a physically meaningful dynamical time evolution of $\hat{\bm x}(t)$ can take place only if\\
      one adjusts the bare mass $m_o^t(\alpha)$ according to the {\color{blue} mass renormalization prescription}
      \be \label{m-ren-d=2}
         {\color{blue}
         m_o^t(\alpha) \; + \; \frac{q_t^2}{4\,\pi\,c^2} \; \ln\m\m\left(\frac{\alpha}{\alpha_0}\right) \; = \; m
         \mez . \mez [d=2]}
      \ee
      Here $m$ represents of course the physical mass of the electron\\
      (which enters also into the atomic Hamiltonian $\hat{H}_{\m{\rm A}}=(\ref{H-A-def})$).\\
      Note that $m_o^t(\alpha)$ diverges to $-\infty$ as $\alpha \to +\infty$.\\
      $[\,$Thus $m_o^t(\alpha)$ should be regarded as mere mathematical tool lacking any physical significance.$\,]$\\
      Having accepted the above peculiar arrangement (\ref{m-ren-d=2}), and having subsequently pushed $\alpha \to +\infty$,\\
      we are able to simplify the Newton-Heisenberg equation of motion (\ref{eom-hat-x-t-take-3-d=2-take-2})\\
      into its final and compelling appearance
      \begin{eqnarray} \label{eom-hat-x-t-take-3-d=2-final}
         {\color{blue} m \, \frac{{\rm d}^2}{{\rm d}t^2} \, \hat{\bm x}(t) } & {\color{blue} = } &
         {\color{blue} -\,\bm\nabla V\m(\hat{\bm x}(t)) \; + \; q_t \; \hat{\bm E}_0\n(t,{\bm 0}) }\\
         & {\color{blue} - } & {\color{blue} \frac{q_t}{2\,\pi\,c^2} \, \int_{-\infty}^{t} \m q_{t'} \, \dddot{\hat{\bm x}}(t'\n) \; \ln\m\m\left(\frac{t-t'}{\tilde{t}}\right) {\rm d}t' \mez . \mez [d=2] } \nonumber
      \end{eqnarray}
      {\color{magenta} {\sl Success! Mass renormalization has just been performed!}}\\
      The integro-differential equation (\ref{eom-hat-x-t-take-3-d=2-final}) should be solved for $\hat{\bm x}(t)$\\
      with the initial conditions (\ref{ics-final-x}), (\ref{ics-final-p}), (\ref{dot-x-infinite-past}).\\
      Recall that $q_t$ equals to the physical charge $q$ of the electron at finite times.\\
      The first line on the r.h.s.~of (\ref{eom-hat-x-t-take-3-d=2-final}) displays\\
      standard scalar potential contribution to the force acting on the electron,\\
      as well as the interaction picture electric force contribution acting on the electron,\\
      which we have already discussed before $[\,$see equation (\ref{electric-force-E0-expand}) and the accompanying remarks$\,]$.\\
      The second line on the r.h.s.~of (\ref{eom-hat-x-t-take-3-d=2-final}) displays\\
      {\color{blue} a finite contribution of the electric RR force which has survived the mass renormalization.\\
      This is the so called Abraham-Lorentz force,
      \be \label{ALF-d=2}
         {\color{blue} \hat{\bm F}_{\rm ALF}\n(t) \; = \; - \; \frac{q_t}{2\,\pi\,c^2} \, \int_{-\infty}^{t} \m q_{t'} \, \dddot{\hat{\bm x}}(t'\n) \; \ln\m\m\left(\frac{t-t'}{\tilde{t}}\right) {\rm d}t' \mez . \mez [d=2]}
      \ee
      Most importantly, for $d=2$, the resulting Abraham-Lorentz force is nonlocal in time,\\
      contanining $\dddot{\hat{\bm x}}(t'\n)$ $[\,t'\n<t\,]$ and a logarithmic memory kernel
      $\ln\m\m\left(\frac{t-t'}{\tilde{t}}\right)$.\\ Hence the quantum dynamical evolution of $\hat{\bm x}(t)$
      is non-Markovian.}\\ $[\,$The term "non-Markovian" means that our equation of motion (\ref{eom-hat-x-t-take-3-d=2-final})
      has a memory, i.e., that\\ \phantom{$[\,$}previous history of the system (including all the time instants $t'$ finitely
      separated from $t$)\\ \phantom{$[\,$}matters when propagating $\hat{\bm x}(t)$ from the time instant $t$ forward.$\,]$\\
      In {\sl Appendix A} we analyze in more detail\\
      mathematical well definedness of the just mentioned Abraham-Lorentz force term (\ref{ALF-d=2}),\\
      and suggest an iterative (perturbative) solution of equation (\ref{eom-hat-x-t-take-3-d=2-final}), starting from\\
      the zeroth order interaction picture approximation
      $\hat{\bm x}(t) \doteq \hat{\bm x}_I\n(t) \; = \; e^{+\frac{i}{\hbar}\hat{H}_{\m{\rm A}}t} \; \hat{\bm x} \; e^{-\frac{i}{\hbar}\hat{H}_{\m{\rm A}}t}$\m.
\item \underline{\bf D.3 Mass renormalization in the case of $\bm d \bm = \bm 3$}\\
      The Newton-Heisenberg equation of motion (\ref{eom-hat-x-t-take-3-d=3}) can be equivalently redisplayed as follows:
      \begin{eqnarray} \label{eom-hat-x-t-take-3-d=3-take-2}
         \hspace*{-1.00cm} \left\{ m_o^t(\alpha) \, + \, \frac{4}{3} \, \frac{q_t^2}{c^2} \, \sqrt{\frac{\alpha}{2\,\pi}}
         \, \right\} \frac{{\rm d}^2}{{\rm d}t^2} \, \hat{\bm x}(t) & = &
         -\,\bm\nabla V\m(\hat{\bm x}(t)) \; + \; q_t \; \hat{\bm E}_0\n(t,{\bm 0}) \; + \;
         \frac{2}{3} \, \frac{q_t^2}{c^3} \, \dddot{\hat{\bm x}}(t) \; + \; {\cal O}(\alpha^{-1/2}) \mz .
      \end{eqnarray}
      Showing that, for large values of $\alpha$,\\
      a physically meaningful dynamical time evolution of $\hat{\bm x}(t)$ can take place only if\\
      one adjusts the bare mass $m_o^t(\alpha)$ according to the {\color{blue} mass renormalization prescription}
      \be \label{m-ren-d=3}
         {\color{blue}
         m_o^t(\alpha) \; + \; \frac{4}{3} \, \frac{q_t^2}{c^2} \, \sqrt{\frac{\alpha}{2\,\pi}} \; = \; m
         \mez . \mez [d=3]}
      \ee
      Here $m$ represents of course the physical mass of the electron\\
      (which enters also into the atomic Hamiltonian $\hat{H}_{\m{\rm A}}=(\ref{H-A-def})$).\\
      Note that $m_o^t(\alpha)$ diverges to $-\infty$ as $\alpha \to +\infty$.\\
      $[\,$Thus $m_o^t(\alpha)$ should be regarded as mere mathematical tool lacking any physical significance.$\,]$\\
      Having accepted the above peculiar arrangement (\ref{m-ren-d=3}), and having subsequently pushed $\alpha \to +\infty$,\\
      we are able to simplify the Newton-Heisenberg equation of motion (\ref{eom-hat-x-t-take-3-d=3-take-2})\\
      into its final and compelling appearance
      \begin{eqnarray} \label{eom-hat-x-t-take-3-d=3-final}
         {\color{blue} m \, \frac{{\rm d}^2}{{\rm d}t^2} \, \hat{\bm x}(t) } & {\color{blue} = } &
         {\color{blue} -\,\bm\nabla V\m(\hat{\bm x}(t)) \; + \; q_t \; \hat{\bm E}_0\n(t,{\bm 0}) \; + \;
         \frac{2}{3} \, \frac{q_t^2}{c^3} \, \dddot{\hat{\bm x}}(t) \mez . \mez [d=3]}
      \end{eqnarray}
      {\color{magenta} {\sl Success! Mass renormalization has just been performed!}}\\
      The differential equation (\ref{eom-hat-x-t-take-3-d=3-final}) should be solved for $\hat{\bm x}(t)$\\
      with the initial conditions (\ref{ics-final-x}), (\ref{ics-final-p}), (\ref{dot-x-infinite-past}).\\
      Recall that $q_t$ equals to the physical charge $q$ of the electron at finite times.\\
      The first term on the r.h.s.~of (\ref{eom-hat-x-t-take-3-d=3-final}) displays\\
      standard scalar potential contribution to the force acting on the electron.\\
      The second term on the r.h.s.~of (\ref{eom-hat-x-t-take-3-d=3-final}) corresponds\\
      to the interaction picture electric force contribution acting on the electron,\\
      which we have already discussed before $[\,$see equation (\ref{electric-force-E0-expand}) and the accompanying remarks$\,]$.\\
      The third term on the r.h.s.~of (\ref{eom-hat-x-t-take-3-d=3-final}) displays\\
      {\color{blue} a finite contribution of the electric RR force which has survived the mass renormalization.\\
      This is the so called Abraham-Lorentz force,
      \be \label{ALF-d=3}
         {\color{blue} \hat{\bm F}_{\rm ALF}\n(t) \; = \; \frac{2}{3} \, \frac{q_t^2}{c^3} \, \dddot{\hat{\bm x}}(t)
         \mez . \mez [d=3]}
      \ee
      Most importantly, for $d=3$, the resulting Abraham-Lorentz force\\
      is local in time and proportional to the third time derivative $\dddot{\hat{\bm x}}(t)$.\\
      Hence the quantum dynamical evolution of $\hat{\bm x}(t)$ is Markovian.}\\
      $[\,$The term "Markovian" means that our equation of motion (\ref{eom-hat-x-t-take-3-d=3-final})
      has no memory, i.e., that\\ \phantom{$[\,$}previous history of the system (including the time instants $t'$ finitely
      separated from $t$)\\ \phantom{$[\,$}does not matter when propagating $\hat{\bm x}(t)$ from the time instant $t$
      forward.$\,]$\\ Physics intuition suggests that equation (\ref{eom-hat-x-t-take-3-d=3-final}) can be solved iteratively
      (perturbatively),\\ starting from the zeroth order interaction picture approximation
      $\hat{\bm x}(t) \doteq \hat{\bm x}_I\n(t) \; = \; e^{+\frac{i}{\hbar}\hat{H}_{\m{\rm A}}t} \; \hat{\bm x} \; e^{-\frac{i}{\hbar}\hat{H}_{\m{\rm A}}t}$\m.\\
      {\color{magenta} \sl We have accomplished the main goal of this course: To introduce the concept of mass renormalization.\\
      Now we shall proceed further and elaborate on important applications.}
\end{itemize}

\vspace*{+0.20cm}

{\color{blue} {\large \bf E. Renormalized perturbation theory for the atomic level shifts}}
\begin{itemize}
\item Perturbative treatment of the atomic energy levels dressed by the radiation field ($d=2,3$).\\
      Vacuum fluctuations in the radiation field modes invoke level shifts\\
      as well as spontaneous emission of photons from the excited atomic levels. 
\item Worked out in {\sl Appendices B.1, B.2, B.3, B.4}.\\
      {\sl Appendix B.1:} Introducing the game.\\
      {\sl Appendix B.2:} Naive perturbation theory, breaking down for $d=2,3$ due to UV divergences.\\
      {\sl Appendix B.3:} Renormalized perturbation theory for $d=2$.\\
      {\sl Appendix B.4:} Renormalized perturbation theory for $d=3$.
\end{itemize}

\vspace*{+0.20cm}

{\color{blue} {\large \bf F. Renormalized mean field theory}}
\begin{itemize}
\item We show explicitly how is the time dependent mean field (Hartree) method properly renormalized.\\
      Such a renormalized mean field approach is even suitable for numerical implementation.
\item Worked out in {\sl Appendix C}.
\end{itemize}

\newpage

{\color{blue} {\large \bf Appendix A: More on the electric RR force for $\bm d \bm = \bm 2$}}
\vspace*{-0.20cm}
\begin{itemize}
\item The purpose of this {\sl Appendix A} is to explore in more detail mathematical well definedness\\
      of the non-Markovian Abraham-Lorentz force term (\ref{ALF-d=2}), that is, of an entity
      \be \label{Appendix-A-take-1}
         \hat{\bm F}_{\rm ALF}\n(t) \; = \; - \;
         \int_{-\infty}^{t} \m q_{t'} \, \dddot{\hat{\bm x}}(t'\n) \, \ln\m\m\left(\frac{t-t'}{t_c}\right) {\rm d}t' \mez ;
      \ee
      which appears in the main text first inside equation (\ref{F-RR-analysis-take-4}),\\
      and which enters later on into the final Newton-Heisenberg equation of motion (\ref{eom-hat-x-t-take-3-d=2-final}) for the electron.
\item Let us look at a convolution
      \be \label{I-t-take-1}
         I_t \; = \; \int_{-\infty}^{t} \m f(t') \; \ln\m\left(\frac{t-t'}{t_c}\right) \, {\rm d}t' \mez ;
      \ee
      with $f(t'\n)$ representing an as yet unspecified $t'$-dependent real valued entity or signal\\
      (or possibly a hermitian operator).\\      
      Assume for a while that the integral of equation (\ref{I-t-take-1}) converges for the chosen $f(t'\n)$,\\
      and write conveniently
      \be \label{I-t-take-2}
         I_t \; = \; \lim_{\varepsilon \to +0} \; \int_{-\infty}^{t} \m f(t') \; e^{-\varepsilon(t-t'\n)} \ln\m\left(\frac{t-t'}{t_c}\right) \, {\rm d}t' \mez .
      \ee
      Assume for a while also that {\color{blue} $f(t'\n)$ is Fourier transformable,}
      \be \label{Appendix-A-condition-1}
         {\color{blue} f(t') \; = \; \int_{-\infty}^{+\infty} \m \tilde{f}(\Omega) \; e^{+i\Omega(t-t')} \, {\rm d}\Omega \mez .}
      \ee
      Real valuedness (or hermiticity) of $f(t')$ implies having
      \be \label{f-t'-real}
         f(t') \; = \; f^*\n(t') \mez , \mez \tilde{f}(\Omega) \; = \; \tilde{f}^*\m(-\Omega) \mez .
      \ee
      Then of course
      \be
         f(t') \; = \; \int_{0}^{+\infty} \m \tilde{f}(\Omega) \; e^{+i\Omega(t-t')} \, {\rm d}\Omega \;\, + \;\, {\rm c.c.} \mez ;
      \ee
      and our convolution formula $I_t=(\ref{I-t-take-2})$ can be rewritten into
      \begin{eqnarray} \label{I-t-take-3}
         \hspace*{-1.00cm} I_t & = & \lim_{\varepsilon \to +0} \; \int_{0}^{+\infty} \m {\rm d}\Omega \;\, \tilde{f}(\Omega) \; 
         \int_{-\infty}^{t} \m e^{+i\Omega(t-t')}\; e^{-\varepsilon(t-t'\n)} \ln\m\left(\frac{t-t'}{t_c}\right) \, {\rm d}t' \; + \; {\rm c.c.} \; = \nonumber\\
         \hspace*{-1.00cm} & = & \lim_{\varepsilon \to +0} \; \int_{0}^{+\infty} \m {\rm d}\Omega \;\, \tilde{f}(\Omega) \, \left\{ \,
         \int_{0}^{+\infty} \m e^{+i\Omega\tau}\; e^{-\varepsilon\tau} \ln\m\left(\frac{\tau}{t_c}\right) \, {\rm d}\tau \right\} \; + \; {\rm c.c.} \; = \nonumber\\
         \hspace*{-1.00cm} & = & \lim_{\varepsilon \to +0} \; \int_{0}^{+\infty} \m {\rm d}\Omega \;\, \tilde{f}(\Omega) \;
         \left\{ \frac{\ln\Bigl((\Omega+i\varepsilon)\,t_c\Bigr) + \gamma - i\,\frac{\pi}{2}}{i\,(\Omega+i\varepsilon)} \right\} \; + \; {\rm c.c.} \mez .
      \end{eqnarray}
      Here $\gamma=0.577\cdots$ is the Euler-Mascheroni constant.\\
      The $\{ \bm\cdots \}$ term of (\ref{I-t-take-3}) diverges to infinity both for $\Omega \to +0$ and for $\Omega \to +\infty$,\\
      this demands an appropriate analysis of convergence conditions for the integral $I_t=(\ref{I-t-take-3})$:
      \vspace*{-0.20cm}
      \begin{itemize}
      \item[$\star$] Let $\Omega_{\rm cut}$ be an arbitrary auxiliary parameter, $0<\Omega_{\rm cut}<\infty$.
      \item[$\star$] The $\Omega \to +0$ singularity of the $\{ \bm\cdots \}$ term is actually weak enough to be integrable,\\
                     provided only that
                     $$
                       {\color{blue} \tilde{f}(\Omega \to +0) \; {\rm admits \; a \; power \; series \; expansion}}
                     $$
                     \be \label{Appendix-A-condition-3}
                        {\color{blue} \tilde{f}(\Omega) \; = \; \tilde{f}(0) \; + \; \Omega \, \tilde{f}'\n(0) \; + \; {\cal O}(\Omega^2)}
                     \ee
                     $$
                       {\color{blue} {\rm with \; finite \; coefficients} \; \tilde{f}(0) \; {\rm and} \; \tilde{f}'\n(0) \; . }
                     $$
                     Valid since the entity
                     \be \label{tilde-f-0-entity}
                        \lim_{\varepsilon \to +0} \; \int_{0}^{\Omega_{\rm cut}} \m {\rm d}\Omega \;\, \tilde{f}(0) \;
                        \left\{ \frac{\ln\Bigl((\Omega+i\varepsilon)\,t_c\Bigr) + \gamma - i\,\frac{\pi}{2}}{i\,(\Omega+i\varepsilon)} \right\} \; + \; {\rm c.c.}
                     \ee
                     is mathematically well defined,\\
                     as one may easily check by exploiting the property $\tilde{f}(0)=\tilde{f}^*\n(0)$ coming from (\ref{f-t'-real}).\\
                     Indeed, direct calculation yields
                     \begin{eqnarray}
                        & & \int_{0}^{\Omega_{\rm cut}} \m {\rm d}\Omega \;\, \tilde{f}(0) \;
                        \left\{ \frac{\ln\Bigl((\Omega+i\varepsilon)\,t_c\Bigr) + \gamma - i\,\frac{\pi}{2}}{i\,(\Omega+i\varepsilon)} \right\} \; + \; {\rm c.c.} \; = \nonumber\\
                        & = & \left\{ \, 2\,{\rm arctan}\m\n\left(\frac{\varepsilon}{\Omega_{\rm cut}}\right) - \pi \, \right\}
                        \left\{ \, \ln\m\n\left(\sqrt{\Omega_{\rm cut}^2+\varepsilon^2}\,t_c\right) \, + \, \gamma \, \right\} \mez ;
                     \end{eqnarray}
                     and hence
                     \be
                        (\ref{tilde-f-0-entity}) \; = \; - \, \pi \, \tilde{f}(0) \, \Bigl( \ln(\Omega_{\rm cut}\,t_c) + \gamma \Bigr) \mez .
                     \ee
      \item[$\star$] The $\Omega \to +\infty$ singularity of the $\{ \bm\cdots \}$ term is encountered when dealing with an integral
                     \be \label{I-t-large-Omega-take-1}
                        \lim_{\varepsilon \to +0} \; \int_{\Omega_{\rm cut}}^{+\infty} \m {\rm d}\Omega \;\, \tilde{f}(\Omega) \;
                        \left\{ \frac{\ln\Bigl((\Omega+i\varepsilon)\,t_c\Bigr) + \gamma - i\,\frac{\pi}{2}}{i\,(\Omega+i\varepsilon)} \right\} \mez .
                     \ee
                     Here the limit of $\varepsilon \to +0$ can be taken inside the $\{ \bm\cdots \}$ term, converting (\ref{I-t-large-Omega-take-1}) into
                     \be \label{I-t-large-Omega-take-2}
                        \int_{\Omega_{\rm cut}}^{+\infty} \m {\rm d}\Omega \;\, \tilde{f}(\Omega) \;
                        \left\{ \frac{\ln\Bigl(\Omega\,t_c\Bigr) + \gamma - i\,\frac{\pi}{2}}{i\,\Omega} \right\} \mez .
                     \ee
                     The integral (\ref{I-t-large-Omega-take-2}) is granted to converge provided only that
                     \be \label{Appendix-A-condition-4}
                        {\color{blue} \tilde{f}(\Omega \to +\infty) \; {\rm is \; falling \; off \; to \; zero \; as} \; {\cal O}(\Omega^{-1}) \; {\rm or \; faster.}}
                     \ee
      \end{itemize}
      \vspace*{-0.20cm}
      We may proceed now in reverse order, starting from (\ref{I-t-take-3}) \& assumptions (\ref{Appendix-A-condition-1}), (\ref{Appendix-A-condition-3}), (\ref{Appendix-A-condition-4}),\\ and going stepwise up to (\ref{I-t-take-1}).
      Such an argument enables us to conclude\\ that the integral $I_t=(\ref{I-t-take-1})$ is a well defined entity\\
      provided only that the conditions (\ref{Appendix-A-condition-1}), (\ref{Appendix-A-condition-3}),
      (\ref{Appendix-A-condition-4}) are satisfied by $f(t')$.\\
      $[\,$Note that most signals $f(t')$ encountered in physics\\
      \phantom{$[\,$}would satisfy conditions as weak as required in (\ref{Appendix-A-condition-1}),
      (\ref{Appendix-A-condition-3}), (\ref{Appendix-A-condition-4}).$\,]$
\item Let us identify $f(t'\n)$ with $q_{t'} \, \dddot{\hat{\bm x}}(t'\n) \, \Theta(t-t'\n)$,
      as appropriate for our starting equation (\ref{Appendix-A-take-1}).\\
      {\color{blue} The basic question is now whether or not $f(t'\n) = q_{t'} \, \dddot{\hat{\bm x}}(t'\n) \, \Theta(t-t'\n)$\\
      matches the above conditions  (\ref{Appendix-A-condition-1}), (\ref{Appendix-A-condition-3}), (\ref{Appendix-A-condition-4}).}
\item Assume that\\
      the Newton-Heisenberg equation of motion (\ref{eom-hat-x-t-take-3-d=2-final})\\ supplemented by the initial conditions
      (\ref{ics-final-x}), (\ref{ics-final-p}), (\ref{dot-x-infinite-past})\\
      has been solved for the electron up to some given time instant $t$.\\
      Assume that the solution $\hat{\bm x}(t'\n)$ (with $t'\n \leq t$)
      behaves in a physically sensible manner,\\
      as appropriate for an electron trapped inside a bound potential well\\
      whose motion is slightly affected by the electric RR force.\\
      $[\,$To apply at least in some subspace ${\mathfrak S}$ of quantum state vectors\\
      \phantom{$[\,$}which correspond to non-relativistic motions of the electron.$\,]$\\
      If so, then the FTs
      \be \label{tilde-f-Omega-exists}
         \hat{\bm \chi}_\nu\n(\Omega) \; = \;
         \frac{1}{2\,\pi} \, \int_{-\infty}^{t} \m q_{t'} \, \frac{{\rm d}^\nu\hat{\bm x}(t'\n)}{{\rm d}t'^\nu} \; e^{-i\Omega(t-t')} \, {\rm d}t'
      \ee
      do exist for all $\Omega \in (-\infty,+\infty)$ as well as for all $\nu=\{1,2,3,4,\bm\cdots\}$, and moreover
      \be \label{chi-nu-falloff}
         \hat{\bm \chi}_\nu\n(\Omega \to \pm\infty) \; = \; 0 \mez . \mez [\,\nu \ge 1\,]
      \ee
      Valid since:
      \vspace*{-0.20cm}
      \begin{itemize}
      \item[$\circ$] The charge $q_{t'}$ possesses the properties (\ref{dot-q-t-zero}) and (\ref{q-t-infinite-past});
      \item[$\circ$] The entity $\frac{{\rm d}^\nu\hat{\bm x}(t'\n)}{{\rm d}t'^\nu}$ is bounded in magnitude, and\\
                     oscillates with $t'$ due to motion of the electron in our strictly bound potential $V\m({\bm x})$, see A.1.\\
                     Such that the average of $\frac{{\rm d}^\nu\hat{\bm x}(t'\n)}{{\rm d}t'^\nu}$ over a sufficiently long time interval is zero.\\
                     Moreover, the oscillations of $\frac{{\rm d}^\nu\hat{\bm x}(t'\n)}{{\rm d}t'^\nu}$ do not contain ultra high frequencies.
      \end{itemize}
      \vspace*{-0.20cm}
      After integrating by parts in (\ref{tilde-f-Omega-exists}) with the help of (\ref{dot-q-t-zero})-(\ref{q-t-infinite-past})
      one finds a neat recursive formula
      \be \label{chi-neat-formula}
         \hat{\bm \chi}_\nu\n(\Omega) \; = \; \frac{q_t}{2\,\pi} \, \frac{{\rm d}^{(\nu-1)}\hat{\bm x}(t)}{{\rm d}t^{(\nu-1)}}
         \; - \; i\,\Omega\,\hat{\bm \chi}_{\nu-1}\n(\Omega) \mez . \mez [\,\nu \ge 2\,]
      \ee
\item Note that $\tilde{f}(\Omega)=\hat{\bm \chi}_3\n(\Omega)$.\\
      {\color{blue} Condition (\ref{Appendix-A-condition-1}) is thus fulfilled by
      $f(t'\n) = q_{t'} \, \dddot{\hat{\bm x}}(t'\n) \, \Theta(t-t'\n)$.}
\item Consider now the regime of $\Omega \to 0$. Straightforward calculation exploiting (\ref{dot-q-t-zero}) and (\ref{q-t-infinite-past}) yields
      \begin{eqnarray}
        \tilde{f}(\Omega) & = &
        \frac{1}{2\,\pi} \, \int_{-\infty}^{t} \m q_{t'} \, \dddot{\hat{\bm x}}(t'\n) \; e^{-i\Omega(t-t')} \, {\rm d}t'
        \; = \nonumber\\
        & = & \frac{1}{2\,\pi} \, \int_{-\infty}^{t} \m q_{t'} \, \dddot{\hat{\bm x}}(t'\n) \,
        \Bigl\{ \, 1 \, - \, i\,\Omega\,(t-t') \, + \, {\cal O}(\Omega^2) \, \Bigr\} \, {\rm d}t' \; = \nonumber\\
        & = & \frac{1}{2\,\pi} \, q_{t} \, \bm\ddot{\hat{\bm x}}(t) \; - \;
        \frac{1}{2\,\pi} \; i\,\Omega \; q_{t} \, \bm\dot{\hat{\bm x}}(t) \; + \; {\cal O}(\Omega^2) \mez .
      \end{eqnarray}
      {\color{blue} Showing thus that the condition (\ref{Appendix-A-condition-3}) is matched by
      $f(t'\n) = q_{t'} \, \dddot{\hat{\bm x}}(t'\n) \, \Theta(t-t'\n)$.}
\item Consider now the regime of $\Omega \to \pm\infty$. Formula (\ref{chi-neat-formula}) implies having
      \be \label{tilde-f-Omega-large}
         \tilde{f}(\Omega) \; = \; \hat{\bm \chi}_{3}\n(\Omega) \; = \;
         \frac{1}{i\,\Omega} \, \frac{q_t}{2\,\pi} \, \dddot{\hat{\bm x}}(t) \; - \;
         \frac{\hat{\bm \chi}_4\n(\Omega)}{i\,\Omega} \mez .
      \ee
      Insights (\ref{tilde-f-Omega-large}) and (\ref{chi-nu-falloff}) enable us to conclude that\\
      {\color{blue} the condition (\ref{Appendix-A-condition-4}) is satisfied by
      $f(t'\n) = q_{t'} \, \dddot{\hat{\bm x}}(t'\n) \, \Theta(t-t'\n)$.}
\item Summarizing the contents of the present {\sl Appendix A},\\
      we have shown that the non-Markovian electric RR force term (\ref{Appendix-A-take-1}) is well defined,\\
      in spite of singularities of the logarithmic memory kernel $\ln\m\m\left(\frac{t-t'}{t_c}\right)$.\\
      This conclusion holds provided only that\\
      the corresponding Heisenberg picture position operator $\hat{\bm x}(t')$ $(t'\n \leq t)$\\
      behaves in a physically sensible manner,\\
      as appropriate for non-relativistic motions of an electron\\
      trapped inside an external bound potential well $V\m({\bm x})$\\
      whose motion is only slightly affected by the electric RR force.\\
      In particular, our non-Markovian electric RR force term (\ref{Appendix-A-take-1}) is well defined\\
      when one approximates $\hat{\bm x}(t')$ by its interaction picture counterpart,
      \be \label{hat-bm-x-I}
         \hat{\bm x}_I\n(t') \; = \; e^{+\frac{i}{\hbar}\hat{H}_{\m{\rm A}}t'} \; \hat{\bm x} \; e^{-\frac{i}{\hbar}\hat{H}_{\m{\rm A}}t'} \mez ;
      \ee
      {\sl cf.}~equation (\ref{ics-final-x}).\\
      Hence the Newton-Heisenberg equation of motion (\ref{eom-hat-x-t-take-3-d=2-final})\\
      supplemented by the initial conditions (\ref{ics-final-x}), (\ref{ics-final-p}), (\ref{dot-x-infinite-past})\\
      can be solved iteratively (perturbatively),\\
      starting from the initial zeroth order approximation $\hat{\bm x}_I\n(t')=(\ref{hat-bm-x-I})$.
\end{itemize}

\vspace*{+0.20cm}

{\color{blue} {\large \bf Appendix B: Perturbation theory for the atomic level shifts}}\\

{\bf Appendix B.1: Introducing the game}
\vspace*{-0.20cm}
\begin{itemize}
\item Consider first our model atom alone.\\
      The corresponding atomic Hamiltonian takes the form $\hat{H}_{\m{\rm A}} = (\ref{H-A-def})$.\\
      Since the potential $V\m(\hat{\bm x})$ is strictly bound, only discrete energy levels are permitted.\\
      We write as usual
      \be
         \hat{H}_{\m{\rm A}} \, | \psi_j \ra_{\m A} \; = \; E_j^A \, | \psi_j \ra_{\m A} \mez ;
      \ee
      with $j=0,1,2,\bm\cdots$.
\item Secondly, consider the quantum radiation field alone.\\
      The corresponding Hamiltonian takes the form $\hat{H}_{{\rm R}}=(\ref{H-R-def})$.\\
      Recall that $\hat{H}_{{\rm R}}$ describes an infinite ensemble of mutually uncoupled harmonic oscillators,\\
      these are labeled by modal indices $({\bm k}\wp)$. An oscillator $({\bm k}\wp)$ possesses the circular frequency $\omega_k$.\\
      The ground state $| {\rm vac} \ra_{\rm R}$ of $\hat{H}_{{\rm R}}$ is annihilated by all the $\hat{a}_{{\bm k}\wp}$\m's.\\
      All the other eigenstates of $\hat{H}_{{\rm R}}$ are built up via repeated actions of the $\hat{a}_{{\bm k}\wp}^\dagger$\m's. Such that
      \be
         | ({\bm k_1}\wp_1) \bm\cdots  ({\bm k_n}\wp_n) \ra_{\m R} \; = \; \frac{1}{\sqrt{n!}} \; \hat{a}_{{\bm k}_1\wp_1}^\dagger
         \bm\cdots \, \hat{a}_{{\bm k}_n\wp_n}^\dagger \, | {\rm vac} \ra_{\rm R} \mez . \mez [n \ge 0]
      \ee
      Clearly,
      \be
         \hat{H}_{{\rm R}} \, | ({\bm k_1}\wp_1) \bm\cdots  ({\bm k_n}\wp_n) \ra_{\m R} \; = \;
         \Bigl( \hbar\omega_{k_1} \, + \, \bm\cdots \, + \, \hbar\omega_{k_n} \Bigr) \, | ({\bm k_1}\wp_1) \bm\cdots  ({\bm k_n}\wp_n) \ra_{\m R} \mez . \mez [n \ge 0]
      \ee
\item Thirdly, consider both our model atom {\sl (subsystem \#1)}\\
      and the quantized radiation field {\sl (subsystem \#2)}.\\
      When the subsystems {\sl \#1} and {\sl \#2} are mutually uncoupled (non-interacting),\\
      the total Hamiltonian of our problem takes of course the form $\hat{H}_{\rm free}=(\ref{H-free-def})$,\\
      and the associated stationary eigenstates result trivially. One has
      \be \label{hat-H-free-eigenproblem}
         \hat{H}_{\rm free} \, | \psi_j \ra_{\m A} \, | ({\bm k_1}\wp_1) \bm\cdots  ({\bm k_n}\wp_n) \ra_{\m R} \; = \;
         \Bigl( E_j^A \, + \, \hbar\omega_{k_1} \, + \, \bm\cdots \, + \, \hbar\omega_{k_n} \Bigr) \, | \psi_j \ra_{\m A} \, | ({\bm k_1}\wp_1) \bm\cdots  ({\bm k_n}\wp_n) \ra_{\m R}
         \mez .
      \ee
\item Fourthly, consider both our model atom {\sl (subsystem \#1)}\\
      and the quantized radiation field {\sl (subsystem \#2)},\\
      coupled together as specified in the Hamiltonian (\ref{H}).\\
      Recall that the Hamiltonian (\ref{H}) is parameterized by a finite (yet arbitrarily large) value of $\alpha>0$.\\
      Moreover, the mass renormalization prescriptions [(\ref{ALF-d=2}) or (\ref{ALF-d=3})] supplement the formula (\ref{H}).\\
      Hereafter we shall be concerned only with physical properties\\
      of our system $[\,$$\equiv$ mutually interacting subsystems {\sl \#1} and {\sl \#2}$\,]$\\
      pertaining to finite time instants $t$.\\
      At finite time instants, the charge $q_t$ of our electron equals to its physical value $q$.\\
      This enables us to update (\ref{H}) \& [(\ref{ALF-d=2}) or (\ref{ALF-d=3})] into
      \be \label{H-no-t}
         \hat{H} \; = \; \frac{1}{2\,m_o(\alpha)} \left( \hat{\bm p} \, - \, \frac{q}{c} \, \hat{\bm A}[\alpha] \right)^{\m\m\m 2} + \; V\m(\hat{\bm x}) \; + \; \hat{H}_{{\rm R}} \mez ;
      \ee
      with
      \be \label{m-ren-d=2-no-t}
         m_o(\alpha) \; + \; \frac{q^2}{4\,\pi\,c^2} \; \ln\m\m\left(\frac{\alpha}{\alpha_0}\right) \; = \; m \mez ; \mez [d=2]
      \ee
      or
      \be \label{m-ren-d=3-no-t}
         m_o(\alpha) \; + \; \frac{4}{3} \, \frac{q^2}{c^2} \, \sqrt{\frac{\alpha}{2\,\pi}} \; = \; m \mez . \mez [d=3]
      \ee
      In the Hamiltonian formula (\ref{H-no-t}), we have adopted for the sake of notational convenience\\ an auxiliary shorthand symbol
      \begin{eqnarray} \label{hat-A-alpha-def-B.1}
         \hat{\bm A}[\alpha] & \equiv & \int_{{\mathbb R}^d} \m \varrho({\bm y}) \; \hat{\bm A}({\bm y}) \; {\rm d}^d\n y \nonumber\\
         & = & \int_{{\mathbb R}^2} \m {\rm d}^d\n k \, \sum_\wp \, \frac{1}{2\,\pi} \, \sqrt{\frac{\hbar\,c^2}{\omega_k}} \;
         e^{-\frac{{\bm k}^2}{4\,\alpha}} \, \hat{a}_{{\bm k}\wp} \; {\bm \varepsilon}_{{\bm k}\wp} \; + \; {\rm c.c.} \mez .
      \end{eqnarray}
      The second line of (\ref{hat-A-alpha-def-B.1}) comes from (\ref{int-varrho-A}).\\
      An eigenvalue problem of the Hamiltonian $\hat{H}=(\ref{H-no-t})$ is formally written down as
      \be \label{hat-H-eigenproblem}
         \hat{H} \, | \Psi \ra \; = \; E \, | \Psi \ra \mez .
      \ee
      In what follows, we shall investigate the eigenproblem (\ref{hat-H-eigenproblem})\\
      in the weak coupling regime\\
      (such that the charge $q$ of the electron is set to be arbitrarily small but nonzero),\\
      using appropriate perturbation expansions in $q$.\\
      $[\,$The case of $q=0$ is resolved trivially, since $\hat{H}|_{q=0}=\hat{H}_{\rm free}$ and
      $(\ref{hat-H-eigenproblem})|_{q=0}$ reduces to $(\ref{hat-H-free-eigenproblem})$.$\,]$\\
      Our interest actually consists in finding only those stationary eigenstates of $\hat{H}=(\ref{H-no-t})$\\
      which reduce in the limit of $q \to 0$ just to
      \be \label{Psi-(0)-unperturbed}
         | \Psi^{(0)} \ra \; = \; | \psi_j \ra_{\m A} \, | {\rm vac} \ra_{\m R} \mez ;
      \ee
      {\sl cf.}~equation $(\ref{hat-H-free-eigenproblem})|_{n=0}$. Here $j$ represents an arbitrary chosen atomic stationary state.\\
      {\color{blue} Loosely speaking, our goal is to investigate those stationary eigenstates of $\hat{H}=(\ref{H-no-t})$\\
      which describe a given stationary state $| \psi_j \ra_{\m A}$ of our model atom\\
      dressed by zero temperature bath of the radiation field modes\\ (i.e., dressed by the electromagnetic vacuum).\\
      Vacuum fluctuations in the radiation field modes invoke atomic level shifts,\\
      as well as spontaneous emission of photons from the excited atomic levels $j \ge 1$.}
\end{itemize}

\vspace*{+0.20cm}

{\bf Appendix B.2: Naive perturbation theory}
\vspace*{-0.20cm}
\begin{itemize}
\item Before implementing explicitly the just outlined program\\
      of renormalized perturbation theory for the atomic level shifts,\\
      it is instructive to try out a naive treatment which ignores the matters of mass renormalization.\\
      Such a naive formulation would replace our starting Hamiltonian $\hat{H}=(\ref{H-no-t})$\\
      by a heuristically plausible (yet deceptive, treacherous) prescription
      \be \label{H-naive}
         \hat{H} \; = \; \frac{1}{2\,m} \left( \hat{\bm p} \, - \, \frac{q}{c} \, \hat{\bm A}({\bm 0}) \right)^{\m\m 2}
         + \; V\m(\hat{\bm x}) \; + \; \hat{H}_{{\rm R}} \mez .
      \ee
      Here $m$ and $q$ stand for the physical mass and charge of our electron,
      and $\hat{\bm A}({\bm 0})$ comes from (\ref{hat-A-def}).\\
      Clearly, the Hamiltonian (\ref{H-naive}) is obtainable from (\ref{H-no-t})\\
      via identifying hastily $m_o(\alpha)$ with $m$ and taking hastily $\alpha \to +\infty$.\\
      $[\,$Such a hasty attitude would hardly seem suspicious to those who never heard of renormalization.$\,]$\\
      {\color{magenta} {\sl As we shall see shortly, the naive Hamiltonian prescription (\ref{H-naive}) leads to an epic failure!\\
      This is not surprising at all for everyone who passed through Sections C and D of the main text.}}
\item We wish to solve the eigenvalue problem (\ref{hat-H-eigenproblem})\\
      by employing the standard Rayleigh-Schr\"{o}dinger perturbation theory (RSPT or just PT)\\
      in the coupling parameter $q$.\\
      It is thus convenient to express the Hamiltonian $\hat{H}=(\ref{H-naive})$ in the form of a sum
      \be
         \hat{H} \; = \; \hat{H}_{\rm free} \; + \; \hat{W} \mez ;
      \ee
      where $\hat{H}_{\rm free}=(\ref{H-free-def})$ is understood as the unperturbed reference part, and
      \be \label{W-naive}
         \hat{W} \; = \; -\,\frac{q}{mc} \, \hat{\bm A}({\bm 0}) \m\bm\cdot\m \hat{\bm p} \; + \; \frac{q^2}{2\,m\,c^2} \, \hat{\bm A}^2\n({\bm 0})
      \ee
      determines the perturbation.\\
      Our unperturbed reference state $| \Psi^{(0)} \ra = (\ref{Psi-(0)-unperturbed})$\\
      corresponds to an unperturbed energy eigenvalue $E^{(0)}=E_j^A$\m,\\
      consistently with relation $(\ref{hat-H-free-eigenproblem})|_{n=0}$.
\item An application of the standard RSPT is now straightforward.\\
      Our PT1 energy correction equals to
      \be
         E^{(1)} \; = \; \la \Psi^{(0)} | \, -\,\frac{q}{mc} \, \hat{\bm A}({\bm 0}) \m\bm\cdot\m \hat{\bm p} \, | \Psi^{(0)} \ra \; = \; 0 \mez .
      \ee
      Our PT1 eigenvector correction reads as
      \be \label{Psi-PT1-take-1-naive}
         | \Psi^{(1)} \ra \; = \; \sum_{j'} \, \int_{{\mathbb R}^d} \m {\rm d}^d\m k \, \sum_\wp \,
         \frac{ _A\m\la \psi_{j'}\n | \, _R\la ({\bm k}\wp) | \,\m\m \left(-\frac{q}{mc} \, \hat{\bm A}({\bm 0}) \m\bm\cdot\m \hat{\bm p}\right) \m\m\,
         | \psi_j \ra_{\m A} \, | {\rm vac} \ra_{\m R} }{E_j^A \, - \, E_{j'}^A \, - \, \hbar\omega_{k} \, + \, i\,\varepsilon} \; | \psi_{j'}\n \ra_{\m A}
         \, | ({\bm k}\wp) \ra_{\m R} \mez .
      \ee
      Accordingly, our PT2 energy correction takes the form
      \begin{eqnarray} \label{E-PT2-take-1-naive}
         E^{(2)} & = & \sum_{j'} \, \int_{{\mathbb R}^d} \m {\rm d}^d\m k \, \sum_\wp \, \frac{ \left| \, _A\m\la \psi_{j'}\n |
         \, _R\la ({\bm k}\wp) | \,\m\m \left(-\frac{q}{mc} \, \hat{\bm A}({\bm 0}) \m\bm\cdot\m \hat{\bm p}\right) \m\m\, | \psi_j \ra_{\m A}
         \, | {\rm vac} \ra_{\m R} \, \right|^2}{E_j^A \, - \, E_{j'}^A \, - \, \hbar\omega_{k} \, + \, i\,\varepsilon} \\
         & + & _A\m\la \psi_j | \, _R\la {\rm vac} | \,\m\m \left( \frac{q^2}{2\,m\,c^2} \, \hat{\bm A}^2\n({\bm 0}) \right) \m\m\, | \psi_j \ra_{\m A} \, | {\rm vac} \ra_{\m R} \mez . \nonumber
      \end{eqnarray}
      Denominators of the formulas (\ref{Psi-PT1-take-1-naive}) and (\ref{E-PT2-take-1-naive}) contain a small imaginary factor
      $(+i\,\varepsilon)$ where $\varepsilon \to +0$.\\ Clearly, the mentioned $(+i\,\varepsilon)$ factor cures singularities\\
      which otherwise would occur in (\ref{Psi-PT1-take-1-naive}) and (\ref{E-PT2-take-1-naive}) whenever $j \ge 1$,\\
      at the resonant modal frequencies $\hbar\omega_k=E_j^A \, - \, E_{j'<j}^A$.\\
      It is important to recall at this point that all the $j \ge 1$ atomic levels are metastable\\
      (decaying due to spontaneous emission).\\
      Metastable states (resonances) are studied by the nonhermitian quantum mechanics, see Ref.~\cite{NHQM}.\\
      The just discussed insertion of the $(+i\,\varepsilon)$ factor into the denominators of (\ref{Psi-PT1-take-1-naive})
      and (\ref{E-PT2-take-1-naive})\\ represents a common practice of the nonhermitian RSPT.\\
      Note that the $(+i\,\varepsilon)$ factor makes $E^{(2)}$ complex valued for $j \ge 1$.\\
      $[\,$Negative imaginary part $\sim$ decay rate of the $j$-th level due to spontaneous emission.$\,]$\\
      Detailed justification for including the $(+i\,\varepsilon)$ term\\
      in the present context of spontaneous emission can be found in my recent work, see Ref.~\cite{SE}.
\item Proceeding further, we wish to evaluate the PT corrections (\ref{Psi-PT1-take-1-naive}) and (\ref{E-PT2-take-1-naive}) more explicitly.\\
      For this purpose we introduce shorthand notation
      \be
         {\bm p}_{j'\n j} \; = \; _A\m\la \psi_{j'}\n | \, \hat{\bm p} \, | \psi_j \ra_{\m A} \, \mez .
      \ee
      We also evaluate an action of $\hat{\bm A}({\bm 0})$ and $\hat{\bm A}^2\n({\bm 0})$ on the vacuum state $| {\rm vac} \ra_{\m R}$. One has
      \be
         \hat{\bm A}({\bm 0}) \, | {\rm vac} \ra_{\m R} \; = \; \int_{{\mathbb R}^d} \m {\rm d}^d\m k \, \sum_\wp \, \frac{1}{2\,\pi} \, \sqrt{\frac{\hbar\,c^2}{\omega_k}} \;
         | ({\bm k}\wp) \ra_{\m R} \; {\bm \varepsilon}_{{\bm k}\wp} \mez ;
      \ee
      and
      \be
         \hat{\bm A}^2\n({\bm 0}) \, | {\rm vac} \ra_{\m R} \; = \; \int_{{\mathbb R}^d} \m {\rm d}^d\m k \, \sum_\wp \, \frac{1}{4\,\pi^2} \, \frac{\hbar\,c^2}{\omega_k} \;
         | {\rm vac} \ra_{\m R} \; + \; {\rm a\phantom{i}two\phantom{i}photon\phantom{i}term} \mez .
      \ee
      Subsequently one may work out the relevant matrix elements
      \be
         _A\m\la \psi_{j'}\n | \, _R\la ({\bm k}\wp) | \,\m\m \left(-\frac{q}{mc} \, \hat{\bm A}({\bm 0}) \m\bm\cdot\m \hat{\bm p}\right) \m\m\, | \psi_j \ra_{\m A}
         \, | {\rm vac} \ra_{\m R} \; = \; -\frac{q}{2\,\pi\,m} \, \sqrt{\frac{\hbar}{\omega_k}} \; {\bm \varepsilon}_{{\bm k}\wp} \m\bm\cdot\m {\bm p}_{j'\n j} \mez ;
      \ee
      and
      \be
         _A\m\la \psi_j | \, _R\la {\rm vac} | \,\m\m \left( \frac{q^2}{2\,m\,c^2} \, \hat{\bm A}^2\n({\bm 0}) \right) \m\m\, | \psi_j \ra_{\m A} \, | {\rm vac} \ra_{\m R} \; = \;
         \frac{q^2}{2\,m\,c^2} \, \int_{{\mathbb R}^d} \m {\rm d}^d\m k \, \sum_\wp \, \frac{1}{4\,\pi^2} \, \frac{\hbar\,c^2}{\omega_k} \mez .
      \ee
\item Returning back to the PT corrections (\ref{Psi-PT1-take-1-naive}) and (\ref{E-PT2-take-1-naive}), we find that
      \be \label{Psi-PT1-take-2-naive}
         | \Psi^{(1)} \ra \; = \; -\frac{q}{2\,\pi\,m} \, \sum_{j'} \, \int_{{\mathbb R}^d} \m {\rm d}^d\m k \, \sum_\wp \,
         \sqrt{\frac{\hbar}{\omega_k}} \, \frac{ {\bm \varepsilon}_{{\bm k}\wp} \m\bm\cdot\m {\bm p}_{j'\n j} }{\left(E_j^A \, - \, E_{j'}^A
         \, - \, \hbar\omega_{k} \, + \, i\,\varepsilon\right)} \; | \psi_{j'}\n \ra_{\m A} \, | ({\bm k}\wp) \ra_{\m R} \mez ;
      \ee
      and
      \begin{eqnarray} \label{E-PT2-take-2-naive}
         E^{(2)} & = & \frac{q^2\,\hbar}{4\,\pi^2\,m^2} \, \sum_{j'} \, \int_{{\mathbb R}^d} \m {\rm d}^d\m k \;\, \omega_k^{-1} \sum_\wp \, \frac{ \Bigl| \, {\bm \varepsilon}_{{\bm k}\wp} \m\bm\cdot\m {\bm p}_{j'\n j} \, \Bigr|^2}{E_j^A \, - \, E_{j'}^A \, - \, \hbar\omega_{k} \, + \, i\,\varepsilon} \\
         & + & \frac{q^2\,\hbar}{4\,\pi^2\,m} \, \int_{{\mathbb R}^d} \m {\rm d}^d\n k \;\, \omega_k^{-1} \mez . \nonumber
      \end{eqnarray}      
      These are the final explicit results of our naive RSPT.
\item From now on, we shall focus on examining more closely the PT2 energy correction $E^{(2)}=(\ref{E-PT2-take-2-naive})$.\\
      The term
      \be \label{E-(2)-discard-naive}
         \frac{q^2\,\hbar}{4\,\pi^2\,m} \, \int_{{\mathbb R}^d} \m {\rm d}^d\n k \;\, \omega_k^{-1}
      \ee
      is obviously UV divergent, yet it does not depend upon the atomic label $j$.\\
      If so, it can be discarded as a physically unimportant constant energy shift\\
      affecting uniformly all the atomic levels.\\
      $[\,$Exactly the same reasoning was implicitly encountered before in A.1,\\
      \phantom{$[\,$}when we eliminated the divergent ZPE part of $\hat{H}_{{\rm R}}=(\ref{H-R-def})$.$\,]$\\
      After discarding the contribution (\ref{E-(2)-discard-naive}), we are left with
      \begin{eqnarray} \label{E-PT2-take-3-naive}
         E^{(2)} & = & \frac{q^2\,\hbar}{4\,\pi^2\,m^2} \, \sum_{j'} \, \int_{{\mathbb R}^d} \m {\rm d}^d\m k \;\, \frac{1}{\omega_k}
         \, \frac{1}{\Bigl( E_j^A \, - \, E_{j'}^A \, - \, \hbar\omega_{k} \, + \, i\,\varepsilon \Bigr)} \; \sum_\wp \, 
         \Bigl| \, {\bm \varepsilon}_{{\bm k}\wp} \m\bm\cdot\m {\bm p}_{j'\n j} \, \Bigr|^2 \mez .
      \end{eqnarray}
      Since $\int_{{\mathbb R}^d} {\rm d}^d\m k = \int_{0}^{\infty} k^{d-1} \, {\rm d}k \int_{{\mathbb S}^{d-1}} {\rm d}^{d-1}\Omega_{\bm k}$, one may even write
      \begin{eqnarray} \label{E-PT2-take-4-naive}
         E^{(2)} & = & \frac{q^2\,\hbar}{4\,\pi^2\,m^2} \, \sum_{j'} \, \lim_{K \to +\infty} \, \int_{0}^{K} k^{d-1} \, {\rm d}k \;\, \frac{1}{\omega_k}
         \, \frac{1}{\Bigl( E_j^A \, - \, E_{j'}^A \, - \, \hbar\omega_{k} \, + \, i\,\varepsilon \Bigr)} \nonumber\\ & &
         \int_{{\mathbb S}^{d-1}} {\rm d}^{d-1}\Omega_{\bm k} \, \sum_\wp \, \Bigl| \, {\bm \varepsilon}_{{\bm k}\wp} \m\bm\cdot\m {\bm p}_{j'\n j} \, \Bigr|^2 \mez .
      \end{eqnarray}
      Observe now that the $k$-integral of equation (\ref{E-PT2-take-4-naive})\\
      refers to an integrand which behaves as $k^{d-3}$ for $k \to +\infty$. Hence:\\
      $\circ$ For $d=3$, our energy correction $E^{(2)}=(\ref{E-PT2-take-4-naive})$ is UV divergent as $K^{+1}$\n.\\
      \phantom{$\circ$} $[\,$In passing we note that $\Im E^{(2)}$ actually does not diverge, only $\Re E^{(2)}$ is a troublemaker.$\,]$\\
      $\circ$ For $d=2$, our energy correction $E^{(2)}=(\ref{E-PT2-take-4-naive})$ is UV divergent as $\ln K$\n.\\
      \phantom{$\circ$} $[\,$In passing we note that $\Im E^{(2)}$ actually does not diverge, only $\Re E^{(2)}$ is a troublemaker.$\,]$\\
      {\color{magenta} Our naively calculated leading order RSPT energy correction $E^{(2)}=(\ref{E-PT2-take-4-naive})$\\ turns out to be UV divergent!\\
      Epic failure of the naive RSPT formulation which ignores mass renormalization!}\\
      The UV divergences just encountered for $d=2,3$ can be cured only\\
      by applying the renormalized perturbation theory, see the {\sl Appendices B.3} and {\sl B.4} below.
\end{itemize}

\vspace*{+0.20cm}

{\bf Appendix B.3: Renormalized perturbation theory for $\bm d \bm = \bm 2$}
\vspace*{-0.20cm}
\begin{itemize}
\item We wish to implement now\\ the renormalized Rayleigh-Schr\"{o}dinger perturbation theory (RSPT)\\
      for the atomic level shifts, assuming $d=2$.\\
      Consider our starting Hamiltonian $\hat{H}=(\ref{H-no-t})|_{d=2}$\\
      supplemented by the mass renormalization prescription (\ref{m-ren-d=2-no-t}).\\
      Our goal is to solve an eigenvalue problem (\ref{hat-H-eigenproblem})\\
      by employing the standard RSPT in the coupling parameter $q$.\\
      It is thus necessary to express $\hat{H}=(\ref{H-no-t})|_{d=2}$ as a power series in $q$.\\
      This is accomplished through writing
      \begin{eqnarray} \label{mass-counterterm-d=2}
         \frac{1}{2\,m_o(\alpha)} & = & \frac{1}{2\,m} \, \frac{m}{m_o(\alpha)} \; = \; \frac{1}{2\,m} \; \frac{1}{1 \, - \,
          \frac{q^2}{4\,\pi\,mc^2} \ln\m\m\left(\frac{\alpha}{\alpha_0}\right)} \; = \nonumber\\
         & = & \frac{1}{2\,m} \; + \; \frac{1}{2\,m} \, \sum_{\ell=1}^{\infty} \left( \frac{q^2}{4\,\pi\,mc^2}
         \ln\m\m\left(\frac{\alpha}{\alpha_0}\right)\right)^{\m\m\ell} \mez .
      \end{eqnarray}
      {\color{blue} Entity $\frac{1}{2\,m} \, \sum_{\ell=1}^{\infty} \left( \frac{q^2}{4\,\pi\,mc^2}\ln\m\m\left(\frac{\alpha}{\alpha_0}\right)\right)^{\m\m\ell}$
      is the so called {\sl mass counterterm}.}\\
      \hspace*{+0.50cm} Note that the just displayed infinite series (\ref{mass-counterterm-d=2})\\
      \hspace*{+0.50cm} is not summable in an ordinary sense for $\alpha \to +\infty$, thus some readers may feel uneasy here.\\
      \hspace*{+0.50cm} In these introductory lecture notes, we avoid delving into\\
      \hspace*{+0.50cm} mathematical peculiarities accompanying the summation (\ref{mass-counterterm-d=2}).\\
      \hspace*{+0.50cm} Such a light hearted attitude is common for vast majority of the present day QFT theorists,\\
      \hspace*{+0.50cm} who keep on using weird expansions of the form similar to (\ref{mass-counterterm-d=2}) daily,\\
      \hspace*{+0.50cm} in their studies of high energy physics within (and beyond) the Standard Model!\\
      The Hamiltonian $\hat{H}=(\ref{H-no-t})|_{d=2}$ is then split into a sum
      \be
         \hat{H} \; = \; \hat{H}_{\rm free} \; + \; \hat{W} \mez ;
      \ee
      where $\hat{H}_{\rm free}=(\ref{H-free-def})$ is understood as the unperturbed reference part, and
      \be \label{W-Appendix-B.3}
         \hat{W} \; = \; -\,\frac{q}{mc} \, \hat{\bm A}[\alpha] \m\bm\cdot\m \hat{\bm p} \; + \; \frac{q^2}{2\,m\,c^2} \, \hat{\bm A}^2\n[\alpha]
         \; + \; \frac{q^2}{8\,\pi\,m^2c^2} \ln\m\m\left(\frac{\alpha}{\alpha_0}\right) \hat{\bm p}^2 \; + \; {\cal O}(q^3)
      \ee
      determines the (adequately renormalized) perturbation.\\
      {\color{blue} Note that the perturbation $\hat{W}=(\ref{W-Appendix-B.3})$ contains all the mass counterterms!\\
      This is what makes $\hat{W}=(\ref{W-Appendix-B.3})$ small compared to $\hat{H}_{\rm free}=(\ref{H-free-def})$ for $q$ small enough.}\\
      $[\,$Our subsequent calculation of the atomic level shifts will confirm the just presented anticipation.$\,]$\\
      {\color{blue} Note that the naive perturbation term $\hat{W}=(\ref{W-naive})$ from B.2 does not contain any mass counterterms!\\
      This is why $\hat{W}=(\ref{W-naive})$ represents an infinite (and in fact ill-defined) contribution to $\hat{H}=(\ref{H-naive})$.}\\
      Much as in B.2, our unperturbed reference state $| \Psi^{(0)} \ra = (\ref{Psi-(0)-unperturbed})$\\
      corresponds to an unperturbed energy eigenvalue $E^{(0)}=E_j^A$\m,\\
      consistently with relation $(\ref{hat-H-free-eigenproblem})|_{n=0}$.
\item An application of the standard RSPT is now straightforward.\\
      Our PT1 energy correction equals to
      \be
         E^{(1)} \; = \; \la \Psi^{(0)} | \, -\,\frac{q}{mc} \, \hat{\bm A}[\alpha] \m\bm\cdot\m \hat{\bm p} \; | \Psi^{(0)} \ra \; = \; 0 \mez .
      \ee
      Our PT1 eigenvector correction reads as      
      \be \label{Psi-PT1-take-1-d=2}
         | \Psi^{(1)} \ra \; = \; \sum_{j'} \, \int_{{\mathbb R}^2} \m {\rm d}^2\n k \, \sum_\wp \,
         \frac{ _A\m\la \psi_{j'}\n | \, _R\la ({\bm k}\wp) | \,\m\m \left(-\frac{q}{mc} \, \hat{\bm A}[\alpha] \m\bm\cdot\m \hat{\bm p}\right) \m\m\,
         | \psi_j \ra_{\m A} \, | {\rm vac} \ra_{\m R} }{E_j^A \, - \, E_{j'}^A \, - \, \hbar\omega_{k} \, + \, i\,\varepsilon} \; | \psi_{j'}\n \ra_{\m A}
         \, | ({\bm k}\wp) \ra_{\m R} \mez .
      \ee
      Accordingly, our PT2 energy correction takes the form      
      \begin{eqnarray} \label{E-PT2-take-1-d=2}
         E^{(2)} & = & \sum_{j'} \, \int_{{\mathbb R}^2} \m {\rm d}^2\n k \, \sum_\wp \, \frac{ \left| \, _A\m\la \psi_{j'}\n |
         \, _R\la ({\bm k}\wp) | \,\m\m \left(-\frac{q}{mc} \, \hat{\bm A}[\alpha] \m\bm\cdot\m \hat{\bm p}\right) \m\m\, | \psi_j \ra_{\m A}
         \, | {\rm vac} \ra_{\m R} \, \right|^2}{E_j^A \, - \, E_{j'}^A \, - \, \hbar\omega_{k} \, + \, i\,\varepsilon} \\
         & + & _A\m\la \psi_j | \, _R\la {\rm vac} | \,\m\m \left( \frac{q^2}{2\,m\,c^2} \, \hat{\bm A}^2\n[\alpha] \right) \m\m\, | \psi_j \ra_{\m A} \, | {\rm vac} \ra_{\m R} \nonumber\\
         & + & \frac{q^2}{8\,\pi\,m^2c^2} \, \ln\m\m\left(\frac{\alpha}{\alpha_0}\right) \,
         _A\m\la \psi_j | \, _R\la {\rm vac} | \, \hat{\bm p}^2 \, | \psi_j \ra_{\m A} \, | {\rm vac} \ra_{\m R} \mez . \nonumber
      \end{eqnarray}
      Denominators of the formulas (\ref{Psi-PT1-take-1-d=2}) and (\ref{E-PT2-take-1-d=2}) contain a small imaginary factor
      $(+i\,\varepsilon)$ where $\varepsilon \to +0$.\\ Clearly, the mentioned $(+i\,\varepsilon)$ factor cures singularities\\
      which otherwise would occur in (\ref{Psi-PT1-take-1-d=2}) and (\ref{E-PT2-take-1-d=2}) whenever $j \ge 1$,\\
      at the resonant modal frequencies $\hbar\omega_k=E_j^A \, - \, E_{j'<j}^A$.\\
      It is important to recall at this point that all the $j \ge 1$ atomic levels are metastable\\
      (decaying due to spontaneous emission).\\
      Metastable states (resonances) are studied by the nonhermitian quantum mechanics, see Ref.~\cite{NHQM}.\\
      The just discussed insertion of the $(+i\,\varepsilon)$ factor into the denominators of (\ref{Psi-PT1-take-1-d=2})
      and (\ref{E-PT2-take-1-d=2})\\ represents a common practice of the nonhermitian RSPT.\\
      Note that the $(+i\,\varepsilon)$ factor makes $E^{(2)}$ complex valued for $j \ge 1$.\\
      $[\,$Negative imaginary part $\sim$ decay rate of the $j$-th level due to spontaneous emission.$\,]$\\
      Detailed justification for including the $(+i\,\varepsilon)$ term\\
      in the present context of spontaneous emission can be found in my recent work, see Ref.~\cite{SE}.
\item Proceeding further, we wish to evaluate the PT corrections (\ref{Psi-PT1-take-1-d=2}) and (\ref{E-PT2-take-1-d=2}) more explicitly.\\
      For this purpose we introduce shorthand notation
      \be
         {\bm p}_{j'\n j} \; = \; _A\m\la \psi_{j'}\n | \, \hat{\bm p} \, | \psi_j \ra_{\m A} \, \mez .
      \ee
      We also evaluate an action of $\hat{\bm A}[\alpha]$ and $\hat{\bm A}^2\n[\alpha]$ on the vacuum state $| {\rm vac} \ra_{\m R}$. One has
      \be
         \hat{\bm A}[\alpha] \, | {\rm vac} \ra_{\m R} \; = \; \int_{{\mathbb R}^2} \m {\rm d}^2\n k \, \sum_\wp \, \frac{1}{2\,\pi} \, \sqrt{\frac{\hbar\,c^2}{\omega_k}} \; e^{-\frac{{\bm k}^2}{4\,\alpha}} \;
         | ({\bm k}\wp) \ra_{\m R} \; {\bm \varepsilon}_{{\bm k}\wp} \mez ;
      \ee
      and
      \be
         \hat{\bm A}^2\n[\alpha] \, | {\rm vac} \ra_{\m R} \; = \; \int_{{\mathbb R}^2} \m {\rm d}^2\n k \, \sum_\wp \, \frac{1}{4\,\pi^2} \, \frac{\hbar\,c^2}{\omega_k} \; e^{-\frac{{\bm k}^2}{2\,\alpha}} \;
         | {\rm vac} \ra_{\m R} \; + \; {\rm a\phantom{i}two\phantom{i}photon\phantom{i}term} \mez .
      \ee
      Subsequently one may work out the relevant matrix elements
      \be
         _A\m\la \psi_{j'}\n | \, _R\la ({\bm k}\wp) | \,\m\m \left(-\frac{q}{mc} \, \hat{\bm A}[\alpha] \m\bm\cdot\m \hat{\bm p}\right) \m\m\, | \psi_j \ra_{\m A}
         \, | {\rm vac} \ra_{\m R} \; = \; -\frac{q}{2\,\pi\,m} \, \sqrt{\frac{\hbar}{\omega_k}} \; e^{-\frac{{\bm k}^2}{4\,\alpha}} \; {\bm \varepsilon}_{{\bm k}\wp} \m\bm\cdot\m {\bm p}_{j'\n j} \mez ;
      \ee
      and
      \be
         _A\m\la \psi_j | \, _R\la {\rm vac} | \,\m\m \left( \frac{q^2}{2\,m\,c^2} \, \hat{\bm A}^2\n[\alpha] \right) \m\m\, | \psi_j \ra_{\m A} \, | {\rm vac} \ra_{\m R} \; = \;
         \frac{q^2}{2\,m\,c^2} \, \int_{{\mathbb R}^2} \m {\rm d}^2\n k \, \sum_\wp \, \frac{1}{4\,\pi^2} \, \frac{\hbar\,c^2}{\omega_k} \; e^{-\frac{{\bm k}^2}{2\,\alpha}} \mez .
      \ee
\item Returning back to the PT corrections (\ref{Psi-PT1-take-1-d=2})-(\ref{E-PT2-take-1-d=2}), we find that
      \be \label{Psi-PT1-take-2-d=2}
         | \Psi^{(1)} \ra \; = \; -\frac{q}{2\,\pi\,m} \, \sum_{j'} \, \int_{{\mathbb R}^2} \m {\rm d}^2\n k \, \sum_\wp \,
         \sqrt{\frac{\hbar}{\omega_k}} \; e^{-\frac{{\bm k}^2}{4\,\alpha}} \; \frac{ {\bm \varepsilon}_{{\bm k}\wp} \m\bm\cdot\m {\bm p}_{j'\n j} }{\left(E_j^A \, - \, E_{j'}^A
         \, - \, \hbar\omega_{k} \, + \, i\,\varepsilon\right)} \; | \psi_{j'}\n \ra_{\m A} \, | ({\bm k}\wp) \ra_{\m R} \mez ;
      \ee
      and
      \begin{eqnarray} \label{E-PT2-take-2-d=2}
         E^{(2)} & = & \frac{q^2\,\hbar}{4\,\pi^2\,m^2} \, \sum_{j'} \, \int_{{\mathbb R}^2} \m {\rm d}^2\n k \;\, \omega_k^{-1} \, e^{-\frac{{\bm k}^2}{2\,\alpha}} \, \sum_\wp \, \frac{ \Bigl| \, {\bm \varepsilon}_{{\bm k}\wp} \m\bm\cdot\m {\bm p}_{j'\n j} \, \Bigr|^2}{E_j^A \, - \, E_{j'}^A \, - \, \hbar\omega_{k} \, + \, i\,\varepsilon} \\
         & + & \frac{q^2\,\hbar}{4\,\pi^2\,m} \, \int_{{\mathbb R}^2} \m {\rm d}^2\n k \;\, \omega_k^{-1} \, e^{-\frac{{\bm k}^2}{2\,\alpha}} \; + \;
         \frac{q^2}{8\,\pi\,m^2c^2} \, \ln\m\m\left(\frac{\alpha}{\alpha_0}\right) \,\m _A\m\la \psi_j | \, \hat{\bm p}^2 \, | \psi_j \ra_{\m A} \mez . \nonumber
      \end{eqnarray}
      Writing down these explicit results\\ serves as an important intermediate step of our renormalized RSPT.
\item From now on, we shall focus on examining more closely the PT2 energy correction $E^{(2)}=(\ref{E-PT2-take-2-d=2})$.\\
      The term
      \be \label{E-(2)-discard-d=2}
         \frac{q^2\,\hbar}{4\,\pi^2\,m} \, \int_{{\mathbb R}^2} \m {\rm d}^2\n k \;\, \omega_k^{-1} \, e^{-\frac{{\bm k}^2}{2\,\alpha}}
         \; = \; \frac{q^2\,\hbar}{2\,\pi\,m\,c} \, \int_{0}^\infty \m {\rm d}k \;\, e^{-\frac{k^2}{2\,\alpha}} \; = \;
         \frac{q^2\,\hbar}{2\,m\,c} \; \sqrt{\frac{\alpha}{2\,\pi}}
      \ee
      is obviously UV divergent for $\alpha \to +\infty$, yet it does not depend upon the atomic label $j$.\\
      If so, it can be discarded as a physically unimportant constant energy shift\\
      affecting uniformly all the atomic levels.\\
      $[\,$Exactly the same reasoning was implicitly encountered before in A.1,\\
      \phantom{$[\,$}when we eliminated the divergent ZPE part of $\hat{H}_{{\rm R}}=(\ref{H-R-def})$.$\,]$\\
      After discarding the contribution (\ref{E-(2)-discard-d=2}), we are left with
      \begin{eqnarray} \label{E-PT2-take-3-d=2}
         E^{(2)} & = & \frac{q^2\,\hbar}{4\,\pi^2\,m^2} \, \sum_{j'} \, \int_{{\mathbb R}^2} \m {\rm d}^2\n k \;\, \omega_k^{-1} \, e^{-\frac{{\bm k}^2}{2\,\alpha}} \, \sum_\wp \, \frac{ \Bigl| \, {\bm \varepsilon}_{{\bm k}\wp} \m\bm\cdot\m {\bm p}_{j'\n j} \, \Bigr|^2}{E_j^A \, - \, E_{j'}^A \, - \, \hbar\omega_{k} \, + \, i\,\varepsilon} \\
         & + & \frac{q^2}{8\,\pi\,m^2c^2} \, \ln\m\m\left(\frac{\alpha}{\alpha_0}\right) \,\m
         _A\m\la \psi_j | \, \hat{\bm p}^2 \, | \psi_j \ra_{\m A} \mez . \nonumber
      \end{eqnarray}
      Formula (\ref{E-PT2-take-3-d=2}) can be redisplayed in an equivalent but more convenient appearance
      \begin{eqnarray} \label{E-PT2-take-4-d=2}
         \hspace*{-1.50cm} E^{(2)} & = & \frac{q^2\,\hbar}{4\,\pi^2\,m^2} \, \sum_{j'} \, \int_{{\mathbb R}^2} \m {\rm d}^2\n k \;\, \omega_k^{-1} \,
         e^{-\frac{{\bm k}^2}{2\,\alpha}} \, \sum_\wp \, \Bigl| \, {\bm \varepsilon}_{{\bm k}\wp} \m\bm\cdot\m {\bm p}_{j'\n j} \, \Bigr|^2 \, \left\{ \,
         \frac{1}{E_j^A \, - \, E_{j'}^A \, - \, \hbar\omega_{k} \, + \, i\,\varepsilon} \, + \, \frac{\chi(k,{\cal K})}{\hbar\omega_{k}} \, \right\} \nonumber\\
         \hspace*{-1.50cm} & - & \frac{q^2\,\hbar}{4\,\pi^2\,m^2} \, \sum_{j'} \, \int_{{\mathbb R}^2} \m {\rm d}^2\n k \;\, \omega_k^{-1} \,
         e^{-\frac{{\bm k}^2}{2\,\alpha}} \, \sum_\wp \, \Bigl| \, {\bm \varepsilon}_{{\bm k}\wp} \m\bm\cdot\m {\bm p}_{j'\n j} \, \Bigr|^2 \, \frac{\chi(k,{\cal K})}{\hbar\omega_{k}} \\
         \hspace*{-1.50cm} & + & \frac{q^2}{8\,\pi\,m^2c^2} \, \ln\m\m\left(\frac{\alpha}{\alpha_0}\right) \,\m
         _A\m\la \psi_j | \, \hat{\bm p}^2 \, | \psi_j \ra_{\m A} \mez . \nonumber
      \end{eqnarray}
      Here ${\cal K}>0$ stands for an as yet unspecified parameter,\\
      the associated factor $\chi(k,{\cal K})$ equals to unity for $|{\bm k}|>{\cal K}$ and vanishes otherwise.\\
      We set also $\tilde{\chi}(k,{\cal K}) = 1 - \chi(k,{\cal K})$.\\
      Instead of (\ref{E-PT2-take-4-d=2}) one may write even more conveniently
      \begin{eqnarray} \label{E-PT2-take-5-d=2}
         \hspace*{-1.00cm} E^{(2)} & = & \frac{q^2\,\hbar}{4\,\pi^2\,m^2} \, \sum_{j'} \, \int_{{\mathbb R}^2} \m {\rm d}^2\n k \;\, \omega_k^{-1} \,
         e^{-\frac{{\bm k}^2}{2\,\alpha}} \, \sum_\wp \, \Bigl| \, {\bm \varepsilon}_{{\bm k}\wp} \m\bm\cdot\m {\bm p}_{j'\n j} \, \Bigr|^2 \,
         \frac{\hbar\omega_{jj'}^A\,\chi(k,{\cal K})+\hbar\omega_k\,\tilde{\chi}(k,{\cal K})}{\hbar\omega_k\,(\hbar\omega_{jj'}^A-\hbar\omega_k+i\varepsilon)} \nonumber\\
         \hspace*{-1.00cm} & - & \frac{q^2}{4\,\pi^2\,m^2\,c^2} \, \sum_{j'} \, \int_{{\mathbb R}^2} \m {\rm d}^2\n k \;\, \chi(k,{\cal K}) \; k^{-2} \,
         e^{-\frac{{\bm k}^2}{2\,\alpha}} \, \sum_\wp \, \Bigl| \, {\bm \varepsilon}_{{\bm k}\wp} \m\bm\cdot\m {\bm p}_{j'\n j} \, \Bigr|^2 \\
         \hspace*{-1.00cm} & + & \frac{q^2}{8\,\pi\,m^2c^2} \, \ln\m\m\left(\frac{\alpha}{\alpha_0}\right) \,\m
         _A\m\la \psi_j | \, \hat{\bm p}^2 \, | \psi_j \ra_{\m A} \mez . \nonumber
      \end{eqnarray}
      We have adopted here a shorthand symbol
      \be
         \hbar\omega_{jj'}^A \; = \; E_j^A \, - \, E_{j'}^A \mez .
      \ee
\item Our next effort is invested into simplifying further the second line of (\ref{E-PT2-take-5-d=2}). We have
      \be
         \sum_{j'} \, \Bigl| \, {\bm \varepsilon}_{{\bm k}\wp} \m\bm\cdot\m {\bm p}_{j'\n j} \, \Bigr|^2 \; = \; \sum_{j'} \, 
         {\bm \varepsilon}_{{\bm k}\wp} \m\bm\cdot\m {\bm p}_{jj'\n} \; {\bm p}_{j'\n j} \m\bm\cdot\m {\bm \varepsilon}_{{\bm k}\wp} \; = \;
         _A\m\la \psi_j | \, {\bm \varepsilon}_{{\bm k}\wp} \m\bm\cdot\m \hat{\bm p} \; \hat{\bm p} \m\bm\cdot\m {\bm \varepsilon}_{{\bm k}\wp} \, | \psi_j \ra_{\m A} \mez ;
      \ee
      and subsequently
      \begin{eqnarray}
         \hspace*{-1.00cm} \sum_\wp \sum_{j'} \, \Bigl| \, {\bm \varepsilon}_{{\bm k}\wp} \m\bm\cdot\m {\bm p}_{j'\n j} \, \Bigr|^2 & = &
         _A\m\la \psi_j | \, \sum_\wp \, \hat{\bm p}^T {\bm \varepsilon}_{{\bm k}\wp} \, {\bm \varepsilon}_{{\bm k}\wp}^T \, \hat{\bm p} \, | \psi_j \ra_{\m A} \; = \;
         _A\m\la \psi_j | \, \hat{\bm p}^T \Bigl( {\mathbb I} - {\bm \kappa} {\bm \kappa}^T \Bigr) \, \hat{\bm p} \, | \psi_j \ra_{\m A} \mez .
      \end{eqnarray}
      The previously established property (\ref{auxiliary-identity-d=2}) implies in turn
      \begin{eqnarray}
         \hspace*{-1.00cm} & & \sum_{j'} \, \int_{{\mathbb R}^2} \m {\rm d}^2\n k \;\, \chi(k,{\cal K}) \; k^{-2} \, e^{-\frac{{\bm k}^2}{2\,\alpha}} \, \sum_\wp \,
         \Bigl| \, {\bm \varepsilon}_{{\bm k}\wp} \m\bm\cdot\m {\bm p}_{j'\n j} \, \Bigr|^2 \; = \; \pi \;
         _A\m\la \psi_j | \, \hat{\bm p}^2 \, | \psi_j \ra_{\m A} \, \int_{{\cal K}}^{\infty} \m k^{-1} \, e^{-\frac{k^2}{2\,\alpha}} \; {\rm d}k \; = \nonumber\\
         \hspace*{-1.00cm} & = & \pi \, _A\m\la \psi_j | \, \hat{\bm p}^2 \, | \psi_j \ra_{\m A} \; \frac{1}{2} \, \ln\m\m\left(\frac{2\,\alpha}{{\cal K}^2\,e^\gamma}\right)
         \; + \; {\cal O}(\alpha^{-1}) \mez .
      \end{eqnarray}
      Here $\gamma=0.577\cdots$ is the Euler-Mascheroni constant encountered already in {\sl Appendix A}.\\
      After setting most conveniently
      \be \label{cal-K-fixing-Appendix-4}
         {\cal K} \; = \; \sqrt{2\;\alpha_0\;e^{-\gamma}} \mez ;
      \ee
      the second line of (\ref{E-PT2-take-5-d=2}) exactly overcompensates the third line,\\
      up to a nonvanishing ${\cal O}(\alpha^{-1})$ contribution.\\
      {\color{blue} This is how the logarithmically UV divergent parts of $E^{(2)}=(\ref{E-PT2-take-5-d=2})$ overkill each other.\\
      Logarithmic UV divergence of the naive RSPT outcome\\ is overcompensated by the logarithmically UV divergent mass counterterm.}\\
      We are thus left with the PT2 energy correction formula
      \begin{eqnarray} \label{E-PT2-take-6-d=2}
         \hspace*{-1.00cm} E^{(2)} & = & \frac{q^2\,\hbar}{4\,\pi^2\,m^2} \, \sum_{j'} \, \int_{{\mathbb R}^2} \m {\rm d}^2\n k \;\, \omega_k^{-1} \,
         e^{-\frac{{\bm k}^2}{2\,\alpha}} \, \sum_\wp \, \Bigl| \, {\bm \varepsilon}_{{\bm k}\wp} \m\bm\cdot\m {\bm p}_{j'\n j} \, \Bigr|^2 \,
         \frac{\hbar\omega_{jj'}^A\,\chi(k,{\cal K})+\hbar\omega_k\,\tilde{\chi}(k,{\cal K})}{\hbar\omega_k\,(\hbar\omega_{jj'}^A-\hbar\omega_k+i\varepsilon)} \\
         \hspace*{-1.00cm} & + & {\cal O}(\alpha^{-1}) \mez . \nonumber
      \end{eqnarray}
\item Expression (\ref{E-PT2-take-6-d=2}) can still be simplified, by using again the property (\ref{auxiliary-identity-d=2}). 
      Indeed, one writes
      \begin{eqnarray}
         \hspace*{-1.50cm} & & \int_{0}^{2\pi} \m {\rm d}\varphi_{\bm k} \;\,
         \sum_\wp \, \Bigl| \, {\bm \varepsilon}_{{\bm k}\wp} \m\bm\cdot\m {\bm p}_{j'\n j} \, \Bigr|^2 \; = \;
         \int_{0}^{2\pi} \m {\rm d}\varphi_{\bm k} \;\,
         {\bm p}_{jj'} \sum_\wp \, {\bm \varepsilon}_{{\bm k}\wp} \, {\bm \varepsilon}_{{\bm k}\wp}^T \; {\bm p}_{j'\n j} \; = \nonumber\\
         \hspace*{-1.50cm} & = & \int_{0}^{2\pi} \m {\rm d}\varphi_{\bm k} \;\,
         {\bm p}_{jj'} \Bigl( {\mathbb I} - {\bm \kappa} {\bm \kappa}^T \Bigr) \, {\bm p}_{j'\n j} \; = \; \phantom{2}\,\pi \; {\bm p}_{jj'} \m\bm\cdot\m {\bm p}_{j'\n j}
         \; = \; \phantom{2}\,\pi \; \Bigl| \, {\bm p}_{jj'} \Bigr|^2 \mez .
      \end{eqnarray}
      Correspondingly, $E^{(2)}=(\ref{E-PT2-take-6-d=2})$ boils down into
      \begin{eqnarray} \label{E-PT2-take-7-d=2}
         \hspace*{-1.00cm} E^{(2)} & = & \frac{q^2\,\hbar}{4\,\pi\,m^2\,c} \, \sum_{j'\n \neq j} \;
         \Bigl| \, {\bm p}_{jj'} \Bigr|^2 \int_0^\infty \m {\rm d}k \;\, e^{-\frac{{\bm k}^2}{2\,\alpha}} \;\,
         \frac{\hbar\omega_{jj'}^A\,\chi(k,{\cal K})+\hbar\omega_k\,\tilde{\chi}(k,{\cal K})}
         {\hbar\omega_k\,(\hbar\omega_{jj'}^A-\hbar\omega_k+i\varepsilon)} \\
         \hspace*{-1.00cm} & + & {\cal O}(\alpha^{-1}) \mez . \nonumber
      \end{eqnarray}
      Most importantly, moment of reflection reveals\\
      that the r.h.s.~of (\ref{E-PT2-take-7-d=2}) remains finite when the limit of $\alpha \to +\infty$ is taken.\\
      Our finalized form of the resulting finite PT2 energy correction reads thus as follows:
      \be \label{E-PT2-take-8-d=2}
         {\color{blue} E^{(2)} \; = \; \frac{q^2\,\hbar}{4\,\pi\,m^2\,c} \, \sum_{j'\n \neq j} \;
         \Bigl| \, {\bm p}_{jj'} \Bigr|^2 \int_0^\infty \m {\rm d}k \;\,
         \frac{\hbar\omega_{jj'}^A\,\chi(k,{\cal K})+\hbar\omega_k\,\tilde{\chi}(k,{\cal K})}
         {\hbar\omega_k\,(\hbar\omega_{jj'}^A-\hbar\omega_k+i\varepsilon)} \mez ; \mez [d=2]}
      \ee
      with ${\cal K}=(\ref{cal-K-fixing-Appendix-4})$.\\
      We may conclude {\sl Appendix B.3} by saying that the renormalized second order RSPT\\
      has provided us finite and physically meaningful atomic level shifts for the case of $d=2$.\\
      {\color{magenta} {\sl Some parts of Appendix B.3 were a bit technical, but the output (\ref{E-PT2-take-8-d=2}) is worthy of an effort!}}
\end{itemize}

\vspace*{+0.20cm}

{\bf Appendix B.4: Renormalized perturbation theory for $\bm d \bm = \bm 3$}
\vspace*{-0.20cm}
\begin{itemize}
\item We wish to implement now\\ the renormalized Rayleigh-Schr\"{o}dinger perturbation theory (RSPT)\\
      for the atomic level shifts, assuming $d=3$.\\
      Consider our starting Hamiltonian $\hat{H}=(\ref{H-no-t})|_{d=3}$\\
      supplemented by the mass renormalization prescription (\ref{m-ren-d=3-no-t}).\\
      Our goal is to solve an eigenvalue problem (\ref{hat-H-eigenproblem})\\
      by employing the standard RSPT in the coupling parameter $q$.\\
      It is thus necessary to express $\hat{H}=(\ref{H-no-t})|_{d=3}$ as a power series in $q$.\\
      This is accomplished through writing
      \begin{eqnarray} \label{mass-counterterm-d=3}
         \hspace*{-1.00cm} \frac{1}{2\,m_o(\alpha)} & = & \frac{1}{2\,m} \, \frac{m}{m_o(\alpha)} \; = \; \frac{1}{2\,m} \; \frac{1}{1 \, - \,
         \frac{4}{3} \, \frac{q^2}{mc^2} \, \sqrt{\frac{\alpha}{2\,\pi}}} \; = \; \frac{1}{2\,m} \; + \; \frac{1}{2\,m} \, \sum_{\ell=1}^{\infty}
         \left( \frac{4}{3} \, \frac{q^2}{mc^2} \, \sqrt{\frac{\alpha}{2\,\pi}}\;\right)^{\m\m\ell} \; . \mz
      \end{eqnarray}
      {\color{blue} Entity $\frac{1}{2\,m} \, \sum_{\ell=1}^{\infty} \m \left(\frac{4}{3} \, \frac{q^2}{mc^2} \, \sqrt{\frac{\alpha}{2\,\pi}}\right)^{\m\ell}$
      is the so called {\sl mass counterterm}.}\\
      \hspace*{+0.50cm} Note that the just displayed infinite series (\ref{mass-counterterm-d=3})\\
      \hspace*{+0.50cm} is not summable in an ordinary sense for $\alpha \to +\infty$, thus some readers may feel uneasy here.\\
      \hspace*{+0.50cm} In these introductory lecture notes, we avoid delving into\\
      \hspace*{+0.50cm} mathematical peculiarities accompanying the summation (\ref{mass-counterterm-d=3}).\\
      \hspace*{+0.50cm} Such a light hearted attitude is common for vast majority of the present day QFT theorists,\\
      \hspace*{+0.50cm} who keep on using weird expansions of the form similar to (\ref{mass-counterterm-d=3}) daily,\\
      \hspace*{+0.50cm} in their studies of high energy physics within (and beyond) the Standard Model!\\
      The Hamiltonian $\hat{H}=(\ref{H-no-t})|_{d=3}$ is then split into a sum
      \be
         \hat{H} \; = \; \hat{H}_{\rm free} \; + \; \hat{W} \mez ;
      \ee
      where $\hat{H}_{\rm free}=(\ref{H-free-def})$ is understood as the unperturbed reference part, and
      \be \label{W-Appendix-B.4}
         \hat{W} \; = \;  -\,\frac{q}{mc} \, \hat{\bm A}[\alpha] \m\bm\cdot\m \hat{\bm p} \; + \; \frac{q^2}{2\,m\,c^2} \, \hat{\bm A}^2\n[\alpha]
         \; + \; \frac{2}{3} \, \frac{q^2}{m^2 c^2} \, \sqrt{\frac{\alpha}{2\,\pi}} \; \hat{\bm p}^2 \; + \; {\cal O}(q^3)
      \ee
      determines the (adequately renormalized) perturbation.\\
      {\color{blue} Note that the perturbation $\hat{W}=(\ref{W-Appendix-B.4})$ contains all the mass counterterms!\\
      This is what should make $\hat{W}=(\ref{W-Appendix-B.4})$ small compared to $\hat{H}_{\rm free}=(\ref{H-free-def})$ for $q$ small enough.}\\
      $[\,$Our subsequent calculation of the atomic level shifts will test the just presented anticipation.$\,]$\\
      {\color{blue} Note that the naive perturbation term $\hat{W}=(\ref{W-naive})$ from B.2 does not contain any mass counterterms!\\
      This is why $\hat{W}=(\ref{W-naive})$ represents an infinite (and in fact ill-defined) contribution to $\hat{H}=(\ref{H-naive})$.}\\
      Much as in B.2 and B.3, our unperturbed reference state $| \Psi^{(0)} \ra = (\ref{Psi-(0)-unperturbed})$\\
      corresponds to an unperturbed energy eigenvalue $E^{(0)}=E_j^A$\m,\\
      consistently with relation $(\ref{hat-H-free-eigenproblem})|_{n=0}$.
\item An application of the standard RSPT is now straightforward.\\
      Our PT1 energy correction equals to
      \be
         E^{(1)} \; = \; \la \Psi^{(0)} | \, -\,\frac{q}{mc} \, \hat{\bm A}[\alpha] \m\bm\cdot\m \hat{\bm p} \; | \Psi^{(0)} \ra \; = \; 0 \mez .
      \ee
      Our PT1 eigenvector correction reads as
      \be \label{Psi-PT1-take-1}
         | \Psi^{(1)} \ra \; = \; \sum_{j'} \, \int_{{\mathbb R}^3} \m {\rm d}^3\n k \, \sum_\wp \,
         \frac{ _A\m\la \psi_{j'}\n | \, _R\la ({\bm k}\wp) | \,\m\m \left(-\frac{q}{mc} \, \hat{\bm A}[\alpha] \m\bm\cdot\m \hat{\bm p}\right) \m\m\,
         | \psi_j \ra_{\m A} \, | {\rm vac} \ra_{\m R} }{E_j^A \, - \, E_{j'}^A \, - \, \hbar\omega_{k} \, + \, i\,\varepsilon} \; | \psi_{j'}\n \ra_{\m A}
         \, | ({\bm k}\wp) \ra_{\m R} \mez .
      \ee
      Accordingly, our PT2 energy correction takes the form
      \begin{eqnarray} \label{E-PT2-take-1}
         E^{(2)} & = & \sum_{j'} \, \int_{{\mathbb R}^3} \m {\rm d}^3\n k \, \sum_\wp \, \frac{ \left| \, _A\m\la \psi_{j'}\n |
         \, _R\la ({\bm k}\wp) | \,\m\m \left(-\frac{q}{mc} \, \hat{\bm A}[\alpha] \m\bm\cdot\m \hat{\bm p}\right) \m\m\, | \psi_j \ra_{\m A}
         \, | {\rm vac} \ra_{\m R} \, \right|^2}{E_j^A \, - \, E_{j'}^A \, - \, \hbar\omega_{k} \, + \, i\,\varepsilon} \\
         & + & _A\m\la \psi_j | \, _R\la {\rm vac} | \,\m\m \left( \frac{q^2}{2\,m\,c^2} \, \hat{\bm A}^2\n[\alpha] \right) \m\m\, | \psi_j \ra_{\m A} \, | {\rm vac} \ra_{\m R} \nonumber\\
         & + & \frac{2}{3} \, \frac{q^2}{m^2 c^2} \, \sqrt{\frac{\alpha}{2\,\pi}} \;\;
         _A\m\la \psi_j | \, _R\la {\rm vac} | \, \hat{\bm p}^2 \, | \psi_j \ra_{\m A} \, | {\rm vac} \ra_{\m R} \mez . \nonumber
      \end{eqnarray}
      Denominators of the formulas (\ref{Psi-PT1-take-1}) and (\ref{E-PT2-take-1}) contain a small imaginary factor
      $(+i\,\varepsilon)$ where $\varepsilon \to +0$.\\ Clearly, the mentioned $(+i\,\varepsilon)$ factor cures singularities\\
      which otherwise would occur in (\ref{Psi-PT1-take-1}) and (\ref{E-PT2-take-1}) whenever $j \ge 1$,\\
      at the resonant modal frequencies $\hbar\omega_k=E_j^A \, - \, E_{j'<j}^A$.\\
      It is important to recall at this point that all the $j \ge 1$ atomic levels are metastable\\
      (decaying due to spontaneous emission).\\
      Metastable states (resonances) are studied by the nonhermitian quantum mechanics, see Ref.~\cite{NHQM}.\\
      The just discussed insertion of the $(+i\,\varepsilon)$ factor into the denominators of (\ref{Psi-PT1-take-1})
      and (\ref{E-PT2-take-1})\\ represents a common practice of the nonhermitian RSPT.\\
      Note that the $(+i\,\varepsilon)$ factor makes $E^{(2)}$ complex valued for $j \ge 1$.\\
      $[\,$Negative imaginary part $\sim$ decay rate of the $j$-th level due to spontaneous emission.$\,]$\\
      Detailed justification for including the $(+i\,\varepsilon)$ term\\
      in the present context of spontaneous emission can be found in my recent work, see Ref.~\cite{SE}.
\item Proceeding further, we wish to evaluate the PT corrections (\ref{Psi-PT1-take-1}) and (\ref{E-PT2-take-1}) more explicitly.\\
      For this purpose we introduce shorthand notation
      \be
         {\bm p}_{j'\n j} \; = \; _A\m\la \psi_{j'}\n | \, \hat{\bm p} \, | \psi_j \ra_{\m A} \, \mez .
      \ee
      We also evaluate an action of $\hat{\bm A}[\alpha]$ and $\hat{\bm A}^2\n[\alpha]$ on the vacuum state $| {\rm vac} \ra_{\m R}$. One has
      \be
         \hat{\bm A}[\alpha] \, | {\rm vac} \ra_{\m R} \; = \; \int_{{\mathbb R}^3} \m {\rm d}^3\n k \, \sum_\wp \, \frac{1}{2\,\pi} \, \sqrt{\frac{\hbar\,c^2}{\omega_k}} \; e^{-\frac{{\bm k}^2}{4\,\alpha}} \;
         | ({\bm k}\wp) \ra_{\m R} \; {\bm \varepsilon}_{{\bm k}\wp} \mez ;
      \ee
      and
      \be
         \hat{\bm A}^2\n[\alpha] \, | {\rm vac} \ra_{\m R} \; = \; \int_{{\mathbb R}^3} \m {\rm d}^3\n k \, \sum_\wp \, \frac{1}{4\,\pi^2} \, \frac{\hbar\,c^2}{\omega_k} \; e^{-\frac{{\bm k}^2}{2\,\alpha}} \;
         | {\rm vac} \ra_{\m R} \; + \; {\rm a\phantom{i}two\phantom{i}photon\phantom{i}term} \mez .
      \ee
      Subsequently one may work out the relevant matrix elements
      \be
         _A\m\la \psi_{j'}\n | \, _R\la ({\bm k}\wp) | \,\m\m \left(-\frac{q}{mc} \, \hat{\bm A}[\alpha] \m\bm\cdot\m \hat{\bm p}\right) \m\m\, | \psi_j \ra_{\m A}
         \, | {\rm vac} \ra_{\m R} \; = \; -\frac{q}{2\,\pi\,m} \, \sqrt{\frac{\hbar}{\omega_k}} \; e^{-\frac{{\bm k}^2}{4\,\alpha}} \; {\bm \varepsilon}_{{\bm k}\wp} \m\bm\cdot\m {\bm p}_{j'\n j} \mez ;
      \ee
      and
      \be
         _A\m\la \psi_j | \, _R\la {\rm vac} | \,\m\m \left( \frac{q^2}{2\,m\,c^2} \, \hat{\bm A}^2\n[\alpha] \right) \m\m\, | \psi_j \ra_{\m A} \, | {\rm vac} \ra_{\m R} \; = \;
         \frac{q^2}{2\,m\,c^2} \, \int_{{\mathbb R}^3} \m {\rm d}^3\n k \, \sum_\wp \, \frac{1}{4\,\pi^2} \, \frac{\hbar\,c^2}{\omega_k} \; e^{-\frac{{\bm k}^2}{2\,\alpha}} \mez .
      \ee
\item Returning back to the PT corrections (\ref{Psi-PT1-take-1})-(\ref{E-PT2-take-1}), we find that
      \be \label{Psi-PT1-take-2}
         | \Psi^{(1)} \ra \; = \; -\frac{q}{2\,\pi\,m} \, \sum_{j'} \, \int_{{\mathbb R}^3} \m {\rm d}^3\n k \, \sum_\wp \,
         \sqrt{\frac{\hbar}{\omega_k}} \; e^{-\frac{{\bm k}^2}{4\,\alpha}} \; \frac{ {\bm \varepsilon}_{{\bm k}\wp} \m\bm\cdot\m {\bm p}_{j'\n j} }{\left(E_j^A \, - \, E_{j'}^A
         \, - \, \hbar\omega_{k} \, + \, i\,\varepsilon\right)} \; | \psi_{j'}\n \ra_{\m A} \, | ({\bm k}\wp) \ra_{\m R} \mez ;
      \ee
      and
      \begin{eqnarray} \label{E-PT2-take-2}
         E^{(2)} & = & \frac{q^2\,\hbar}{4\,\pi^2\,m^2} \, \sum_{j'} \, \int_{{\mathbb R}^3} \m {\rm d}^3\n k \;\, \omega_k^{-1} \, e^{-\frac{{\bm k}^2}{2\,\alpha}} \, \sum_\wp \, \frac{ \Bigl| \, {\bm \varepsilon}_{{\bm k}\wp} \m\bm\cdot\m {\bm p}_{j'\n j} \, \Bigr|^2}{E_j^A \, - \, E_{j'}^A \, - \, \hbar\omega_{k} \, + \, i\,\varepsilon} \\
         & + & \frac{q^2\,\hbar}{4\,\pi^2\,m} \, \int_{{\mathbb R}^3} \m {\rm d}^3\n k \;\, \omega_k^{-1} \, e^{-\frac{{\bm k}^2}{2\,\alpha}} \; + \;
         \frac{2}{3} \, \frac{q^2}{m^2 c^2} \, \sqrt{\frac{\alpha}{2\,\pi}} \;\;
         _A\m\la \psi_j | \, \hat{\bm p}^2 \, | \psi_j \ra_{\m A} \mez . \nonumber
      \end{eqnarray}
      Writing down these explicit results\\ serves as an important intermediate step of our renormalized RSPT.
\item From now on, we shall focus on examining more closely the PT2 energy correction $E^{(2)}=(\ref{E-PT2-take-2})$.\\
      The term
      \be \label{E-(2)-discard-d=3}
         \frac{q^2\,\hbar}{4\,\pi^2\,m} \, \int_{{\mathbb R}^3} \m {\rm d}^3\n k \;\, \omega_k^{-1} \, e^{-\frac{{\bm k}^2}{2\,\alpha}}
         \; = \; \frac{q^2\,\hbar}{\pi\,m\,c} \, \int_{0}^\infty \m {\rm d}k \;\, k \, e^{-\frac{k^2}{2\,\alpha}} \; = \;
         \frac{q^2\,\hbar}{\pi\,m\,c} \; \alpha
      \ee
      is obviously UV divergent for $\alpha \to +\infty$, yet it does not depend upon the atomic label $j$.\\
      If so, it can be discarded as a physically unimportant constant energy shift\\
      affecting uniformly all the atomic levels.\\
      $[\,$Exactly the same reasoning was implicitly encountered before in A.1,\\
      \phantom{$[\,$}when we eliminated the divergent ZPE part of $\hat{H}_{{\rm R}}=(\ref{H-R-def})$.$\,]$\\
      After discarding the contribution (\ref{E-(2)-discard-d=3}), we are left with
      \begin{eqnarray} \label{E-PT2-take-3}
         E^{(2)} & = & \frac{q^2\,\hbar}{4\,\pi^2\,m^2} \, \sum_{j'} \, \int_{{\mathbb R}^3} \m {\rm d}^3\n k \;\, \omega_k^{-1} \, e^{-\frac{{\bm k}^2}{2\,\alpha}} \, \sum_\wp \, \frac{ \Bigl| \, {\bm \varepsilon}_{{\bm k}\wp} \m\bm\cdot\m {\bm p}_{j'\n j} \, \Bigr|^2}{E_j^A \, - \, E_{j'}^A \, - \, \hbar\omega_{k} \, + \, i\,\varepsilon} \\
         & + & \frac{2}{3} \, \frac{q^2}{m^2 c^2} \, \sqrt{\frac{\alpha}{2\,\pi}} \;\; _A\m\la \psi_j | \, \hat{\bm p}^2 \, | \psi_j \ra_{\m A} \mez . \nonumber
      \end{eqnarray}
      Formula (\ref{E-PT2-take-3}) can be redisplayed in an equivalent but more convenient appearance
      \begin{eqnarray} \label{E-PT2-take-4}
         \hspace*{-1.00cm} E^{(2)} & = & \frac{q^2\,\hbar}{4\,\pi^2\,m^2} \, \sum_{j'} \, \int_{{\mathbb R}^3} \m {\rm d}^3\n k \;\, \omega_k^{-1} \,
         e^{-\frac{{\bm k}^2}{2\,\alpha}} \, \sum_\wp \, \Bigl| \, {\bm \varepsilon}_{{\bm k}\wp} \m\bm\cdot\m {\bm p}_{j'\n j} \, \Bigr|^2 \, \left\{ \,
         \frac{1}{E_j^A \, - \, E_{j'}^A \, - \, \hbar\omega_{k} \, + \, i\,\varepsilon} \, + \, \frac{1}{\hbar\omega_{k}} \, \right\} \nonumber\\
         \hspace*{-1.00cm} & - & \frac{q^2\,\hbar}{4\,\pi^2\,m^2} \, \sum_{j'} \, \int_{{\mathbb R}^3} \m {\rm d}^3\n k \;\, \omega_k^{-1} \,
         e^{-\frac{{\bm k}^2}{2\,\alpha}} \, \sum_\wp \, \Bigl| \, {\bm \varepsilon}_{{\bm k}\wp} \m\bm\cdot\m {\bm p}_{j'\n j} \, \Bigr|^2 \, \frac{1}{\hbar\omega_{k}} \\
         \hspace*{-1.00cm} & + & \frac{2}{3} \, \frac{q^2}{m^2 c^2} \, \sqrt{\frac{\alpha}{2\,\pi}} \;\; _A\m\la \psi_j | \, \hat{\bm p}^2 \, | \psi_j \ra_{\m A} \mez ;\nonumber
      \end{eqnarray}
      or even more conveniently as
      \begin{eqnarray} \label{E-PT2-take-5}
         \hspace*{-1.00cm} E^{(2)} & = & \frac{q^2\,\hbar}{4\,\pi^2\,m^2} \, \sum_{j'} \, \int_{{\mathbb R}^3} \m {\rm d}^3\n k \;\, \omega_k^{-1} \,
         e^{-\frac{{\bm k}^2}{2\,\alpha}} \, \sum_\wp \, \Bigl| \, {\bm \varepsilon}_{{\bm k}\wp} \m\bm\cdot\m {\bm p}_{j'\n j} \, \Bigr|^2 \,
         \frac{\hbar\omega_{jj'}^A}{\hbar\omega_k\,(\hbar\omega_{jj'}^A-\hbar\omega_k+i\varepsilon)} \nonumber\\
         \hspace*{-1.00cm} & - & \frac{q^2}{4\,\pi^2\,m^2\,c^2} \, \sum_{j'} \, \int_{{\mathbb R}^3} \m {\rm d}^3\n k \;\, k^{-2} \,
         e^{-\frac{{\bm k}^2}{2\,\alpha}} \, \sum_\wp \, \Bigl| \, {\bm \varepsilon}_{{\bm k}\wp} \m\bm\cdot\m {\bm p}_{j'\n j} \, \Bigr|^2 \\
         \hspace*{-1.00cm} & + & \frac{2}{3} \, \frac{q^2}{m^2 c^2} \, \sqrt{\frac{\alpha}{2\,\pi}} \;\; _A\m\la \psi_j | \, \hat{\bm p}^2 \, | \psi_j \ra_{\m A} \mez . \nonumber
      \end{eqnarray}
      We have adopted here a shorthand symbol
      \be
         \hbar\omega_{jj'}^A \; = \; E_j^A \, - \, E_{j'}^A \mez .
      \ee
\item Our next effort is invested into simplifying further the second line of (\ref{E-PT2-take-5}). We have
      \be
         \sum_{j'} \, \Bigl| \, {\bm \varepsilon}_{{\bm k}\wp} \m\bm\cdot\m {\bm p}_{j'\n j} \, \Bigr|^2 \; = \; \sum_{j'} \, 
         {\bm \varepsilon}_{{\bm k}\wp} \m\bm\cdot\m {\bm p}_{jj'\n} \; {\bm p}_{j'\n j} \m\bm\cdot\m {\bm \varepsilon}_{{\bm k}\wp} \; = \;
         _A\m\la \psi_j | \, {\bm \varepsilon}_{{\bm k}\wp} \m\bm\cdot\m \hat{\bm p} \; \hat{\bm p} \m\bm\cdot\m {\bm \varepsilon}_{{\bm k}\wp} \, | \psi_j \ra_{\m A} \mez ;
      \ee
      and subsequently
      \begin{eqnarray}
         \hspace*{-0.50cm} \sum_\wp \sum_{j'} \, \Bigl| \, {\bm \varepsilon}_{{\bm k}\wp} \m\bm\cdot\m {\bm p}_{j'\n j} \, \Bigr|^2 & = &
         _A\m\la \psi_j | \, \sum_\wp \, \hat{\bm p}^T {\bm \varepsilon}_{{\bm k}\wp} \, {\bm \varepsilon}_{{\bm k}\wp}^T \, \hat{\bm p} \, | \psi_j \ra_{\m A} \; = \;
         _A\m\la \psi_j | \, \hat{\bm p}^T \Bigl( {\mathbb I} - {\bm \kappa} {\bm \kappa}^T \Bigr) \, \hat{\bm p} \, | \psi_j \ra_{\m A} \mez .
      \end{eqnarray}
      The previously established property (\ref{auxiliary-identity-d=3}) implies in turn
      \begin{eqnarray}
         \hspace*{-1.00cm} & & \sum_{j'} \, \int_{{\mathbb R}^3} \m {\rm d}^3\n k \;\, k^{-2} \, e^{-\frac{{\bm k}^2}{2\,\alpha}} \, \sum_\wp \,
         \Bigl| \, {\bm \varepsilon}_{{\bm k}\wp} \m\bm\cdot\m {\bm p}_{j'\n j} \, \Bigr|^2 \; = \; \frac{8\,\pi}{3} \,
         _A\m\la \psi_j | \, \hat{\bm p}^2 \, | \psi_j \ra_{\m A} \, \int_{0}^{\infty} \m e^{-\frac{k^2}{2\,\alpha}} \; {\rm d}k \; = \nonumber\\
         \hspace*{-1.00cm} & = & \frac{8\,\pi}{3} \, \sqrt{\frac{\pi\,\alpha}{2}} \, _A\m\la \psi_j | \, \hat{\bm p}^2 \, | \psi_j \ra_{\m A} \mez .
      \end{eqnarray}
      If so, then the second line of (\ref{E-PT2-take-5}) exactly overcompensates the third line.\\
      {\color{blue} This is how the most abruptly UV divergent parts of $E^{(2)}=(\ref{E-PT2-take-5})$ overkill each other.\\
      The most abrupt UV divergence of our naive RSPT outcome (of order ${\cal O}(\alpha^{1/2})$)\\ is overcompensated by the accordingly UV divergent mass counterterm.}\\
      We are thus left with the PT2 energy correction formula
      \begin{eqnarray} \label{E-PT2-take-6}
         \hspace*{-1.00cm} E^{(2)} & = & \frac{q^2}{4\,\pi^2\,m^2\,c^2} \, \sum_{j'} \, \int_{{\mathbb R}^3} \m {\rm d}^3\n k \;\, k^{-2} \,
         e^{-\frac{{\bm k}^2}{2\,\alpha}} \, \sum_\wp \, \Bigl| \, {\bm \varepsilon}_{{\bm k}\wp} \m\bm\cdot\m {\bm p}_{j'\n j} \, \Bigr|^2 \,
         \frac{\hbar\omega_{jj'}^A}{(\hbar\omega_{jj'}^A-\hbar\omega_k+i\varepsilon)} \mez .
      \end{eqnarray}
\item Expression (\ref{E-PT2-take-6}) can still be simplified, by using again the property (\ref{auxiliary-identity-d=3}). Indeed, one writes
      \begin{eqnarray}
         \hspace*{-1.50cm} & & \int_{0}^{2\pi} \m {\rm d}\varphi_{\bm k} \, \int_{0}^{\pi} \m \sin\vartheta_{\bm k} \; {\rm d}\vartheta_{\bm k} \;\,
         \sum_\wp \, \Bigl| \, {\bm \varepsilon}_{{\bm k}\wp} \m\bm\cdot\m {\bm p}_{j'\n j} \, \Bigr|^2 \; = \;
         \int_{0}^{2\pi} \m {\rm d}\varphi_{\bm k} \, \int_{0}^{\pi} \m \sin\vartheta_{\bm k} \; {\rm d}\vartheta_{\bm k} \;\,
         {\bm p}_{jj'} \sum_\wp \, {\bm \varepsilon}_{{\bm k}\wp} \, {\bm \varepsilon}_{{\bm k}\wp}^T \; {\bm p}_{j'\n j} \; = \nonumber\\
         \hspace*{-1.50cm} & = & \int_{0}^{2\pi} \m {\rm d}\varphi_{\bm k} \, \int_{0}^{\pi} \m \sin\vartheta_{\bm k} \; {\rm d}\vartheta_{\bm k} \;\,
         {\bm p}_{jj'} \Bigl( {\mathbb I} - {\bm \kappa} {\bm \kappa}^T \Bigr) \, {\bm p}_{j'\n j} \; = \; \frac{8\,\pi}{3} \; {\bm p}_{jj'} \m\bm\cdot\m {\bm p}_{j'\n j}
         \; = \; \frac{8\,\pi}{3} \; \Bigl| \, {\bm p}_{jj'} \Bigr|^2 \mez .
      \end{eqnarray}
      Our finalized form of the resulting PT2 energy correction reads thus as follows:
      \be \label{E-PT2-take-7}
         {\color{blue} \hspace*{-0.50cm} E^{(2)} \; = \; \frac{2}{3\,\pi} \, \frac{q^2}{m^2 c^2} \, \sum_{j'} \, \Bigl| \, {\bm p}_{jj'} \Bigr|^2 \,
         \int_{0}^{\infty} \m {\rm d}k \;\, e^{-\frac{k^2}{2\,\alpha}} \, \frac{\hbar\omega_{jj'}^A}{(\hbar\omega_{jj'}^A-\hbar\omega_k+i\varepsilon)} \mez . \mez [d=3] }
      \ee
      The $k$-integral of equation (\ref{E-PT2-take-7}) diverges for $\alpha \to +\infty$ as $\ln\alpha$.\\
      Showing that our PT2 energy correction $E^{(2)}=(\ref{E-PT2-take-7})$\\
      turns out to be logarithmically UV divergent even after the mass renormalization.\\
      Note however that $E^{(2)}=(\ref{E-PT2-take-7})$ is robustly insensitive with respect to changing $\ln\alpha$\\
      on the scales $\alpha \sim \lambda^{-2}$ where $\lambda=(\ref{lambda-Compton})$ is the Compton wavelength of the electron.\\
      In fact, our above derived formula $E^{(2)}=(\ref{E-PT2-take-7})$ does provide physically meaningful results\\
      for the atomic level shifts, provided only that one employs any finite cutoff value of $\alpha \sim \lambda^{-2}$\m.\\
      The just discussed logarithmic divergence of $E^{(2)}=(\ref{E-PT2-take-7})$\\
      represents a characteristic feature/artefact of the non-relativistic QED theory.\\
      As a matter of fact, the studied atomic level shifts (more precisely, their relative differences)\\
      come out finite for $d=3$ only within the fully relativistic QED treatment,\\
      where the vacuum polarization effects are properly accounted for,\\
      and where also the charge $q$ of the electron is renormalized.\\
      $[\,$Such a treatment is presented e.g.~in Ref.~\cite{JZ}.\\
      \phantom{$[\,$}The underlying technical details are painfully tedious,\\
      \phantom{$[\,$}much as in any other calculation involving relativistic QFT.\\
      \phantom{$[\,$}Yet computational complexity is the price one needs to pay.$\,]$\\
      We may conclude {\sl Appendix B.4} by saying that the renormalized second order RSPT\\
      has provided us a logarithmically divergent (yet still useful) result\\ for the atomic level shifts in the case of $d=3$.\\
      {\color{magenta} {\sl We are done with introducing the renormalized RSPT,\\ and ready to proceed into Appendix C which deals with the renormalized mean field theory!}}
\end{itemize}

\vspace*{+0.20cm}

{\color{blue} {\large \bf Appendix C: Renormalized mean field theory}}\\

{\bf Appendix C.1: Introducing the game}
\vspace*{-0.20cm}
\begin{itemize}
\item Suppose that we wish to explore in more detail the quantum dynamics of our model problem\\
      by calculating {\sl explicitly} the time evolution of some chosen physical observables.\\
      $[\,$Like e.g.~the time dependent velocity $\bm\dot{\hat{\bm x}}(t)$ or acceleration $\bm\ddot{\hat{\bm x}}(t)$ of the electron,\\
      \phantom{$[\,$}and the electric field $\hat{\bm E}(t,{\bm y})$ affected by motion of the electron.$\,]$\\
      One possible strategy would be to solve explicitly\\
      the renormalized Newton-Heisenberg equation of motion (\ref{eom-hat-x-t-take-3-d=2-final}) or (\ref{eom-hat-x-t-take-3-d=3-final}),\\
      starting from the initial conditions (\ref{ics-final-x}), (\ref{ics-final-p}), (\ref{dot-x-infinite-past}).\\
      Such an approach, based upon the Heisenberg picture, seems however to be\\ impractical or even untractable from the computational (numerical) point of view.\\
      $[\,$Unless one decides to apply perturbation expansion in the electronic charge $q_t$.\\
      {\color{blue}
      \phantom{$[\,$}Yet the power of perturbation methods is not unlimited.\\
      \phantom{$[\,$}It is important to explore also nonperturbative aspects of the theory.}$\,]$
\item Another possibility is to switch into the Schr\"{o}dinger picture,\\
      and aim at resolving explicitly the time dependent Schr\"{o}dinger equation (TDSCHE)
      \be \label{TDSCHE}
         i\hbar \, \partial_t \, | \Psi_t \ra \; = \; \hat{H}_t \, | \Psi_t \ra \mez ;
      \ee
      starting from an appropriate initial condition $| \Psi_{t_0} \ra$ of our interest.\\
      Here $\hat{H}_t=(\ref{H})$ represents of course the full Hamiltonian defining our model system in Section A.\\
      Recall that formula (\ref{H}) is supplemented by the mass renormalization prescription (\ref{m-ren-d=2}) or (\ref{m-ren-d=3}).\\
      Recall also that equations (\ref{H}) \& (\ref{m-ren-d=2}) or (\ref{m-ren-d=3}) contain an UV regulator $\alpha$, which needs to be pushed\\
      towards $+\infty$ at the very end of every computation of a physically relevant observable quantity.
\item The Hamiltonian $\hat{H}_t=(\ref{H})$ can be redisplayed in an equivalent form
      \begin{eqnarray} \label{Appendix-6-hat-H-def}
         \hat{H}_t & = & \frac{\hat{\bm p}^2}{2\,m_o^t(\alpha)}  \; - \; \frac{q_t}{m_o^t(\alpha)\,c} \; \hat{\bm A}[\alpha] \m\bm\cdot\m \hat{\bm p} \; + \;
         \frac{q_t^2}{2\,m_o^t(\alpha)\,c^2} \; \hat{\bm A}^2\n[\alpha] \; + \; V\m(\hat{\bm x}) \; + \; \hat{H}_{{\rm R}} \mez .
      \end{eqnarray}
      Here
      \be \label{Appendix-6-hat-A-def}
         \hat{\bm A}[\alpha] \; = \; \int_{{\mathbb R}^d} \m {\rm d}^d\n k \, \sum_\wp \, \frac{1}{2\,\pi} \,
         \sqrt{\frac{\hbar\,c^2}{\omega_k}} \;\, e^{-\frac{{\bm k}^2}{4\,\alpha}} \, \hat{a}_{{\bm k}\wp} \; {\bm \varepsilon}_{{\bm k}\wp} \; + \; {\rm c.c.} \mez ;
      \ee
      exactly as in equation (\ref{hat-A-alpha-def-B.1}) of {\sl Appendix B}, and $\hat{H}_{{\rm R}}=(\ref{H-R-def})$.\\
      The squared vector potential term $\hat{\bm A}^2\n[\alpha]$ can conveniently be normal ordered, one gets
      \be
         \hat{\bm A}^2\n[\alpha] \; = \; \bm: \hat{\bm A}^2\n[\alpha] \bm: \; + \;\, \varsigma \, \hat{1} \mez ;
      \ee
      where by definition
      \be
         \varsigma \; = \; \int_{{\mathbb R}^d} \m {\rm d}^d\n k \, \sum_\wp \, \frac{1}{4\,\pi^2} \, \frac{\hbar\,c^2}{\omega_k} \;\,
         e^{-\frac{{\bm k}^2}{2\,\alpha}} \mez .
      \ee
      The constant contribution $\varsigma \, \hat{1}$ (which is UV divergent for $\alpha \to +\infty$)\\
      is discarded consonantly with our common practice,\\
      {\sl cf.}~our elimination of the divergent ZPE part of $\hat{H}_{{\rm R}}=(\ref{H-R-def})$ in A.1,\\
      and similar steps taken in {\sl Appendix B} $[\,$equations (\ref{E-(2)-discard-naive}), (\ref{E-(2)-discard-d=2}), (\ref{E-(2)-discard-d=3})$\,]$.\\
      Correspondingly, our finalized Hamiltonian $[\,$entering into the TDSCHE (\ref{TDSCHE})$\,]$ reads as follows:
      \begin{eqnarray} \label{Appendix-6-hat-H-def-2}
         \hat{H}_t & = & \frac{\hat{\bm p}^2}{2\,m_o^t(\alpha)}  \; - \; \frac{q_t}{m_o^t(\alpha)\,c} \; \hat{\bm A}[\alpha] \m\bm\cdot\m \hat{\bm p} \; + \;
         \frac{q_t^2}{2\,m_o^t(\alpha)\,c^2} \; \bm: \hat{\bm A}^2\n[\alpha] \bm: \; + \; V\m(\hat{\bm x}) \; + \; \hat{H}_{{\rm R}} \mez .
      \end{eqnarray}
\item An explicit numerical solution of the TDSCHE (\ref{TDSCHE}) represents a worthy challenge.\\
      Difficulties do arise not only due to an {\color{blue} enormous number of the degrees of freedom involved}\\
      (in fact, the radiation field possesses an infinite number of degrees of freedom),\\
      but also due to the {\color{blue} UV divergencies\\
      entering into the TDSCHE (\ref{TDSCHE}) via (\ref{H}) \& (\ref{m-ren-d=2}) or (\ref{m-ren-d=3}) in the limit of $\alpha \to +\infty$.}\\
      In the present lecture notes, we prefer not to pursue an ambitious (yet feasible) program\\
      of resolving the TDSCHE (\ref{TDSCHE}) in a numerically exact manner without approximations.\\
      Instead, we attack equation (\ref{TDSCHE}) by employing the {\color{blue} approximative mean field (Hartree) method.
      The mean field formalism must of course be properly adapted,\\
      in order to cope with the UV divergencies just mentioned above.}\\
      This is explained in a self contained fashion below, see {\sl Appendix C.2} and {\sl Appendix C.3}.\\
      As we shall see shortly,\\
      the resulting renormalized mean field method (in which the limit of $\alpha \to +\infty$ is already taken)\\
      does not contain any UV divergencies,\\ and the corresponding algorithm is suitable for numerical implementation.\\
      See the summary given in {\sl Appendix C.4}.\\
      $[\,$It is not the purpose of {\sl Appendix C} to comment on an (in)capability of the mean field approach\\
      \phantom{$[\,$}to describe various physical phenomena accommodated by our studied model.\\
      \phantom{$[\,$}Our goal is mainly to {\color{blue} illustrate how is the concept of mass renormalization implemented\\
      \phantom{$[\,$}into a nonperturbative (and approximative) numerical method}.$\,]$
\end{itemize}

\vspace*{+0.20cm}

{\bf Appendix C.2: Standard developments of the mean field theory}
\vspace*{-0.20cm}
\begin{itemize}
\item Within the mean field (Hartree) method,\\
      the sought solution $| \Psi_t \ra$ of the TDSCHE (\ref{TDSCHE}) is approximated by a product ansatz
      \be \label{Appendix-6-Psi-t-ansatz}
         | \Psi_t \ra \; = \; | \psi_t \ra_{\m A} \; | \beta_t \ra_{\m R} \mez .
      \ee
      Here $| \psi_t \ra_{\m A}$ stands for the pertinent unit normalized dynamical atomic state vector,\\
      and $| \beta_t \ra_{\m R}$ represents an (unit normalized) coherent state of the radiation field.\\
      Recall that $| \beta_t \ra_{\m R}$ is an eigenstate of all the modal annihilation operators, one has
      \be \label{beta-CS-eigenproperty}
         \hat{a}_{{\bm k}\wp} \, | \beta_t \ra_{\m R} \; = \; \beta_{{\bm k}\wp}\n(t) \, | \beta_t \ra_{\m R} \mez .
      \ee
      The corresponding modal eigenvalues $\beta_{{\bm k}\wp}\n(t)$ characterize completely $| \beta_t \ra_{\m R}$.\\
      More information on the concept of coherent states can be found e.g.~in Ref.~\cite{CS}.\\
      $[\,$In passing we note that an ansatz (\ref{Appendix-6-Psi-t-ansatz}) is almost always well justified
      in the (semi)classical regime,\\ \phantom{$[\,$}where our model atom undergoes driving by a laser pulse.
      Yet important exceptions do occur either.\\ \phantom{$[\,$}See Ref.~\cite{CS} for details.$\,]$\\
      Appropriate equations of motion for $| \psi_t \ra_{\m A}$ and $| \beta_t \ra_{\m R}$ will be determined below\\
      by employing the Dirac--Frenkel variational principle \cite{DF}, which demands having
      \be \label{Appendix-6-Dirac-Frenkel}
         \delta \, \int_{t_1}^{t_2} \m \la \Psi_t | \, \hat{H}_t \, - \, i\hbar\,\partial_t \, | \Psi_t \ra \; {\rm d}t \; = \; 0 \mez .
      \ee
      Recall that in (\ref{Appendix-6-Dirac-Frenkel}) one tacitly assumes $\delta \, | \Psi_{t_1} \ra \, = \, | \emptyset \ra \, = \, \delta \, | \Psi_{t_2} \ra$ and similarly for the bras.
\item It is a straightforward matter to write down an explicit formula for our Dirac--Frenkel functional
      \be \label{Appendix-6-Dirac-Frenkel-functional}
         {\cal S} \; = \; \int_{t_1}^{t_2} \m \la \Psi_t | \, \hat{H}_t \, - \, i\hbar\,\partial_t \, | \Psi_t \ra \; {\rm d}t \mez .
      \ee
      One just combines (\ref{Appendix-6-hat-H-def-2}), (\ref{Appendix-6-Psi-t-ansatz}), (\ref{beta-CS-eigenproperty}),
      (\ref{Appendix-6-Dirac-Frenkel})\\ and takes advantage of the well known textbook properties of the coherent states\\
      listed e.g.~in Ref.~\cite{CS}. This yields
      \be \label{Appendix-6-Dirac-Frenkel-functional-take-2}
         {\cal S} \; = \; \int_{t_1}^{t_2} \m {\cal F}(t) \; {\rm d}t \mez ;
      \ee
      with
      \begin{eqnarray} \label{Appendix-6-cal-F-t-def}
         {\cal F}(t) & \equiv & \la \Psi_t | \, \hat{H}_t \, - \, i\hbar\,\partial_t \, | \Psi_t \ra \; = \nonumber\\
         & = & _A\m\la \psi_t | \left\{ \frac{\hat{\bm p}^2}{2\,m_o^t(\alpha)} \; + \; V\m(\hat{\bm x}) \; - \; i\hbar\,\partial_t \right\} | \psi_t \ra_{\m A}
         \; - \; \frac{q_t}{m_o^t(\alpha)\,c} \; {\bm A}[\alpha,\beta_t] \m\bm\cdot\m \,\m\m _A\m\la \psi_t | \,\hat{\bm p} | \psi_t \ra_{\m A} \nonumber\\
         & + & \frac{q_t^2}{2\,m_o^t(\alpha)\,c^2} \; {\bm A}^2\n[\alpha,\beta_t] \, _A\m\la \psi_t | \psi_t \ra_{\m A} \; + \;
         \int_{{\mathbb R}^d} \m {\rm d}^d\n k \, \sum_\wp \, \hbar\omega_k \; \beta_{{\bm k}\wp}^*\n(t) \, \beta_{{\bm k}\wp}\n(t) \\
         & - & \frac{i\,\hbar}{2} \, \int_{{\mathbb R}^d} \m {\rm d}^d\n k \, \sum_\wp \, \left\{ \, \beta_{{\bm k}\wp}^*\n(t) \, \bm\dot{\beta}_{{\bm k}\wp}\n(t)
         \, - \, \bm\dot{\beta}_{{\bm k}\wp}^*\n(t) \, \beta_{{\bm k}\wp}\n(t) \, \right\} \mez . \nonumber
      \end{eqnarray}
      Here by definition
      \be \label{Appendix-6-bm-A-alpha-beta-def}
         {\bm A}[\alpha,\beta_t] \; = \; \int_{{\mathbb R}^d} \m {\rm d}^d\n k \, \sum_\wp \, \frac{1}{2\,\pi} \,
         \sqrt{\frac{\hbar\,c^2}{\omega_k}} \;\, e^{-\frac{{\bm k}^2}{4\,\alpha}} \, \beta_{{\bm k}\wp}\n(t) \; {\bm \varepsilon}_{{\bm k}\wp} \; + \; {\rm c.c.} \mez ;
      \ee
      this is the classical vector potential field corresponding to the complex amplitudes $\beta_{{\bm k}\wp}\n(t)$
      of $| \beta_t \ra_{\m R}$.\\ The Dirac--Frenkel variational principle (\ref{Appendix-6-Dirac-Frenkel})\\
      is now equivalent to the following set of stationarity conditions:
      \begin{eqnarray}
         \label{Appendix-6-DF-condition-1} \frac{\delta \, {\cal S}}{\delta\,\m\m_A\m\la \psi_t |} & = & | \emptyset \ra_{\m A} \mez ;\\
         \label{Appendix-6-DF-condition-2} \frac{\delta \, {\cal S}}{\delta\,\beta_{{\bm k}\wp}^*\n(t)} & = & 0 \hspace*{+0.975cm} .
      \end{eqnarray}
      As usual, the involved variations $\delta\,\m\m_A\m\la \psi_t |$ and $\delta\,\beta_{{\bm k}\wp}^*\n(t)$\\
      are allowed to be arbitrary yet vanishing at the edges $t=t_1$ or $t=t_2$ of our trial time interval.\\
      $[\,$The same applies of course also for their adjoint counterparts $\delta \, | \psi_t \ra_{\m A}$ and $\delta\,\beta_{{\bm k}\wp}\n(t)$.$\,]$\\
      An explicit calculation converts (\ref{Appendix-6-DF-condition-1}) into
      \be \label{Appendix-6-EOM-psi-take-1}
         {\color{blue} i \hbar \, \partial_t \, | \psi_t \ra_{\m A} \; = \;
         \left\{ \frac{1}{2\,m_o^t(\alpha)} \m \left( \hat{\bm p} \, - \, \frac{q_t}{c} \, {\bm A}[\alpha,\beta_t] \right)^{\m\n 2} \; + \; V\m(\hat{\bm x}) \right\} | \psi_t \ra_{\m A} \mez .}
      \ee
      {\color{blue} This is the sought mean field equation of motion for $| \psi_t \ra_{\m A}$.\\
      Its appearance is consistent with physics intuition.}\\
      Furthermore, condition (\ref{Appendix-6-DF-condition-2}) is converted into
      \be \label{Appendix-6-EOM-beta-take-1-prelim}
         {\color{blue} i \hbar \, \partial_t \, \beta_{{\bm k}\wp}\n(t) \; = \; \hbar\omega_k \; \beta_{{\bm k}\wp}\n(t) \; - \; \frac{q_t}{2\,\pi} \,
         \sqrt{\frac{\hbar}{\omega_k}} \; e^{-\frac{{\bm k}^2}{4\,\alpha}} \, {\bm \varepsilon}_{{\bm k}\wp} \m\bm\cdot\m \,\m\m _A\m\la \psi_t | \;
         \frac{1}{m_o^t(\alpha)} \m \left( \hat{\bm p} \, - \, \frac{q_t}{c} \, {\bm A}[\alpha,\beta_t] \right) | \psi_t \ra_{\m A} \mez .}
      \ee
      {\color{blue} This is the sought mean field equation of motion for $\beta_{{\bm k}\wp}\n(t)$.\\
      Its appearance is again consistent with physics intuition}, since it closely resembles formula (\ref{Heisenberg-a}).\\
      The just derived mean field equations of motion (\ref{Appendix-6-EOM-psi-take-1}) and (\ref{Appendix-6-EOM-beta-take-1-prelim}) are mutually coupled,\\
      and can be propagated in time once the appropriate initial conditions are specified at $t=t_0$.\\
      Both equations (\ref{Appendix-6-EOM-psi-take-1}) and (\ref{Appendix-6-EOM-beta-take-1-prelim}) still contain the bare mass $m_o^t(\alpha)$ in the denominator.\\ Yet $m_o^t(\alpha)$ becomes infinitely large for $\alpha \to +\infty$, see (\ref{m-ren-d=2}) or (\ref{m-ren-d=3}).\\
      If so, {\color{blue} it is not obviously seen whether or not\\
      a well behaved mean field dynamics of $| \psi_t \ra_{\m A}$ and $\beta_{{\bm k}\wp}\n(t)$
      emerges in the limit of $\alpha \to +\infty$.}\\ 
      It is the aim of our subsequent elaborations\\
      to provide an adequate clarification and an affirmative answer.\\      
      The mean field equations of motion (\ref{Appendix-6-EOM-psi-take-1}) and (\ref{Appendix-6-EOM-beta-take-1-prelim}),
      as they stand now,\\ are of no practical use for numerical computations.\\
      It is the aim of our subsequent elaborations\\
      to transform (\ref{Appendix-6-EOM-psi-take-1}) and (\ref{Appendix-6-EOM-beta-take-1-prelim})
      into a practically useful appearance (by taking the limit $\alpha \to +\infty$).
\end{itemize}

\vspace*{+0.20cm}

{\bf Appendix C.3: Radiation reaction and mass renormalization in the mean field theory}
\vspace*{-0.20cm}
\begin{itemize}
\item The matrix element
      \be \label{Appendix-6-velocity-def}
         {\bm v}(t) \; = \; _A\m\la \psi_t | \; \frac{1}{m_o^t(\alpha)} \m \left( \hat{\bm p} \, - \, \frac{q_t}{c} \,
         {\bm A}[\alpha,\beta_t] \right) | \psi_t \ra_{\m A}
      \ee
      possesses the meaning of an expectation value of the velocity of the electron.\\
      In order to validate the just presented statement, we define the position expectation value
      \be \label{Appendix-6-position-def}
         {\bm x}(t) \; = \; _A\m\la \psi_t | \, \hat{\bm x} \, | \psi_t \ra_{\m A} \mez ;
      \ee
      and examine its time derivative $\bm\dot{\bm x}(t)$. Direct calculation yields
      \be
         \bm\dot{\bm x}(t) \; = \; _A\m\la \bm\dot{\psi}_t | \, \hat{\bm x} \, | \psi_t \ra_{\m A} \; + \; _A\m\la \psi_t | \, \hat{\bm x} \, | \bm\dot{\psi}_t \ra_{\m A}
         \; = \; _A\m\la \psi_t | \, \frac{1}{i\,\hbar} \, \Bigl[ \, \hat{\bm x} \, , \hat{H}_{\m A}\n(t) \Bigr] \, | \psi_t \ra_{\m A} \; = \; (\ref{Appendix-6-velocity-def}) \mez ;
      \ee
      exactly as claimed.\\ $[\,$Here $\hat{H}_{\m A}\n(t)$ stands for the effective atomic Hamiltonian appearing on the r.h.s.~of (\ref{Appendix-6-EOM-psi-take-1}).$\,]$
\item Having established the physical meaning of entity (\ref{Appendix-6-velocity-def}),\\ we can replace (\ref{Appendix-6-EOM-beta-take-1-prelim}) by a more concise equation of motion
      \be \label{Appendix-6-EOM-beta-take-1}
         i \hbar \, \partial_t \, \beta_{{\bm k}\wp}\n(t) \; = \; \hbar\omega_k \; \beta_{{\bm k}\wp}\n(t) \; - \; \frac{q_t}{2\,\pi} \,
         \sqrt{\frac{\hbar}{\omega_k}} \; e^{-\frac{{\bm k}^2}{4\,\alpha}} \, {\bm \varepsilon}_{{\bm k}\wp} \m\bm\cdot\m {\bm v}(t) \mez ;
      \ee
      which possesses an analytic solution
      \be \label{Appendix-6-EOM-beta-solution}
         \beta_{{\bm k}\wp}\n(t) \; = \; \beta_{{\bm k}\wp}\n(t_0) \; e^{-i\omega_k t} \; - \; \frac{1}{2\,\pi} \,
         \sqrt{\frac{\hbar}{\omega_k}} \; e^{-\frac{{\bm k}^2}{4\,\alpha}} \int_{t_0}^{t} \m {\rm d}t' \;\,
         e^{-i\omega_k(t-t'\n)} \; q_{t'} \, {\bm \varepsilon}_{{\bm k}\wp} \m\bm\cdot\m {\bm v}(t'\n) \mez .
      \ee
      Note that equations analogous to (\ref{Appendix-6-EOM-beta-take-1})-(\ref{Appendix-6-EOM-beta-solution})\\
      have already been encountered in our previous elaborations,\\
      see formulas (\ref{Heisenberg-a}) and (\ref{hat-a-Heisenberg-explicit}) of the main text.
\item We may now follow further an analogy with Section C of the main text,\\
      and evaluate explicitly the associated electric field
      \be \label{Appendix-6-bm-E-t-def}
         {\bm E}[\alpha,\beta_t] \; = \; -\,\frac{1}{c} \; \partial_t \, {\bm A}[\alpha,\beta_t] \mez ;
      \ee
      while pushing $\alpha \to +\infty$. Similarly as in (\ref{hat-E-t-y-explicit-splitting}) one expresses
      ${\bm E}[\alpha,\beta_t]$ as a sum
      \be \label{Appendix-6-bm-E-sum}
         {\bm E}[\alpha,\beta_t] \; = \; {\bm E}_0\n[\alpha,\beta_{t_0},t] \; + \; {\bm E}_{\rm RR}[\alpha,\beta_t] \mez .
      \ee
      Here ${\bm E}_0\n[\alpha,\beta_{t_0},t]$ stands for the incoming freely propagating electric field (generated e.g.~by a laser),\\
      which is characterized by its time independent modal amplitudes $\beta_{{\bm k}\wp}\n(t_0)$.\\ Stated mathematically, we have
      \be \label{Appendix-6-bm-E-0-def}
         {\bm E}_0\n[\alpha,\beta_{t_0},t] \; = \; \frac{i}{2\,\pi} \, \int_{{\mathbb R}^d} \m {\rm d}^d\n k \, \sum_\wp
         \, \sqrt{\hbar\,\omega_k} \;\, e^{-i\omega_k (t-t_0)} \, e^{-\frac{{\bm k}^2}{4\,\alpha}} \, \beta_{{\bm k}\wp}\n(t_0) \;
         {\bm \varepsilon}_{{\bm k}\wp} \; + \; {\rm c.c.} \mez ;
      \ee
      much like in (\ref{bm-F-0-el-t}).\\
      On the other hand, ${\bm E}_{\rm RR}[\alpha,\beta_t]$ represents electric field of the radiation reaction (RR),\\
      which can be treated in much the same way as done in the main text (see in particular C.4-C.5).\\
      Importantly, it has turned out in C.4-C.5 that an explicit appearance of the electric RR field\\
      (thus, of ${\bm E}_{\rm RR}[\alpha,\beta_t]$ in our present context)\\
      depends crucially upon the pertinent spatial dimension $d$.\\
      For $d=2$, one finds that
      \begin{eqnarray} \label{Appendix-6-bm-E-RR-d=2}
         \hspace*{-0.50cm} {\bm E}_{\rm RR}[\alpha,\beta_t] & = & -\,\frac{q_t}{4\,\pi\,c^2} \; \bm\dot{\bm v}(t) \, \ln\m\m\left(\frac{\alpha}{\alpha_0}\right)
         \; - \; \frac{1}{2\,\pi\,c^2} \, \int_{t_0}^{t} \m q_{t'} \, \bm\ddot{\bm v}(t'\n) \; \ln\m\m\left(\frac{t-t'}{\tilde{t}}\right)
         {\rm d}t' \\
         \hspace*{-0.50cm} & + & {\rm an}\;\,\alpha{\rm -dependent \;\, correction \;\, vanishing \;\, as} \;\, \alpha \to +\infty \hspace*{+2.00cm} ; \nonumber
      \end{eqnarray}
      {\sl cf.}~equation (\ref{F-RR-analysis-take-6}). 
      We refer to the main text for a detailed explanation of the involved symbols.\\
      For $d=3$, one finds that
      \begin{eqnarray} \label{Appendix-6-bm-E-RR-d=3}
         {\bm E}_{\rm RR}[\alpha,\beta_t] & = &  - \, \frac{4}{3} \, \frac{q_t}{c^2} \, \sqrt{\frac{\alpha}{2\,\pi}} \; \bm\dot{\bm v}(t)
         \; + \; \frac{2}{3} \, \frac{q_t}{c^3} \; \bm\ddot{\bm v}(t) \; + \; {\cal O}(\alpha^{-\frac{1}{2}}) \mez ;
      \end{eqnarray}
      {\sl cf.}~equation (\ref{electric-force-ERR-d=3-final}).
\item Having resolved exhaustively the mean field dynamics of our radiation field,\\
      let us move on towards examining further the mean field dynamics of our atom.\\
      Formula (\ref{Appendix-6-EOM-psi-take-1}) refers to the Schr\"{o}dinger picture description\\
      of the dynamical time evolution of $| \psi_t \ra_{\m A}$. It is a straightforward matter\\
      to write down the corresponding Heisenberg picture equation of motion for $\hat{\bm x}(t)$.\\
      We proceed analogously as in Section B of the main text, and find that
      \be
         m_o^t(\alpha) \; \bm\ddot{\hat{\bm x}}(t) \; = \; -{\bm \nabla}V\m(\hat{\bm x}(t)) \; + \; q_t\,{\bm E}[\alpha,\beta_t]
         \mez ;
      \ee
      where ${\bm E}[\alpha,\beta_t]=(\ref{Appendix-6-bm-E-t-def})$. Subsequently, after taking an expectation value over
      $| \psi_{t_0} \ra_{\m A}$, one gets
      \be
         m_o^t(\alpha) \; _A\m\la \psi_{t_0} | \, \bm\ddot{\hat{\bm x}}(t) \, | \psi_{t_0} \ra_{\m A} \; = \;
         _A\m\la \psi_{t_0} | \, -\m\m{\bm \nabla}V\m(\hat{\bm x}(t)) \, | \psi_{t_0} \ra_{\m A}
         \; + \; q_t\,{\bm E}[\alpha,\beta_t] \mez ;
      \ee
      or, equivalently,
      \be \label{Appendix-6-bm-v-EOM-prelim}
         m_o^t(\alpha) \; \bm\dot{\bm v}(t) \; = \;
         _A\m\la \psi_t | \, -\m\m{\bm \nabla}V\m(\hat{\bm x}) \, | \psi_t \ra_{\m A}
         \; + \; q_t\,{\bm E}[\alpha,\beta_t] \mez .
      \ee
      Here ${\bm v}(t)=(\ref{Appendix-6-velocity-def})$, we recall also the definition (\ref{Appendix-6-position-def}) in this context.\\
      Equation (\ref{Appendix-6-bm-v-EOM-prelim}) can be now combined with (\ref{Appendix-6-bm-E-sum}), (\ref{Appendix-6-bm-E-0-def})
      and (\ref{Appendix-6-bm-E-RR-d=2}) or (\ref{Appendix-6-bm-E-RR-d=3}).\\ This yields more explicit outcomes
      \begin{eqnarray} \label{Appendix-6-bm-v-EOM-d=2}
         \hspace*{-1.50cm} m_o^t(\alpha) \; \bm\dot{\bm v}(t) & = &
         _A\m\la \psi_t | \, -\m\m{\bm \nabla}V\m(\hat{\bm x}) \, | \psi_t \ra_{\m A}
         \; + \; q_t\,{\bm E}_0\n[\alpha,\beta_{t_0},t] \\
         \hspace*{-1.50cm} & - & \frac{q_t^2}{4\,\pi\,c^2} \; \bm\dot{\bm v}(t) \,
         \ln\m\m\left(\frac{\alpha}{\alpha_0}\right) \; - \; \frac{q_t}{2\,\pi\,c^2} \,
         \int_{t_0}^{t} \m q_{t'} \, \bm\ddot{\bm v}(t'\n) \; \ln\m\m\left(\frac{t-t'}{\tilde{t}}\right)
         {\rm d}t' \mez \mez [\,d=2\,] \nonumber\\
         \hspace*{-1.50cm} & + & {\rm an}\;\,\alpha{\rm -dependent \;\, correction \;\, vanishing \;\, as} \;\, \alpha \to +\infty \hspace*{+1.00cm} ; \nonumber
      \end{eqnarray}
      and
      \begin{eqnarray} \label{Appendix-6-bm-v-EOM-d=3}
         \hspace*{-1.50cm} m_o^t(\alpha) \; \bm\dot{\bm v}(t) & = &
         _A\m\la \psi_t | \, -\m\m{\bm \nabla}V\m(\hat{\bm x}) \, | \psi_t \ra_{\m A}
         \; + \; q_t\,{\bm E}_0\n[\alpha,\beta_{t_0},t] \\
         \hspace*{-1.50cm} & - & \frac{4}{3} \, \frac{q_t}{c^2} \, \sqrt{\frac{\alpha}{2\,\pi}}
         \; \bm\dot{\bm v}(t) \; + \; \frac{2}{3} \, \frac{q_t}{c^3} \; \bm\ddot{\bm v}(t)
         \; + \; {\cal O}(\alpha^{-\frac{1}{2}})\mez . \mez [\,d=3\,] \nonumber
      \end{eqnarray}
      Importantly, the just displayed relations (\ref{Appendix-6-bm-v-EOM-d=2})-(\ref{Appendix-6-bm-v-EOM-d=3}) possess a well defined limit for
      $\alpha \to +\infty$,\\ provided only that the mass renormalization is made,\\
      following our already established prescriptions (\ref{m-ren-d=2}) and (\ref{m-ren-d=3}).\\
      Having done so, equations (\ref{Appendix-6-bm-v-EOM-d=2})-(\ref{Appendix-6-bm-v-EOM-d=3}) boil down into
      \begin{eqnarray} \label{Appendix-6-bm-v-EOM-d=2-ren}
         \hspace*{-1.50cm} m \; \bm\dot{\bm v}(t) & = &
         _A\m\la \psi_t | \, -\m\m{\bm \nabla}V\m(\hat{\bm x}) \, | \psi_t \ra_{\m A} \; + \; q_t\,{\bm E}_0\n[\beta_{t_0},t]
         \; - \; \frac{q_t}{2\,\pi\,c^2} \, \int_{t_0}^{t} \m q_{t'} \, \bm\ddot{\bm v}(t'\n) \;
         \ln\m\m\left(\frac{t-t'}{\tilde{t}}\right) {\rm d}t' \mz ; \mz [\,d=2\,]
      \end{eqnarray}
      and
      \begin{eqnarray} \label{Appendix-6-bm-v-EOM-d=3-ren}
         \hspace*{-0.50cm} m \; \bm\dot{\bm v}(t) & = &
         _A\m\la \psi_t | \, -\m\m{\bm \nabla}V\m(\hat{\bm x}) \, | \psi_t \ra_{\m A} \; + \; q_t\,{\bm E}_0\n[\beta_{t_0},t]
         \; + \; \frac{2}{3} \, \frac{q_t}{c^3} \; \bm\ddot{\bm v}(t) \mez . \mez [\,d=3\,]
      \end{eqnarray}
      Here by definition
      \be \label{Appendix-6-bm-E-0-ren}
         {\bm E}_0\n[\beta_{t_0},t] \; = \; \frac{i}{2\,\pi} \, \int_{{\mathbb R}^d} \m {\rm d}^d\n k \, \sum_\wp
         \, \sqrt{\hbar\,\omega_k} \;\, e^{-i\omega_k (t-t_0)} \, \beta_{{\bm k}\wp}\n(t_0) \;
         {\bm \varepsilon}_{{\bm k}\wp} \; + \; {\rm c.c.} \mez ;
      \ee
      consistently with (\ref{Appendix-6-bm-E-0-def}).
      The just derived formulas (\ref{Appendix-6-bm-v-EOM-d=2-ren}) and (\ref{Appendix-6-bm-v-EOM-d=3-ren})\\
      do not contain any divergent or (near)singular terms,\\
      as opposed to their not-yet-renormalized counterparts (\ref{Appendix-6-bm-v-EOM-d=2}) and (\ref{Appendix-6-bm-v-EOM-d=3}).
\item Let us focus our attention now on simplifying the Schr\"{o}dinger equation (\ref{Appendix-6-EOM-psi-take-1}) for $| \psi_t \ra_{\m A}$\\
      in the limit of $\alpha \to +\infty$. Equations (\ref{Appendix-6-bm-A-alpha-beta-def}), (\ref{Appendix-6-bm-E-t-def}) and (\ref{Appendix-6-bm-E-sum})
      motivate us to set
      \be \label{Appendix-6-bm-A-sum}
         {\bm A}[\alpha,\beta_t] \; = \; {\bm A}_0\n[\alpha,\beta_{t_0},t] \; + \; {\bm A}_{\rm RR}\n[\alpha,\beta_t] \mez .
      \ee      
      Here by definition
      \be \label{Appendix-6-bm-A-0-def}
         {\bm A}_0\n[\alpha,\beta_{t_0},t] \; = \; \int_{{\mathbb R}^d} \m {\rm d}^d\n k \, \sum_\wp \, \frac{1}{2\,\pi} \,
         \sqrt{\frac{\hbar\,c^2}{\omega_k}} \;\, e^{-i\omega_k(t-t_0)} \, e^{-\frac{{\bm k}^2}{4\,\alpha}} \, \beta_{{\bm k}\wp}\n(t_0) \; {\bm \varepsilon}_{{\bm k}\wp} \; + \; {\rm c.c.} \mez ;
      \ee
      consistently with (\ref{Appendix-6-bm-E-0-def}). Clearly, ${\bm A}_0\n[\alpha,\beta_{t_0},t]$ stands for\\
      the incoming freely propagating vector potential field (generated e.g.~by a laser),\\
      which is characterized by its time independent modal amplitudes $\beta_{{\bm k}\wp}\n(t_0)$.\\
      On the other hand, ${\bm A}_{\rm RR}\n[\alpha,\beta_t]$ represents the vector potential field of the radiation reaction (RR).\\
      It is defined unambiguously by an initial value problem
      \be \label{A-RR-IVP}
         -\,\frac{1}{c} \; \partial_t \, {\bm A}_{\rm RR}\n[\alpha,\beta_t] \; = \; {\bm E}_{\rm RR}\n[\alpha,\beta_t]
         \mez , \mez {\bm A}_{\rm RR}\n[\alpha,\beta_t]\,\Bigr|_{t=t_0} \; = \; {\bm 0} \mez .
      \ee
      The initial time instant $t_0$ has been left as yet arbitrary.\\
      To simplify our subsequent considerations, let us conveniently push now $t_0 \to -\infty$.\\
      This converts an initial condition of the problem (\ref{A-RR-IVP}) into
      \be
         {\bm A}_{\rm RR}\n[\alpha,\beta_t]\,\Bigr|_{t \to -\infty} \; = \; {\bm 0} \mez .
      \ee
      Our previously discussed explicit results (\ref{Appendix-6-bm-E-RR-d=2}) and (\ref{Appendix-6-bm-E-RR-d=3}) for
      ${\bm E}_{\rm RR}\n[\alpha,\beta_t]$ enable us then to infer that
      \begin{eqnarray} \label{Appendix-6-bm-A-RR-d=2}
         \hspace*{-1.50cm} {\bm A}_{\rm RR}[\alpha,\beta_t] & = & \frac{q_t}{4\,\pi\,c} \; {\bm v}(t)
         \, \ln\m\m\left(\frac{\alpha}{\alpha_0}\right) \; + \; \frac{1}{2\,\pi\,c} \, \int_{-\infty}^{t}
         {\rm d}t' \int_{-\infty}^{t'} \m {\rm d}t'' \, q_{t''} \, \bm\ddot{\bm v}(t''\n) \; \ln\m\m\left(\frac{t'\n-t''}{\tilde{t}}
         \right) \mez \mez [\,d=2\,] \\
         \hspace*{-1.50cm} & + & {\rm an}\;\,\alpha{\rm -dependent \;\, correction \;\, vanishing \;\, as} \;\, \alpha \to +\infty \hspace*{+2.80cm} ; \nonumber
      \end{eqnarray}
      and
      \begin{eqnarray} \label{Appendix-6-bm-A-RR-d=3}
         {\bm A}_{\rm RR}[\alpha,\beta_t] & = &  \frac{4}{3} \, \frac{q_t}{c} \, \sqrt{\frac{\alpha}{2\,\pi}}
         \; {\bm v}(t) \; - \; \frac{2}{3} \, \frac{q_t}{c^2} \; \bm\dot{\bm v}(t) \; + \;
         {\cal O}(\alpha^{-\frac{1}{2}}) \mez . \mez [\,d=3\,]
      \end{eqnarray}
\item The just obtained expansions (\ref{Appendix-6-bm-A-RR-d=2}) and (\ref{Appendix-6-bm-A-RR-d=3}) represent a crucial ingredient/insight\\
      which ultimately rationalizes the kinetic energy operator
      \be
         \frac{1}{2\,m_o^t(\alpha)} \m \left( \hat{\bm p} \, - \, \frac{q_t}{c} \, {\bm A}[\alpha,\beta_t] \right)^{\m\n 2}
         \; = \; \frac{\hat{\bm p}^2}{2\,m_o^t(\alpha)} \; - \; \frac{q_t}{m_o^t(\alpha)\,c} \, {\bm A}[\alpha,\beta_t] \m\bm\cdot\m
         \hat{\bm p} \; + \; \frac{q_t^2}{2\,m_o^t(\alpha)\,c^2} \, {\bm A}^{\n 2}\n[\alpha,\beta_t]
      \ee
      appearing in the Schr\"{o}dinger equation (\ref{Appendix-6-EOM-psi-take-1}).
      Indeed, one observes that:
      \vspace*{-0.20cm}
      \begin{itemize}
      \item[$\circ$] For $\alpha \to +\infty$,\\ the ${\bm A}_0\n[\alpha,\beta_{t_0},t]$ contribution to
                     ${\bm A}[\alpha,\beta_t]$ becomes negligible compared to ${\bm A}_{\rm RR}\n[\alpha,\beta_t]$;
      \item[$\circ$] The only relevant part of ${\bm A}_{\rm RR}\n[\alpha,\beta_t]$ is given by its leading ($\alpha$-divergent) 
                     order,\\ namely by the ${\cal O}(\ln\alpha)$ term for $d=2$ and by the ${\cal O}(\alpha^{1/2})$ term for $d=3$;
      \item[$\circ$] The $\frac{\hat{\bm p}^2}{2\,m_o^t(\alpha)}$ term disappears
                     since $m_o^t(\alpha) \to +\infty$ as $\alpha \to +\infty$;
      \item[$\circ$] The $\frac{q_t^2}{2\,m_o^t(\alpha)\,c^2} \, {\bm A}^{\n 2}\n[\alpha,\beta_t]$ term is a multiple of $\hat{1}$,\\
                     and can be thus eliminated (even for any finite $\alpha$)\\
                     via attaching an extra time dependent global phase to $| \psi_t \ra_{\m A}$.\\
                     Written mathematically, we set
                     \be \label{Appendix-6-tilde-psi-def}
                        | \psi_t \ra_{\m A} \; = \; e^{-\frac{i}{\hbar} \, \frac{q_t^2}{2\,m_o^t(\alpha)\,c^2} \,
                        \int_{t_0}^{t} {\bm A}^{\n 2}\n[\alpha,\beta_{t'\n}] \, {\rm d}t'} \, | \tilde{\psi}_t \ra_{\m A}
                        \mez ;
                     \ee
                     consonantly with common practice of theorists dealing with matter-laser interaction.
      \item[$\circ$] What remains is the $- \, \frac{q_t}{m_o^t(\alpha)\,c} \, {\bm A}[\alpha,\beta_t] \m\bm\cdot\m\hat{\bm p}$ term,\\
                     which simplifies dramatically in the limit of $\alpha \to +\infty$, due to (\ref{Appendix-6-bm-A-RR-d=2}) and (\ref{Appendix-6-bm-A-RR-d=3}),\\ and due to our established mass renormalization prescriptions
                     (\ref{m-ren-d=2}) and (\ref{m-ren-d=3}).\\ Indeed, the mentioned term boils down simply into
                     \be
                        {\bm v}(t) \m\bm\cdot\m\hat{\bm p} \mez ;
                     \ee
                     valid both for $d=2$ and $d=3$.
      \end{itemize}
      \vspace*{-0.20cm}
      Meaning in turn that, {\color{blue} after taking the limit of $\alpha \to +\infty$ accompanied by mass renormalization,\\
      our Schr\"{o}dinger equation (\ref{Appendix-6-EOM-psi-take-1}) takes the following compelling appearance:
      \be \label{Appendix-6-EOM-psi-take-2}
         i \hbar \, \partial_t \, | \tilde{\psi}_t \ra_{\m A} \; = \;
         \Bigl\{ {\bm v}(t) \m\bm\cdot\m\hat{\bm p} \; + \; V\m(\hat{\bm x}) \Bigr\} \, | \tilde{\psi}_t \ra_{\m A} \mez .
      \ee
      Most importantly, the just derived (mass renormalized) Schr\"{o}dinger equation (\ref{Appendix-6-EOM-psi-take-2})\\
      does not contain any divergent or (near)singular terms,\\ as opposed to its original counterpart (\ref{Appendix-6-EOM-psi-take-1}).}\\
      Before proceeding further, it is worth noting that\\
      the Heisenberg picture equivalent of the Schr\"{o}dinger picture atomic dynamics (\ref{Appendix-6-EOM-psi-take-2})\\
      takes apparently the form $\bm\dot{\hat{\bm x}}(t) \, = \, {\bm v}(t)\,\hat{1}$ and
      $\bm\dot{\hat{\bm p}}(t) \, = \, -{\bm\nabla}V\m(\hat{\bm x}(t))$. Subsequently one has
      \be \label{Appendix-6-dot-x-v-consistency}
         _A\m\la \psi_t | \, \bm\dot{\hat{\bm x}}(t) \, | \psi_t \ra_{\m A} \; = \;
         _A\m\la \tilde{\psi}_t | \, \bm\dot{\hat{\bm x}}(t) \, | \tilde{\psi}_t \ra_{\m A} \; = \; {\bm v}(t) \mez ;
      \ee
      and this is an useful consistency check.
\end{itemize}

\vspace*{+0.20cm}

{\bf Appendix C.4: Summarizing the renormalized mean field theory}
\vspace*{-0.20cm}
\begin{itemize}
\item All in all, in {\sl Appendix C.2} we have shown that\\
      our mean field ansatz (\ref{Appendix-6-Psi-t-ansatz}) leads towards the equations of motion
      (\ref{Appendix-6-EOM-psi-take-1}) and (\ref{Appendix-6-EOM-beta-take-1-prelim}).\\
      These equations of motion are however dependent upon the UV regulator $\alpha$\\
      and contain terms which become singular for $\alpha \to +\infty$.\\
      It was then not obviously seen whether or not\\
      a well behaved mean field dynamics of $| \psi_t \ra_{\m A}$ and $\beta_{{\bm k}\wp}\n(t)$
      emerges in the limit of $\alpha \to +\infty$.\\
      An adequate clarification was subsequently elaborated in {\sl Appendix C.3}. We are now able\\
      to answer the fundamental question regarding the $\alpha \to +\infty$ limit of equations
      (\ref{Appendix-6-EOM-psi-take-1}) and (\ref{Appendix-6-EOM-beta-take-1-prelim})\\ in a concise and sharp fashion:\\
      {\color{blue} {\sl It turns out that our original equations of motion (\ref{Appendix-6-EOM-psi-take-1}) and (\ref{Appendix-6-EOM-beta-take-1-prelim})
      boil down for $\alpha \to +\infty$\\ to solving the (mass renormalized) Schr\"{o}dinger equation (\ref{Appendix-6-EOM-psi-take-2})\\
      which is coupled with the (mass renormalized) classically looking equation (\ref{Appendix-6-bm-v-EOM-d=2-ren}) or
      (\ref{Appendix-6-bm-v-EOM-d=3-ren})\\ for the velocity expectation value ${\bm v}(t)$, {\sl cf.}~property (\ref{Appendix-6-dot-x-v-consistency}).}}\\
      We recall that in (\ref{Appendix-6-bm-v-EOM-d=2-ren}) and (\ref{Appendix-6-bm-v-EOM-d=3-ren}) the incoming electric field
      ${\bm E}_0\n[\beta_{t_0},t]=(\ref{Appendix-6-bm-E-0-ren})$.\\ We also point out that the gradient term of (\ref{Appendix-6-bm-v-EOM-d=2-ren}) and (\ref{Appendix-6-bm-v-EOM-d=3-ren})\\ is expressible in terms of the redefined atomic state vector $| \tilde{\psi}_t \ra_{\m A}$ simply as
      \be \label{Appendix-6-gradient}
         _A\m\la \psi_t | \, -\m\m{\bm \nabla}V\m(\hat{\bm x}) \, | \psi_t \ra_{\m A} \; = \;
         _A\m\la \tilde{\psi}_t | \, -\m\m{\bm \nabla}V\m(\hat{\bm x}) \, | \tilde{\psi}_t \ra_{\m A} \mez ;
      \ee
      {\sl cf.}~the global phase transformation (\ref{Appendix-6-tilde-psi-def}).\\
      The just described {\color{blue} {\sl renormalized mean field equations of motion (\ref{Appendix-6-EOM-psi-take-2}) \& (\ref{Appendix-6-bm-v-EOM-d=2-ren}) or
      (\ref{Appendix-6-bm-v-EOM-d=3-ren})}}\\ do not contain any divergent or (near)singular terms.\\
      Hence the coupled problem (\ref{Appendix-6-EOM-psi-take-2}) \& (\ref{Appendix-6-bm-v-EOM-d=2-ren}) or (\ref{Appendix-6-bm-v-EOM-d=3-ren}) possesses\\
      well behaved particular solutions $| \psi_t \ra_{\m A}$ and ${\bm v}(t)$\\
      which are even obtainable {\color{blue} numerically via standard methods of computational physics}.\\
      In passing we note that the Abraham-Lorentz force\\
      is explicitly accounted for in the last term of (\ref{Appendix-6-bm-v-EOM-d=2-ren}) and (\ref{Appendix-6-bm-v-EOM-d=3-ren}).\\
      We also recall that the field part of the problem possesses an explicit solution\\
      described by formulas (\ref{Appendix-6-EOM-beta-solution}) and (\ref{Appendix-6-bm-E-sum}), (\ref{Appendix-6-bm-E-0-def}), (\ref{Appendix-6-bm-E-RR-d=2}),
      (\ref{Appendix-6-bm-E-RR-d=3}), and (\ref{Appendix-6-bm-A-sum}), (\ref{Appendix-6-bm-A-0-def}), (\ref{Appendix-6-bm-A-RR-d=2}), (\ref{Appendix-6-bm-A-RR-d=3}).\\
      {\color{magenta} \sl Success, the renormalized mean field theory is now properly formulated for our model problem.}
\end{itemize}

\newpage

{\bf Appendix C.5: Two additional remarks}
\vspace*{-0.20cm}
\begin{itemize}
\item Let us find out how does the mean field approach break down\\
      if one tries to implement a naive approach without mass renormalization.\\
      In that case one would start from the Hamiltonian (\ref{Appendix-6-hat-H-def-2}),\\
      in which the bare mass $m_o^t(\alpha)$ is replaced by the physical mass $m$,\\
      and in which the limit of $\alpha \to +\infty$ is taken in $\hat{\bm A}[\alpha]$. Thus
      \begin{eqnarray} \label{Appendix-6-hat-H-def-2-naive}
         \hat{H}_t & = & \frac{\hat{\bm p}^2}{2\,m}  \; - \; \frac{q_t}{m\,c} \; \hat{\bm A}[\infty] \m\bm\cdot\m \hat{\bm p} \; + \;
         \frac{q_t^2}{2\,m\,c^2} \; \bm: \hat{\bm A}^2\n[\infty] \bm: \; + \; V\m(\hat{\bm x}) \; + \; \hat{H}_{{\rm R}} \mez .
      \end{eqnarray}
      The corresponding mean field equations of motion\\
      would then be obtained as appropriate modifications of (\ref{Appendix-6-EOM-psi-take-1}), (\ref{Appendix-6-velocity-def}), (\ref{Appendix-6-EOM-beta-take-1}).\\
      Namely, one would have
       \be \label{Appendix-6-EOM-psi-take-1-naive}
         i \hbar \, \partial_t \, | \psi_t \ra_{\m A} \; = \;
         \left\{ \frac{1}{2\,m} \m \left( \hat{\bm p} \, - \, \frac{q_t}{c} \, {\bm A}[\infty,\beta_t] \right)^{\m\n 2} \; + \; V\m(\hat{\bm x}) \right\} | \psi_t \ra_{\m A} \mez ;
      \ee
      \be \label{Appendix-6-velocity-def-naive}
         {\bm v}(t) \; = \; _A\m\la \psi_t | \; \frac{1}{m} \m \left( \hat{\bm p} \, - \, \frac{q_t}{c} \,
         {\bm A}[\infty,\beta_t] \right) | \psi_t \ra_{\m A} \mez ;
      \ee
      \be \label{Appendix-6-EOM-beta-take-1-naive}
         i \hbar \, \partial_t \, \beta_{{\bm k}\wp}\n(t) \; = \; \hbar\omega_k \; \beta_{{\bm k}\wp}\n(t)
         \; - \; \frac{q_t}{2\,\pi} \, \sqrt{\frac{\hbar}{\omega_k}} \;
         {\bm \varepsilon}_{{\bm k}\wp} \m\bm\cdot\m {\bm v}(t) \mez .
      \ee
      Subsequently, an equation of motion for ${\bm v}(t)$ would take the form
      \be \label{Appendix-6-bm-v-EOM-prelim-naive}
         m \; \bm\dot{\bm v}(t) \; = \; _A\m\la \psi_t | \, -\m\m{\bm \nabla}V\m(\hat{\bm x}) \, | \psi_t \ra_{\m A}
         \; + \; q_t\,{\bm E}[\infty,\beta_t] \mez ;
      \ee
      following a close analogy with (\ref{Appendix-6-bm-v-EOM-prelim}).\\
      The anticipated disaster emerges after (\ref{Appendix-6-EOM-beta-take-1-naive}) is solved for $\beta_{{\bm k}\wp}\n(t)$
      as we did above in (\ref{Appendix-6-EOM-beta-solution}).\\
      Indeed, when one aims at evaluating the resulting fields ${\bm E}[\infty,\beta_t]$ and ${\bm A}[\infty,\beta_t]$,\\
      infinities are encountered. This can be seen immediately from (\ref{Appendix-6-bm-E-0-def}), (\ref{Appendix-6-bm-E-RR-d=2}) and
      (\ref{Appendix-6-bm-A-RR-d=2}), (\ref{Appendix-6-bm-A-RR-d=3}).\\
      Hence the equations of motion (\ref{Appendix-6-EOM-psi-take-1-naive}) and (\ref{Appendix-6-bm-v-EOM-prelim-naive}) are recognized to be unphysical,\\
      and the whole naive mean field approach thus breaks down hopelessly.
\item Let us examine how does our renormalized mean field theory work\\
      when applied on studying the atomic ground state dressed by the (vacuum) radiation field.\\
      Recall that we are supposed to propagate in time the coupled problem
      (\ref{Appendix-6-EOM-psi-take-2}) \& (\ref{Appendix-6-bm-v-EOM-d=2-ren}) or (\ref{Appendix-6-bm-v-EOM-d=3-ren})\\
      where the gradient term is determined by formula (\ref{Appendix-6-gradient}).\\
      The corresponding initial conditions, set up at the time instant $t_0 \to -\infty$,\\
      look for our particular situation as follows:
      \be \label{Appendix-6-ground-ics}
         | \tilde{\psi}_{t_0} \ra_{\m A} \; = \; | \varphi_{\rm ground} \ra_{\m A} \mez , \mez {\bm v}(t \leq t_0) \; = \; {\bm 0}
         \mez , \mez \beta_{{\bm k}\wp}\n(t_0) \; = \; 0 \mez .
      \ee
      Here $| \varphi_{\rm ground} \ra_{\m A}$ stands for the ground state of the free atomic Hamiltonian
      $\frac{\hat{\bm p}^2}{2\,m}+V\m(\hat{\bm x})$.\\ Moment of reflection reveals that
      ${\bm v}(t)={\bm 0}$ at all the time instants, while (\ref{Appendix-6-EOM-psi-take-2}) boils down into
      \be \label{Appendix-6-EOM-psi-take-2-ground}
         i \hbar \, \partial_t \, | \tilde{\psi}_t \ra_{\m A} \; = \; V\m(\hat{\bm x}) \; | \tilde{\psi}_t \ra_{\m A} \mez ;
      \ee
      giving immediately
      $| \tilde{\psi}_t \ra_{\m A} \, = \, e^{-\frac{i}{\hbar}V\n(\hat{\bm x})(t-t_0)} \, | \varphi_{\rm ground} \ra_{\m A}$.\\
      Showing that the resulting position probability density $|_A\m\la {\bm x} | \tilde{\psi}_t \ra_{\m A}|^2$
      remains constant in time\\ and coincides with $|_A\m\la {\bm x} | \varphi_{\rm ground} \ra_{\m A}|^2$.\\
      In summary, for our specific initial conditions (\ref{Appendix-6-ground-ics}),\\
      the mean field theory provides a rather trivial (though physically plausible) solution.\\
      $[\,$The extra ${\bm x}$-dependent phase term $e^{-\frac{i}{\hbar}V\n(\hat{\bm x})(t-t_0)}$
      carried by $| \tilde{\psi}_t \ra_{\m A}$ can arguably be interpreted here\\
      \phantom{$[\,$}as a rudiment of radiation reaction dressing.$\,]$\\
      A much more physically interesting dynamics would follow e.g.\\
      in the case of atomic resonances\\ (excited atomic levels becoming unstable due to spontaneous emission),\\
      or in the situation when laser driving is present\\
      (laser light would be brought into the game via setting $\beta_{{\bm k}\wp}\n(t_0) \neq 0$).\\
      $[\,$Working out such examples more explicitly can actually be a nice exercise or a mini-project.$\,]$
\end{itemize}

\vspace*{+8.00cm}

\end{document}